\documentclass[
11pt, 
english, 
singlespacing, 
headsepline, 
]{ClassFile} 

\usepackage[utf8]{inputenc} 
\usepackage[T1]{fontenc} 

\usepackage{mathpazo} 

\usepackage[backend=biber,style=numeric-comp,sortcites,sorting=none,natbib=true]{biblatex} 

\usepackage[all]{xy} 
\usepackage{listings} 
\usepackage{algorithm} 
\usepackage{algorithmic} 
\usepackage[autostyle=true]{csquotes} 

\usepackage{enumitem}

\usepackage{amsmath,amssymb}

\usepackage{color}
\usepackage{soul}

\thesistitle{Fibration, Nexus and Cosmological Composite Topological Defects in Uniaxially Disordered Superfluid $^3$He} 
\examiner{} 
\degree{Doctor of Philosophy} 
\author{Kuang. Zhang} 
\addresses{} 
\university{{University of Helsinki}}
\department{{Department of Mathematics and Statistics, University of Helsinki}}
\group{{Topological Quantum Fluids (ROTA) Group, Aalto University}}
\faculty{{School of Applied Physics, Aalto University}}

\AtBeginDocument{
\hypersetup{pdftitle=\ttitle} 
\hypersetup{pdfauthor=\authorname} 
\hypersetup{pdfkeywords=\keywordnames} 
}

\begin{document}

\frontmatter 

\pagestyle{plain} 


\begin{titlepage}
\begin{center}

\vspace*{.06\textheight}
{\scshape\LARGE  }\vspace{1.5cm} 
\textsc{}\\[0.5cm] %

\HRule \\[0.4cm] 
{\huge \bfseries \ttitle\par}\vspace{0.4cm} 
\HRule \\[1.5cm] 
 
\begin{minipage}[t]{0.4\textwidth}
\begin{center} \large
\emph{Author:}\\
\authorname 
\end{center}
\end{minipage}
 
\vfill

\textit{ }\\[0.4cm]
\groupname\\\deptname\\[2cm] 
 
\vfill

{\large \today}\\[4cm] 

\vfill
\end{center}
\end{titlepage}


\begin{abstract}
\addchaptertocentry{\abstractname} 
The liquid Helium-3 is the unique member in the many bodies systems and condensed matter systems. This is not only because it 
never freezes to solid even in zero temperature by the existence of strong enough quantum fluctuation, but also because it has complicated enough symmetry breaking patterns to different superfluid phases. The related discussions started from 1960s till now, and lots of intriguing physics have been proposed or been observed, such as novel topological defects and chiral anomaly of quisaiparticle induced by the textures of order parameters. 

In last decade years, the impurity with nanometer length scales, i.e., aerogel was introduced into the liquid Helium-3 to modify the scattering properties of Helium-3 quasiparticles. One of them, the nafen, which is collection of thin $Al_{2}O_{3}$ strands is used in the ROTA's experiment, and generates a series of axially polarized new phases, in which the half quantum vortices have been observed.  

The half quantum vortices are firstly stabilized in polar phase, which is also the first observed nafen distorted superfluid phase of Helium-3. This novel string defects are analogy of the cosmological strings which probably appear in the early universe. The existence of half quantum vortices provides the clue about the existence of other cosmological objects in the nafen distorted superfluid. To solve this question, we analyzed the possible symmetry breaking patterns in detail by using the algebraic topology and group theory. It turns out that the fibrations of the degenerate parameter spaces of symmetry breaking patterns dominate the existence of composite cosmological defects in the successive symmetry breaking of nafen distorted Helium-3.     

In this review, the fibrations of degenerate parameter spaces in the successive symmetry breaking phase transition of nafen distorted superfluid Helium-3 are reveled by using the homotopy theory and group theory. The results of this deep mathematics i.e., the composite string monople (Numbu monopole) and the string wall (Kibble-Lazarides-Shafi domain wall) are described topologically by using the exact sequences of homotopy groups. To compare our model with ROTA's experiment of string wall, we demonstrate how the topological defects with coherent length scales extend to mesoscopic spin solitons in equilibrium states. After this, the equilibrium free energies, which determines the extended equilibrium configurations, are evaluated by non-linear numerical optimization algorithm. Based on these equilibrium configurations, we calculated the spectrum of spin dynamical response of system under weak magnetic driving. The results exactly coincide with the experimental observations.

In addition, the implements of the numerical algorithms, which were used to evaluated the equilibrium configurations of spin solitions and the spin dynamical response, also generate a useful programs library. The massive applications of the functional analysis and theory of integral operators in the algorithms and their implements provide a instance, through which the physical expressions can be mapped to data objects of computer and evolve as physical expressions require.  
\end{abstract}


\begin{acknowledgements}
\addchaptertocentry{\acknowledgementname} 
We especially thank the instructive and inspiring discussions and comments from professor Grigory.~E. Volovik during the process of this work. We thank Hiromitsu Takeuchi, Mikhail Silaev, Jaakko Nissinen, Vladislav Zavyalov and professor Erkki. V. Thuneberg for important discussions. We also thank professor Vladimir. B. Eltsov, Jere. T. M\"akinen and Juho. Rysti for instructive discussions about experiments of polar distorted B-phase.  This work has been supported by the European
Research  Council  (ERC)  under  the  European  Union’s
Horizon 2020 research and innovation programme (Grant Agreement No. 694248).
\end{acknowledgements}

\tableofcontents 




\mainmatter 

\pagestyle{thesis} 


\chapter{Introduction} 

\label{Chapter1} 


\newcommand{\keyword}[1]{\textbf{#1}}
\newcommand{\tabhead}[1]{\textbf{#1}}
\newcommand{\code}[1]{\texttt{#1}}
\newcommand{\file}[1]{\texttt{\bfseries#1}}
\newcommand{\option}[1]{\texttt{\itshape#1}}

The North American Nonahertz Observatory for Gravitational Wave (NANOGrav) team reported its finding from the accumulated data in 12.5 years at the end of 2020 \cite{Zaven2020, Simone2021, EllisLewicki2021, Luca2021, VaskonenVeermae2021}. Similar to the famous LIGO/Virgo, NANOGrav is a gravitational wave observatory. What the significant difference between them is that the NANOGrav is sensitive to gravitational wave with frequency in nanohertz scales. To implement this, scientists actually setup the observatory by utilizing our galaxy. Forty fives pulsars were chosen to form the so called pulsar timing array, and their time ticks will change if a gravitational wave signal passes between the earth and the pulsars array. By analyzing the 12.5 years observation data, NANOGrav team reveals the possible existence of a stochastic background of gravitational wave, and this discovery takes the cosmological strings to the table of candidates of the sources for gravitational wave \cite{Kibble1976, VilenkinShellard1994, HindmarshKibble1995}.\\[0.1cm]

The cosmological strings typically appear in many different Grand Unify Theory (GUT) models and String-Theories \cite{ BattyeRobinsonAlbrecht1998, AvelinoShellardWuAllen1998, MajumdarDavis2002, SarangiTye2002, PogosianTyeWassermanWyman2003,DvaliVilenkin2004, JeannerotRocherSakellariadou2003}. Usually, these topological defects are proposed to appear when the primary GUT symmetry groups spontaneously breaks into its subgroups in the early age of universe after big bang \cite{ArkaniHamedDimopoulosDvali1998, ArkaniHamedDimopoulosDvali1999, RandallRSundrum1999}. Even the existence of cosmological strings is controversial \cite{Kibble2004, Durrer2002, ContaldiHindmarshMagueijo1999, BouchetPeterRiazueloSakellariadou2002}, some of them, which have intriguing and novel properties, still have been discussed intensively in last decades, not only in cosmology, but also in condensed matter physics \cite{Hindmarsh1995,VilenkinBook2000,Cui2018,Matsunami2019,Auclair2020,Kibble1982a,Kibble1982b,Ruutu1996,KibbleVolovik1997,SirThomasWalterKibble2021}.  The Alice strings or half quantum vortices (HQVs) are one example of this kind of string defect \cite{Kiskis1978, Schwarz1982}. Alice strings appear when the residual symmetry of a symmetry breaking phase is disconnected and contains a $\mathbb{Z}_{2}$ subgroup, such that its two elements belong to the disconnected parts of the residual symmetry group respectively \cite{Kiskis1978}. In the experiment of ROTA group of Aalto University, this is implemented in the polar phase of $^3$He, which results from the modifications of microscopic scattering properties by axially polarized impurity i.e., nafen aerogel \cite{Autti2016}.\\[0.1cm] 

The idea, that introducing aerogel to modify the microscopic scattering properties of $^3$He quasiparticles, was discussed and intensively practiced around late 1990s to 2000s by using isotropic $SiO_{2}$ aerogel \cite{Porto1995,Thuneberg1998,Sprague1995,Alles1999,Barker2000,Baumgardner2004,
Gervais2001,Fisher2003,Kotera2003,Fomin2003,Halperin2004,Aoyama2005,Higashitani2005,Bunkov2005,Fomin2005,Kazushi2006,
Volovik2008,Fomin2008,Surovtsev2008,Bradley2008,Dmitriev2008,Kunimatsu2008,
Nakagawa2008}. Because the characteristic length of aerogel is less than coherent length of Cooper paring of superfluid $^3$He, this impurity may dramatically modify the stable phase and the equilibrium phase diagram of superfluid $^3$He \cite{Sauls2013,Fomin2013,Fomin2014,YangIkeda2014,Ikeda2015,Halperin2014, Halperin2015}. In the case of nafen aerogel, which consists of randomly distributed parallel $Al_{2}O_{3}$ strands, the polar phase was predicted as the possible new phase and later be identified experimentally \cite{Kazushi2006,YangIkeda2014,Askhadullin2012,Yudin2015}. This new phase dominates the most part of the phase diagram, and the reason of its domination was explained by the Anderson-Fomin theorem \cite{Anderson1959,Fomin2018,Fomin2020,Ikeda2019b}. Moreover, the new phase diagram shows the two-step successive symmetry breaking phase transitions via polar phase is possible \cite{Kazushi2006,Fomin2013,Yudin2012,Yudin2014}. When the temperature of polar phase superfluid reaches the transition temperature of polar distorted B-phase (PdB), the second time symmetry breaking phase transition occurs \cite{Yudin2014,Makinen2019}. If this transition happens just below the transition temperature of phase transition from polar phase to PdB phase, a very fantastic composite topological object -- the Kibble-Lazarides-Shafi (KLS) string domain wall will appear in the distorted sufperfluid $^3$He \cite{Makinen2019}.\\[0.1cm]

This novel composite cosmological object and the corresponding symmetry breaking pattern was introduced in the 1980s. As an example, Kibble, Lazarides and Shafi discussed the two-step symmetry breaking of $spin(10)$ gauge theory to $H=\{H_{0}, K\}$, where $H_{0}=spin(6){\otimes}spin(4)$ and $K=H_{0}i\sigma_{67}$ \cite{Kibble1982a, Kibble1982b, KibbleBook2000}. Because the charge conjugation $C = i\sigma_{67}{i \sigma_{23}}$ is an element of $K$, this symmetry breaking generates Alice string as we have mentioned. At the second stage of symmetry breaking form $H$ to the $SU(3){\times}SU(2){\times}U(1)$, the charge conjugation symmetry $C$ is broken, as a result, domain wall appears and is terminated on the Alice string. Following the idea of KLS, similar two-step symmetry breaking pattern were discussed in different unified gauge models. And physicists soon found that this mechanism induces  domain wall problem of the axion solution of the CP violation in QCD \cite{ZeldovichSoviet1974,Zeldovich1974,Vilenkin1982,Sikivie1982,Sikivie1985,Sikivie2008,Kim2010}. In the axion solution, two phase transitions successively occur in our universe during its temperature cools down. In the first time transition, the $U(1)_{PQ}$ symmetry of Peccei-Quinn mechanism  spontaneously breaks, then the axion and string defect appear. When the cosmic temperature reaches the QCD temperature, the $U(1)_{PQ}$ symmetry breaks to discrete symmetry and then the domain wall appears. As a result, the cosmic strings formed in the first time symmetry breaking convert to string domain wall under QCD temperature \cite{Vilenkin1982,Sikivie2008}. The universe which has this structure will be very different with what we have observed. In order to solve this problem, a lots of ideas have been reported, and the corresponding decay dynamics of the string wall system also be researched \cite{Lazarides1982,Lazarides1985,Sato2018,Chatterjee2019,Andrea2019}.\\[0.1cm]

In spite of the controversy about the route of the string walls decay, the story looks running well until we take the Nambu monopole (string monopole) into account \cite{Nambu1977}. This unusual monopole may be generated by different mechanism beside the two-step successive symmetry breaking pattern \cite{VolovikBook2009}. However, it has an unanticipated similarity with the KLS string domain wall, that is, Nambu monopole is the composite cosmological object formed by combination between zero dimensional object (monopole) and one dimensional object (string). While the KLS string domain wall is formed by combination between one dimensional object (string) and two dimensional object (domain wall). Considering about that the polar phase, though which the successive symmetry breaking occurs, also has monopole because its $\pi_{2}$ groups is non-trivial, we started to get the idea that the monopole of polar phase may convert to Nambu monopole in a similar way, in which Alice string converts to KLS string domain wall when the second symmetry breaking happens. Fortunately, the answer is yes. In the system with two-step successive symmetry breaking pattern, there is general mechanism to guarantee the cosmological objects appearing in the first time symmetry breaking convert to composite topological objects during the second time symmetry breaking occurs. And this guarantee has a deep mathematical origin in algebraic topology \cite{HatcherBook2002}.\\[0.1cm]  

In this review, we demonstrate in details what this guarantee is by using the algebraic topology and group theory. The significant tools are the relative homotopy groups and the exact sequences consisting of homotopy groups and relative homotopy groups of the degenerate parameter spaces of PdB order parameters. We will see how the composite topological defects are classified by relative homotopy groups and characterized by boundary homormophism of the corresponding exact sequences. Originally the classification  
in terms of the relative homotopy groups has been used if there is the hierarchy of the energy scale or length scales in physical system \parencite{MineyevVolovik1978,Mermin1979,Michel1980}, when each energy scale has its own well defined vacuum manifold $R_i$ -- the space of the degenerate states. In our case of two-step successive symmetry breaking transitions, two energy scales arise in the vicinity of the second transition. There, the coherence length related to the first symmetry breaking is much smaller than the coherence length related to the second symmetry breaking. This gives rise to two well defined degenerate parameter spaces, and allows us to apply the relative homotopy groups. By directly calculating the exact sequences and relative homotopy groups $\pi_n(R_1,R_2)$, where $R_{1}$ and $R_{2}$ are degenerate spaces of PdB phase generated from different vacua, we found the string monopole (Nambu monopoles) and KLS string wall are classified by $\pi_2(R_1,R_2)$ and $\pi_1(R_1,R_2)$ groups respectively. Moreover, we found an amazing fact, that is $\pi_{n}(R_{1},R_{2})$ are always isomorphic to $\pi_{n}(R_{P})$, which are the homotopy groups of topological defects in polar phase, i.e.,
\begin{equation}
\pi_{n}(R_{P}) = \pi_{n}(R_{1},R_{2}),
\end{equation}      
where $n \in \mathbb{Z}^{+}$ and $n \geq 1$. This means the topological defects in polar phase convert to composite cosmological defects when the second time symmetry breaking occurs! Facing to this amazing fact, all physicists must ask why? Our work reveal that there is fibration between  PdB phase vacuum manifold and  polar phase vacuum manifold, and this fibration determines that $\pi_{n}(R_{1},R_{2})$ must equal to $\pi_{n}(R_{P})$. \\[0.1cm]

The rest parts of review are organized as follows. In Chapter.~\ref{Chapter2} we consider the conventional scheme of the symmetry breaking and the vacuum manifolds of different superfluid phases appeared in the two-step successive transitions in the polar distorted $^3$He. The topological defects in these phases are described in terms of the conventional homotopy groups.
In Chapter.~\ref{Chapter3} we discuss the composite topological objects in the vicinity of the second transition. We use the relative homotopy groups and corresponding  exact sequence of homomorphisms to classify the composite objects, which are topologically stable in the vicinity of the transition. We demonstrate how fibration of the degenerate parameter spaces happens and what it results in. 
In order to test our theory with experiment about the KLS string wall of ROTA group, we further discuss the equilibrium extended structures of the KLS string wall with length scale around the dipole length in Chapter.~\ref{Chapter4}. In Chapter.~\ref{Chapter5}, we calculate NMR frequency shift -- the experimental observable of the extended structures of KLS string wall by using the linear response theory. The results of numerical simulations exactly coincide with the experimental observations. We summarize the researches and discuss the future work which deserve to be focused on in Chapter.~\ref{Chapter6}. In the appendices, we provide some essential background discussions about technical details. \\[0.1cm]


\chapter{Symmetry Breaking Phase Transitions and Topological Defects in Polar Distorted Superfluid $^3$He} 

\label{Chapter2} 



\section{The Primary Symmetry Groups of Normal Phase Vacuum of Nafen Distorted $^3$He}
The continuous phase transition is understood as spontaneous symmetry breaking by order parameters about a primary symmetry group $G$. In most case, order parameters are elements of (complex) vector spaces in which the primary symmetry group $G$ is represented. For example, the magnetization vector of ferromagnetic materials is element of 3-dimensional Euclidean space, where the $SO(3)$ spin rotation is represented. And the order parameter of s-wave superconductor is element of complex number set, which provides $S=0$ and $L=0$ representation of $SO_{S}(3) \times SO_{L}(3) \times U(1)$ group, where $S$ and $L$ are quantum numbers of spin and orbital angular momentums. \\[0.1cm]

In $^3$He liquid at low temperature, the basis vectors of vector spaces are eigen-functions of $S=1$ and $L=1$, thus the order parameters space consists of two 3-dimensional vector spaces and the phase space. This gives rise to the bilinear complex-valued order parameter i.e., dyadic tensor $A_{{\alpha}i}$ \cite{VollhardtWolfle1990}, which transforms under the action of spin, orbital and phase rotations of the primary group $G$. Stabilizer of those actions, which is collection of the residual symmetry transformations of given order parameter, forms the residual symmetry group $H$ of superfluid phase of $^3$He. \\[0.1cm] 

In our case, the symmetry group  $G$ of  normal liquid $^3$He in the "nematically ordered" aerogel with the uniaxial anisotropy is different from that in the bulk  $^3$He \cite{VollhardtWolfle1990}. In ROTA's experiment, the nafen aerogel consisting of parallel-distributed $Al_{2}O_{3}$ strands with diameter around $8nm$ is immersed in to liquid $^3$He \parencite{Makinen2019,Yudin2012,Askhadullin2008,Askhadullin2012}. The coupling energy between the orbital vector of order parameter and nafen strand is proportional to 
\begin{equation}
\eta_{ij} A_{{\alpha}i} A_{{\alpha}j}^{*},
\label{NafenCoupling}
\end{equation}
where $\eta_{ij}$ is tensor describing the coupling between aerogel and orbital degree of freedom (DoF) for Cooper paring. In the case of spatially isotropic aerogel, $\eta_{ij}$ is an isotropic tensor and can be written as
\begin{equation}
\eta^{isotrpic}_{ij} = \eta_{0} \delta_{ij},
\label{etaiso}
\end{equation}
where $\eta_{0}$ is the eigenvalue of $\eta^{isotrpic}_{ij}$. When the aerogel turns to be axially polarized like nafen aerogel, the anisotropic traceless part appears and then $\eta_{ij}$ can be written as   
\begin{equation}
\eta_{ij} = \eta_{0} \delta_{ij} + 3\kappa (\hat{u}_{i} \hat{u}_{j} -\frac{1}{3}\delta_{ij}),
\label{eta}
\end{equation}
where $\kappa = (\eta_{0} - \eta_{\bot})$ and $\eta_{\bot}$ is the eigenvalue of $\eta_{ij}$ along the directions perpendicular to the axially polarized axis i.e., direction of nafen strands \parencite{Siegfried2015}.
The dimensionless parameters $\eta_{0}$ and $\kappa$ describe the common gap shifts of all angular momentum states in the presence of aerogel and the discrepancies of the gap shifts between different angular momentum states because of the uniaxial anisotropy respectively \cite{Yudin2014,Fomin2003,Fomin2005,Fomin2013}. The unit vector $\hat{\mathbf{u}}$ represents the direction of nafen strands. \\[0.1cm]

In the case with $\kappa < 0$, the $l_{z} = 0$ angular momentum state has biggest gap and highest superfluid transition temperature \cite{Fomin2013,Fomin2014}. As a result, polar phase with uniaxially polarized orbital vector $\hat{\mathbf{z}}$ is system-favorable when temperature decreases \cite{Autti2016}. The relative direction between $\hat{\mathbf{z}}$ and $\hat{\mathbf{u}}$ of bulk equilibrium state is determined by minimum of Eq.~(\ref{NafenCoupling}). For polar phase with order parameter $A_{\alpha i} \sim  \Delta_{P}  \hat{d}_{\alpha} \hat{z}_{i}$, Eq.~(\ref{NafenCoupling}) suggests $\hat{\mathbf{z}} \parallel \hat{\mathbf{u}}$ immediately.  
These influences induced by the nafen aerogel to superfluid $^3$He significantly modify the symmetry of this p-wave system. In fact the orbital three dimensional rotation symmetry group $SO_{L}(3)$ of pure $^3$He is reduced to $O_{L}(2)$ because the orbital $\hat{\mathbf{z}}$ vector is parallel to the nafen strands in all cases \cite{Makinen2019, Zhang2020}.   
 This situation is equivalent to say the normal phase vacuum has the following primary symmetries:
\begin{equation}
O_L(2)\times SO_S(3)\times U(1) \times T \times P
 \,,
\label{Ggeneral}
\end{equation}
where
$SO_{S}(3)$ is the group of spin rotations; $U(1)$ is the global gauge group of phase degree of freedom, which is broken in superfluid states; $T$ is time reversal symmetry; $P$ is parity; $O_{L}(2) \cong SO_{L}(2){\rtimes} C_{2x}^L $ where $C_{2x}^L $ is $\pi$ rotation in orbital space about the traverse axis. \\[0.1cm]

Microscopic theories have predicted that the polar phase is the stable phase in this nafen distorted system \cite{Kazushi2006,YangIkeda2014} and  this predictions was experimentally identified later \cite{Askhadullin2012, Yudin2015}. However, the unexpected part of story is polar phase dominates a huge part of phase diagram.  Later it became clear that the main reason of the domination of the polar phase in nafen distorted $^3$He is the extension of the Anderson theorem \cite{Anderson1959} -- the Anderson-Fomin theorem of the polar phase with columnar impurities i.e., the transition temperature to the polar phase is practically not suppressed by the strands of nafen \cite{Fomin2018,Fomin2020,Ikeda2019b}, as distinct from the other superfluid phases.  Similar extension of the Anderson theorem was also discussed in multi-orbital superconductors \cite{Ramires2018}.
Recently, another signature of the Anderson-Fomin theorem in polar distorted $^3$He is the detected $T^3$ dependence of the gap amplitude that results from the non-suppression of Dirac nodal line of the spectrum of Bogoliubov quasipartilces in the polar phase \cite{Eltsov2019}. \\[0.1cm]

In what follows, we ignore the time reversal symmetry, since it is not broken in the polar and in PdB phases, and also ignore the parity $P$ which is reduced to $Pe^{i\pi}$ in all $p$-wave superfluid phases, where $e^{i\pi}$ is the $\pi$-rotation in phase space. Also, because we focus on the topological objects related to the spin and $U(1)$ gauge parts of the order parameter, the $\mathbb{Z}_{2} = \{0,C_{2x}^{L}\}$ symmetry coming from $C_{2x}^L$ could be neglected in the rest parts of this review. Then the referring starting group $G$ of symmetry breaking schemes in this review is
\begin{equation}
G = SO_L(2)\times SO_S(3)\times U(1)
 \,.
\label{G}
\end{equation}
Starting from this normal phase vacuum, we discuss three different of phase transition: (a) from the normal phase to the polar phase; (b)  from the polar phase to the PdB phase;
and (c) Direct transition from the normal phase to the PdB phase. In Sec.~\ref{SecConventional2} we consider the topological objects related to these symmetry breaking scenarios, using the conventional approaches of homotopy groups. \\[0.1cm]

\section{Conventional Symmetry Breaking Scheme of Nafen Distorted $^3$He and Vacuum Manifolds}
\label{SecConventional}

\subsection{Transition from normal phase to polar phase}

As we mentioned, the order parameter in the $p$-wave spin-triplet superfluids is the dyadic tensor $A_{{\alpha}i}$, which transforms as a vector under spin rotation  (the greek index) and as a vector under orbital rotations (the latin index). In the polar phase it has the form:
\begin{equation}
A^{P}_{{\alpha}i}=\Delta_P \hat{d}_{\alpha} \hat{z}_{i}e^{i\Phi}
 \,,
\label{AP}
\end{equation}
where $\Phi$ is the phase, $\hat{d}_{\alpha}(\equiv {\hat{\mathbf{d}}})$ and $\hat{z}_{i}$($\equiv \mathbf{u}$) are unit vectors of spin and orbital uniaxial anisotropy  respectively, 
and $\Delta_P$ is the gap amplitude. The residual symmetry group of the polar phase, the stabilizer of the order parameter in Eq.~(\ref{AP}),  is
\begin{equation}
H_{\rm P} = SO_L(2)\times SO_S(2) {\rtimes} \mathbb{Z}_{2}^{S-{\Phi}} \subset G
 \,.
\label{HP}
\end{equation}
Here $\mathbb{Z}_{2}^{S-{\Phi}} = \{1,C_{2x}^Se^{i\pi}\}$, where $C_{2x}^S$ is $\pi$-rotation of the  vector $\hat{\mathbf{d}}$ about perpendicular axis and $e^{i\pi}$ is  the phase rotation by $\pi$, i.e.  $\Phi \rightarrow \Phi +\pi$. Then the vacuum manifold of the polar phase is given as
\begin{equation}
R_{\rm P} \cong G/H_{\rm P} \cong (S^2\times  U(1))/{\mathbb{Z}}_2 
\,.
\label{RP}
\end{equation}
The coherence length $\xi ={\hslash}v_F/\Delta_P$ in the polar phase is the smallest length scale in our question, which determines the size of singular (hard core) topological defects in the polar phase.

\subsection{From polar phase to PdB phase}

Let us now consider the second symmetry breaking phase transition: from the polar phase vacuum with fixed $\hat{\mathbf{d}}$ and $\Phi$ to the PdB phase.  In the vicinity of this transition the order parameter in Eq.~(\ref{AP}) acquires the symmetry breaking term with amplitude $q\ll 1$:
\begin{equation}
A^{PdB}_{{\alpha}i}=e^{i\Phi}\Delta_P[\hat{d}_{\alpha} \hat{z}_{i} + q (\hat{\mathbf{e}}^{1}_{\alpha} \hat{x}_{i} + \hat{\mathbf{e}}^{2}_{\alpha} \hat{y}_{i})]
 \,.
\label{ApdbP}
\end{equation}
Here $\hat{\mathbf{e}}^{1}$, $\hat{\mathbf{e}}^{2}$ and $\hat{\mathbf{d}}$ are three orthogonal vectors in spin space. The corresponding coherence length of the second transition $\xi/q$ is large in the vicinity of this transition. This provides the hierarchy of the length scales, $\xi$ and $\xi/q\gg \xi$. 

The residual symmetry subgroup of the PdB phase in the symmetry breaking from the polar phase is
\begin{equation}
H_{\rm PdB} = SO_{S-L}(2) \subset H_{\rm P} 
 \,,
\label{HPdB}
\end{equation}
where $SO_{S-L}(2)$ represents the common rotations of spin and orbital spaces.
 The manifold of the vacuum states, which characterizes the second symmetry breaking is:
\begin{equation}
R_2 \equiv R_{{\rm P}\rightarrow {\rm PdB}} = H_{\rm P}/H_{\rm PdB} = \cong SO_{L-S}(2) \times  \mathbb{Z}_{2}^{S-\Phi}. 
\label{PtoPDBQuientSpaceMotherFucker}
\end{equation}
Here PdB phase breaks one of $SO(2)$ symmetries of $H_{P}$ respecting to relative rotations of spin and orbital spaces.

\subsection{From normal phase to PdB phase}
\label{NormaToPdB}

Here we consider the two-step symmetry breaking form normal phase to PdB phase with a general view point, which the parameter $q$ is not necessarily small. In this general case there is only a single length scale which is relevant i.e., the coherent length $\xi_{PdB}$ of Cooper paring in PdB phase, and thus this situation becomes similar to that of the direct transition from the normal state to the PdB phase, $G\rightarrow H_{\rm PdB}$. The order parameter Eq.~(\ref{ApdbP}) of the PdB phase could be written as
\begin{equation}
 A^{PdB}_{{\alpha}i} = e^{i\Phi}[\Delta_\| \hat{d}_{\alpha} \hat{z}_{i} + \Delta_{\bot} (\hat{\mathbf{e}}^{1}_{\alpha} \hat{x}_{i} + \hat{\mathbf{e}}^{2}_{\alpha} \hat{y}_{i})]
 \,,
\label{ApdbPa}
\end{equation}
where $\Delta_{\bot}\leq\Delta_{\|}$. The corresponding residual symmetry group is still Eq. (\ref{HPdB}),
and the vacuum manifold of PdB phase in this scenario  of symmetry breaking is:
\begin{equation}
R_1 \equiv R_{{\rm normal}\rightarrow {\rm PdB}} \cong G/H_{PdB} \cong SO_{S-L}(3)\times U(1)
 \,.
\label{RPdB}
\end{equation}

\begin{figure*}[htbp] 
\centerline{\includegraphics[width=9cm,height=7cm]{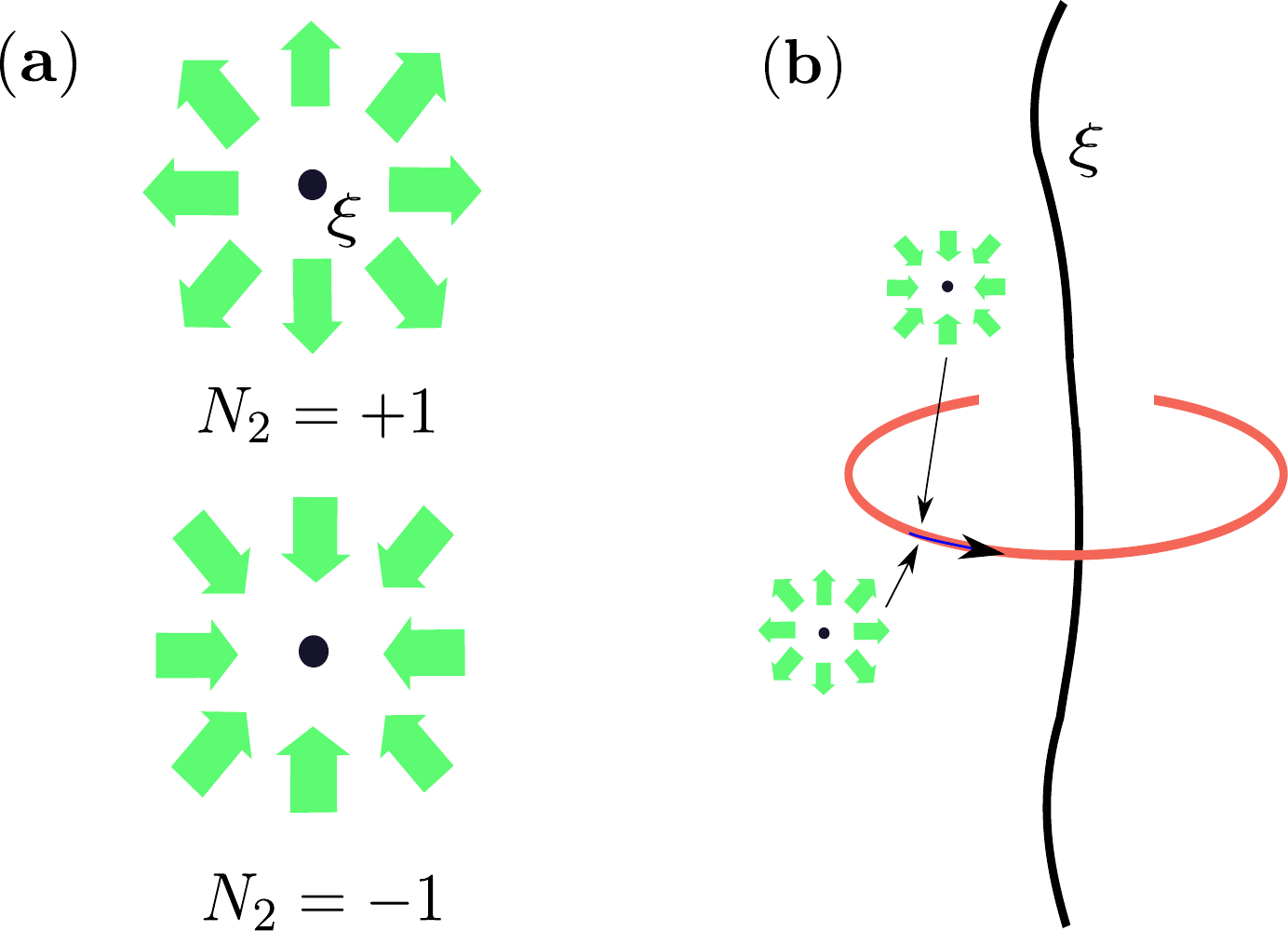}}
\caption{Topological defects in the polar phase. We omit the free conventional vortices of group $\mathbb{Z}$. (a) $\hat{\mathbf{d}}$-vector monopole and anti-monopole described by the homotopy group $\pi_{2}(R_{P})$ with topological charges $N_{2}=\pm 1$ respectively. Their core size are on the order of the coherence length $\xi$.  The green arrows depict the configuration of $\hat{\mathbf{d}}$ vectors. (b) Half-quantum vortex (HQV) described by the $\mathbb{Z}_{2}$ subgroup of $\pi_{1}(R_{P})$ with core size $\sim \xi$. This object also got the name "Alice string", because the $\hat{\mathbf{d}}$ monopole transforms to the anti-monopole after going around the HQV,  in the same manner as it happens for the charge going around Alice string \cite{Schwarz1982}. The red circle shows the path.
}
\label{polar_defect}
\end{figure*} 
    
\section{Topological Objects Generated from Different Symmetry Breakings}
\label{SecConventional2}
In this Section we consider the topologically stable defects, which emerge at each of three symmetry breaking transitions discussed in Sec. \ref{SecConventional}.

\subsection{Defects in polar phase}
\label{SubNtoP}

The polar phase vacuum manifold Eq. (\ref{RP}) has the homotopy groups
\begin{equation}
\pi_1(R_{\rm P}) = \tilde{\mathbb{Z}}\,\,, \,\, \pi_2(R_{\rm P}) = \mathbb{Z} \,\,, \,\, \pi_0(R_{\rm P}) = 0
 \,.
\label{HomotopyP}
\end{equation}
And the calculation details of $\pi_{n}(R_{P})$ are put into Appendix.~\ref{LESAndSESofRP}. The group $\pi_1(R_{\rm P}) = \tilde{\mathbb{Z}}=\{n/2|n\in\mathbb{Z}\}$ includes the integers of the group $\mathbb{Z}$ via inclusion map: $n\in\mathbb{Z} \hookrightarrow n \in\tilde{\mathbb{Z}}$, which describes the free quantized vortices with integer winding number. More significantly, $\pi_1(R_{\rm P})$ contains the set of half-odd integers, i.e., $\{n+1/2|n\in\mathbb{Z}\}$. This subset describes vortices containing HQV, which has one-half circulation of a conventional quantized vortex. The reason which polar phase has HQVs can be retrospected to the discussion about the disconnected residual symmetry  of a given symmetry breaking phase in Chapter.~\ref{Chapter1}. From the Eq.~(\ref{HP}), we can see the $\mathbb{Z}_{2} = \{1,C_{2x}^Se^{i\pi}\}$ is subgroup of $H_{P}$, then there are two possible ways to move around linear defect in polar phase, one is moving back to the starting point state, while the other one is continuously moving to the $\pi$-rotation state (analog to Charge conjugation state) as shown in Fig.~\ref{polar_defect}(b), which is transferred from the starting point state by action of $C_{2x}^Se^{i\pi}$. This means the HQVs with the topological charges $N=\pm 1/2$ are the analogs of the cosmological Alice strings \parencite{Kiskis1978,Schwarz1982}. \\[0.1cm] 

The group $\pi_2(R_{\rm P})=\mathbb{Z}$ describes the hedgehogs (monopoles) in the $\hat{\bf d}$-field, as shown in Fig. \ref{polar_defect}(a). The core sizes of  vortices and monopoles in polar phase are on the order of the coherence length $\xi = {\hslash}v_{F}/\Delta_{P}$. 
In the presence of HQVs, The topological classification of hedgehogs is modified to $\mathbb{Z}_{2}$ from $\mathbb{Z}$ because of the Charge conjugation symmetry. The monopole transforms to anti-monopole when circling around the Alice string (HQV), and thus in the presence of HQVs the hedgehogs (monopoles) of the group $\mathbb{Z}$ is degenerated with the anti-monopole of group $\mathbb{Z}$.  
This phenomenon is an example that $\pi_2(R)$ group may be influenced by $\pi_1(R)$ topology \cite{VolovikMineev1977}. \\[0.1cm]

The HQVs have been identified in NMR experiments of polar phase \cite{Autti2016}. By applying static magnetic field tilted with respect to nafen strands, one creates the soliton as extended structure, which has dipole length scale, attached to the HQVs. This mesoscopic extended structure produces the measured frequency shift in the NMR spectrum under RF magnetic drive. Unfortunately, Hedgehogs (monopoles) are still not identified in superfluid $^3$He. 
\subsection{Defects in PdB phase with $R_{1}$}
\label{Sec:defectsPdBfromN}
The vacuum manifold $R_{1}$ of the PdB phase in Eq. (\ref{RPdB}) has homotopy groups
\begin{equation}
\pi_1(R_1) = \mathbb{Z} \times \mathbb{Z}_{2} \,\,, \,\, \pi_2(R_1) = 0 \,\,, \,\, \pi_0(R_1) = 0
 \,.
\label{HomotopyPdB}
\end{equation}
Different with the polar phase, which has QHVs and monopoles, the topologically protected defects of PdB phase are integer-quantized phase vortices of $\mathbb{Z}$ and spin vortices with $\mathbb{Z}_{2}$ in the general case as we mentioned in Sec.~\ref{NormaToPdB}. 
The hard core (sized by $\xi$) defects (HQVs and hedgehogs) of polar phase are not supported by topology any more. Moreover, the new topological object -- the  $\mathbb{Z}_{2}$ spin vortices are similar to that which have been observed in the bulk B-phase \cite{Kondo1992}. 
\subsection{Defects in PdB phase with $R_{2}$}
The vacuum manifold $R_{2}$ of the PdB phase generating at the symmetry breaking transition from the polar phase vacuum in Eq.~(\ref{PtoPDBQuientSpaceMotherFucker}) has homotopy groups:
\begin{equation}
\pi_1(R_2) = \mathbb{Z} \,\,, \,\, \pi_2(R_2) = 0\,\,, \,\, \pi_0(R_2) = \mathbb{Z}_2
 \,.
\label{HomotopyPdBfrom Polar}
\end{equation}
These homotopy groups are responsible for the topological defects formed in the symmetry breaking transition from the fixed degenerate vacuum of the polar phase (with $\hat{\bf d}={\rm const}$ and $\Phi={\rm const}$) to the PdB phase. These topological defects are such important because they connect to defects with higher spatial dimension and then form the composite topological defects in the vicinity of the second time symmetry breaking transition. We will discuss this soon in Chapter.~\ref{Chapter3}. Before that, Let us consider them in details separately.

\subsubsection{Spin vortices}

The homotopy group $\pi_1(R_{2}) = \mathbb{Z}$ describes the spin vortices with $2\pi n$ rotation of vectors ${\hat{\bf e}}^1$ and  ${\hat{\bf e}}^2$ about the fixed $\hat{\bf d}$ vector of the polar phase. The winding number is:
\begin{equation}
n_1 = \frac{1}{2\pi}  \oint dx^i \,\hat{\bf e}^1 \cdot \nabla_i\hat{\bf e}^2  
 \,.
\label{SpinVortexInvariant}
\end{equation}
This relation is analogy with the winding number of integer-quantized vortices of superfluid A-phase, in which the differential 1-form is formed by the two perpendicular unit vectors in orbital degree of freedom \cite{VolovikBook1992}. In the vicinity of the transition, these spin vortices get the soft core of size $\xi/q  \gg \xi$, which corresponds to the coherent length of symmetry breaking from the polar to the PdB phase. As distinct from the topological defects in the polar  phase, which have the "normal phase" core of size $\xi$, the spin vortices in the PdB phase with $R_{2}$ have the "polar phase" core. Then we get a multicomponent systems, in which the order parameter is not necessarily equal to zero on the axis of the topological defects. This fact may modify the model of continuous phase transition of symmetry breaking to the model of first order phase transition via the appearance of the nucleation of new phase \cite{Pikin1993,Gleiser1994,Arkin1999}. When the PdB phase with $R_{2}$ transform to the polar phase, the proliferation of spin vortices of $\pi_{1}(R_{2})$ in PdB marks this kind of modification.\\[0.1cm]

\subsubsection{Unstable monopoles and spin vortices}

Since in the PdB phase $\pi_2(R_1) = \pi_2(R_2) = 0$, there is no way to protect hedgehog (monopole) of polar phase topologically when the second time symmetry breaking to PdB phase occurs. In fact, the $\hat{\mathbf{d}}$-vector hedgehog converts to the termination point of the spin vortices of $\pi_{1}(R_{2})$ with two quanta, as we will discuss in detail in the Chapter. \ref{Chapter3}. As a result the spin-vortices-dressed $\hat{\bf d}$-hedgehog in PdB phase, which is generated by two-step successive symmetry breaking transition, becomes the analog of Nambu monopole, which terminates the electroweak cosmic string \cite{Nambu1977}.


\subsubsection{HQVs and the KLS domain wall}
\label{KLSwallSec}

Similar situation takes place with the HQVs of polar phase, which are not topologically stable in the PdB phase. They become the termination lines of the domain walls of $\pi_{0}(R_{2})$. We will discuss this mechanism in Chapter.~\ref{Chapter3}, and we will see this process actually is analogy of the mechanism, though which the Kibble-Lazarides-Shafi (KLS) domain wall appears \cite{Kibble1982a}. In $^3$He experiments, after transition from the polar phase to the PdB phase in the presence of HQVs, the KLS walls appear between the neighboring vortices, and
in spite of the tension of  domain walls, the HQVs remain pinned by the nafen strands \cite{Dmitriev2008,Volovik2008,Makinen2019,Zhang2020}. \\[0.1cm]

In general the KLS wall is not topologically stable, and can be stabilized only due to symmetry reasons \cite{SalomaaVolovik1988}. 
However, in the vicinity of the transition to PdB phase from the polar phase vacuum, the KLS wall becomes topological by the meaning of fibration of between vacuum manifolds $R_{1}$ and $R_{P}$. The topological domain wall of the thickness $\xi/q$ is described 
by the nonzero element of the homotopy group $\pi_0(R_{{\rm P}\rightarrow {\rm PdB}}) = \mathbb{Z}_{2}$. Example of such a wall is the domain wall between the domains with $A_{\alpha i}=\Delta_P{\rm Diag}(1, q,q)$ and $A_{\alpha i}=\Delta_P{\rm Diag}(1, q,-q)$.
 

\chapter{Fibration and Composite Cosmological Objects} 

\label{Chapter3} 



\begin{figure*}
\centerline{\includegraphics[width=15cm,height=8cm]{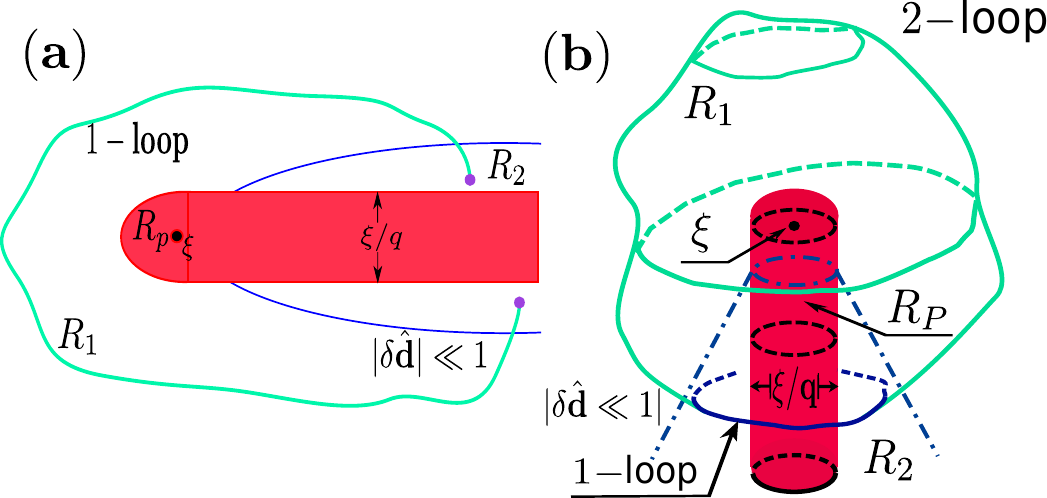}}
\caption{Illustration of relative $n$-loops describing the combined defects, which emerge in the two-step symmetry breaking transition in polar distorted $^3$He.
(a) KLS string wall.  In general the KLS wall is non-topological, but it acquires the nontrivial topology in the vicinity of the second phase transition. In this limit case there are two well separated length scales: the coherence length $\xi$ of the first transition, which determines the  size of the hard core of  string (the black dot),  and the much larger coherence length $\xi/q \gg \xi$ of the second transition, which determines the soft core size of the wall (the pink region).
The hierarchy of scales gives rise to two types of the vacuum manifolds in the PdB phase, $R_{1}$ and $R_{2}$. The $R_{1}$ includes all the degenerate vacua of the PdB phase, while the $R_2$ consists of those, which are obtained from the fixed order parameter of the polar phase, i.e. at fixed $\hat{\mathbf{d}}$ and $\Phi$ in Eq. (\ref{AP}). There exists spatial region, where the asymptotic condition $|\delta \hat{\mathbf{d}}| \ll 1$ is achieved.  The blue line shows the characteristic border between the regions of two classes of vacuum manifolds. The topology of the string wall is determined by relative homotopy group $\pi_1(R_1,R_2)$, in which the green relative $1$-loop is mapped to the space $R_1$, with the ends of the loop mapped to $R_{2}$ \cite{NashBook1988}. 
(b) 
The string monopole is described by relative homotopy group $\pi_2(R_1,R_2)$. In this case the black dot shows the core of the hedgehog in the $\hat{\mathbf{d}}$-field and the pink region is the core of $4\pi$ spin vortex, which is terminated by the hedgehog. The green $2$-loop is mapped to the space $R_1$, with its $1$-loop edge mapped to $R_{2}$ \cite{NashBook1988}. The spatial boundary between $R_{1}$ and $R_{2}$ is shown as a cone-surface by blue-dash line.}  
\label{R1R2loop}
\end{figure*}

In this Chapter, we discuss how to practice the method of relative homotopy group within the case of two-step successive symmetry breaking transition in polar distorted superfluid $^3$He. We also discuss how fibration between different vacuum manifolds emerges and significantly influence physics of topological defects. The nexus object consisting of vortex skyrmion and string monopole in the presence of orientation energy --- the magnetic energy, is discussed in Sec.~\ref{VorticesSkyrmion}.


\section{Relative Homotopy Groups and Fibration between Vacuum Manifolds}
\label{FibrationAndRelativeHomotpy}


As mentioned before, in the vicinity of the second transition, the system has two well separated length scales (gap energy scales), $\xi$ and $\xi/q \gg \xi$. This leads to the new classes of  objects, which combine the topology of both vacuum spaces, $R_{1}$ and $R_{2}$. That is because the order parameter fields are mapped into different degenerate vacuum manifolds of PdB phase at different spatial regions when $\xi$ and $\xi/q$ are well separated, thus the homotopy classes of these unusual $n$-loops of order parameters constitute $\pi_n(R_1,R_2)$ \cite{GoloMonastyrky1978,MineyevVolovik1978,MMV1978,Volovik1978,GoloMonastyrky1978}. \\[0.1cm] 

This combined topology can be illustrated by the following example of the KLS string wall.  At small distances $\xi\ll r \ll \xi/q$  from the core of HQV, the HQV is described by the homotopy group $\pi_1(R_{P})$. However, at larger distances $r\gg \xi/q$, the HQV becomes the termination line of the wall, which is described by the $\pi_0(R_2)$ topology and has size of $\xi/q$. As shown in Fig.~\ref{R1R2loop}(a), in this case, the 1-loop with big enough size ($\gg \xi/q$) is continuous mapping of $R_{1}$, such that the two end-points of the 1-loop are mapped into $R_{2}$. This kind of 1-loop is named as relative 1-loop \cite{NashBook1988}. All the equvlence classes of relative 1-loops constitute the relative homotopy group $\pi_{1}(R_{1},R_{2})$. \\[0.1cm] 

The similar physics takes place for string monopoles \cite{Nambu1977}. At small distances $\xi\ll r \ll \xi/q$  
from the core of the hedgehog, it is described by the homotopy group $\pi_2(R_{P})$, while at larger distances $r\gg \xi_q$, the monopole becomes the termination point of spin vortices described by the $\pi_1(R_2)$ topology, see Fig. \ref{R1R2loop}(b). The relative 2-loop maps the order parameter fields into $R_{1}$, of which one of the boundary manifold $S^{1}$ is mapped in to $R_{2}$ in the region with constant $\hat{\mathbf{d}}$ vector. Again, all the homotopy classes of relative 2-loops constitute the $\pi_{2}(R_{1},R_{2})$ group. \\[0.1cm]

In order to calculate $\pi_{n}(R_{1},R_{2})$ and properly understand the composite objects described by them, we need the help of exact sequence of homormophisms of relative homotopy groups. All calculation details about these are put into Appendix.~\ref{AppendixA}. Even though $\pi_{n}(R_{1},R_{2})$ could be directly calculated, it is always good that we can get understanding of them from different viewpoint. For our case of two-step symmetry breaking phase transition, this part of story is the most excited. \\[0.1cm]

Recalling the Eq.~(\ref{HP}) and Eq.~(\ref{HPdB}) in last chapter, we have $H_{PdB} \subset H_{P}$ and $H_{P} \subset G$, then we get 
\begin{equation}
\frac{G/H_{PdB}}{H_{P}/H_{PdB}} = R_{1}/R_{2} \cong R_{P} = G/H_{P},
\label{RIsomophsm}
\end{equation} 
by the meaning of the third isomorphism theorem \cite{MartinBook2009}. Equation.~(\ref{RIsomophsm}) suggests that there is continuous map
\begin{equation}
p:R_{1} \rightarrow R_{P},
\label{CoveringMap}
\end{equation}
through which the cosets $kR_{2}$ are mapped to elements of $R_{P}$ isomorphically, where $k \in R_{1}$. Then (i) $p$ is surjection from $R_{1}$ to $R_{P}$. Moreover, because $p$ maps cosets of $R_{1}$  to elements of $R_{P}$, (ii) the image of inverse mapping $p^{-1}(U)$ with $U \subset R_{P}$ are disjoint union of cosets in $R_{1}$ because all cosets are disjoint \cite{suzuki1982,Jones1998}. These two properties of $p$ indicate the isomorphism $p: R_{1} \rightarrow R_{P}$ is covering mapping and $R_{1}$ is covering space of $R_{P}$ \cite{Nakahara2003}. This important fact gives rise to fibration sequence
\begin{equation}
\xymatrix@1@R=10pt@C=13pt{
R_{2} \,\, \ar@{^{(}->}[r] \,\, & \,\, R_{1} \,\, \ar[r]^-{p} \,\, & \,\, R_{P}\\
}\,
\label{RFibration}
\end{equation}
between vacuum manifolds $R_{1}$ and $R_{P}$, where $R_{2}$ is mapped into $R_{1}$ by inclusion. Fibration Eq.~(\ref{RFibration}) immediately suggests
\begin{equation}
\pi_{n}(R_1, R_2) \cong \pi_{n}(R_P)
 \,
\label{RelativeGroup}
\end{equation} 
by applying an theorem in Appendix.~\ref{FibrationAndTheorems} \cite{HatcherBook2002}. In this case, we can see $R_{2}$ actually is fiber of this fibration. This means the vacuum manifold $R_{2}$ of PdB phase with fixed $\hat{\mathbf{d}}$ is generated from given vacuum state of polar phase, which corresponds to element of $R_{P}$, because $p^{-1}(e)$ with $e \in R_{P}$ homotopicly equivalent to $R_{2}$ for fibration Eq.~\ref{RFibration}. Then we can get a natural and self-consistent result that
\begin{equation}
\pi_{n}(R_{2}^{e}) = \pi_{n}(R_{2}^{f}),
\end{equation}
where $e,f \in R_{P}$ correspond to polar phase vacua with different $\hat{\mathbf{d}}$. More details see Appendix.~\ref{FibrationAndTheorems}. \\[0.1cm]   

\begin{figure*}
\centerline{\includegraphics[width=1.0\linewidth, height=0.5\linewidth]{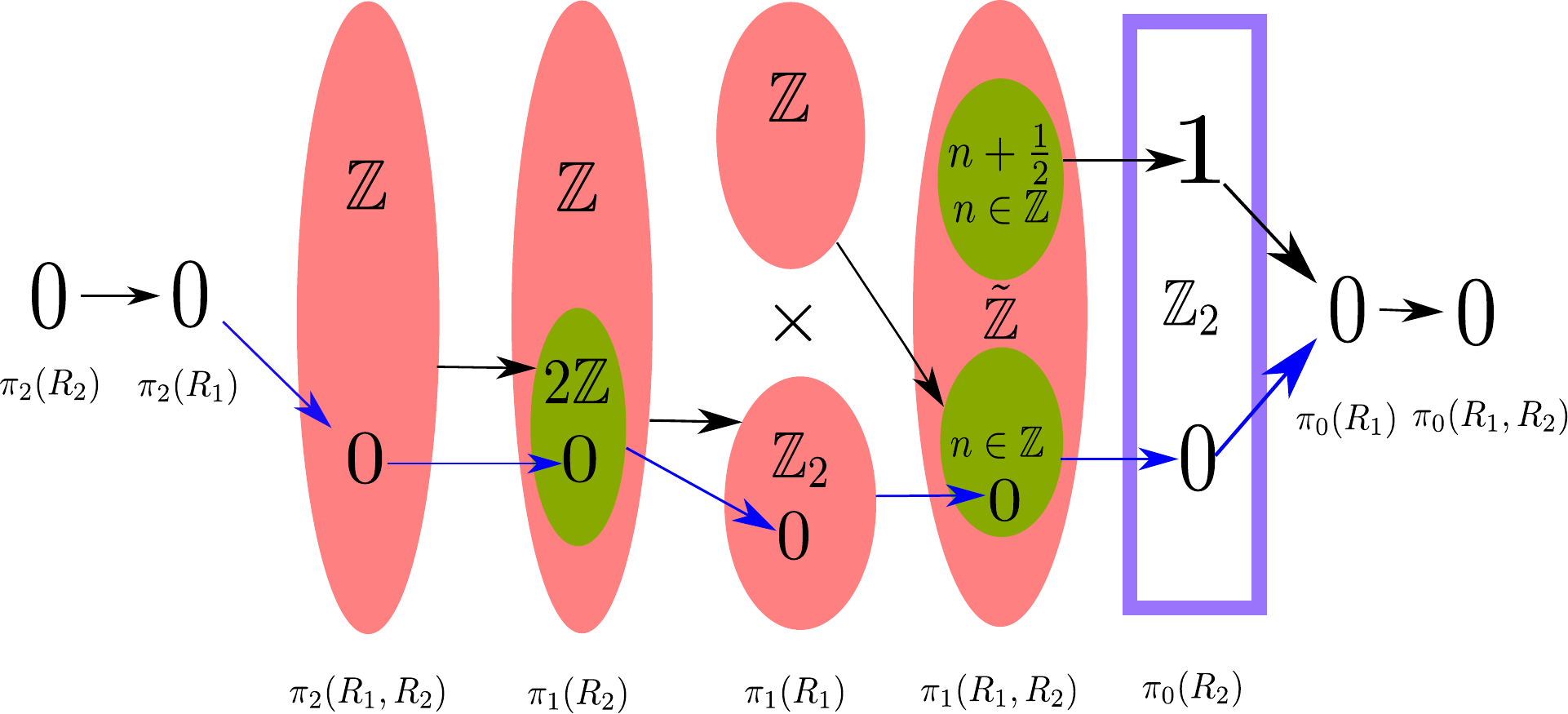}}
\caption{Illustration of the exact sequence of homomorphisms for $\pi_n( R_1, R_2)$ in Eq.~(\ref{sequence}). This demonstrates that the elements of $\pi_n( R_1, R_2)$ group have two sources: the 
kernel of the mapping $\pi_{n-1}( R_2) \rightarrow \pi_{n-1}( R_1)$  and the quotient space of $\pi_n(R_1)$ over the image 
of the mapping $\pi_{n}( R_2) \rightarrow \pi_{n}( R_1)$. The black arrows represent the image of homomorphisms, while the blue arrows represent the kernal of every homomorphsim. This mapping diagram prescribes the relation between the elements of the composite topological defects. In particular,  it shows that  the relative homotopy group $\pi_2( R_1, R_2)$ is determined by the 
kernel of the mapping $\pi_{1}( R_2) \rightarrow \pi_{1}( R_1) =2\mathbb{Z} \cong \mathbb{Z}$. It demonstrates that 
the nontrivial monopoles are termination points of spin vortices with the total winding number being even. 
As we will see in next section, this mapping relations is identical with that describing vortex skyrmions. 
On the other hand, the relative homotopy group $\pi_{1}( R_1, R_2) \cong \tilde{\mathbb{Z}}$ is determined by both sources: 
$\mathbb{Z}_2$ which is the kernel of the homomorphism $\pi_{0}( R_2) \rightarrow \pi_{0}( R_1)$ and the quotient group of $\pi_{1}(R_{1})/{\operatorname{im}m_{*}}$
with $m_{*}:\pi_{1}(R_{2}) \rightarrow \pi_{1}(R_{1})$. 
As a result, there are two different kinds of phase vortices, that terminate and do not terminate the KLS wall. Those two classes of vortices consist of the two cosets of quotient $\tilde{\mathbb{Z}}/\mathbb{Z} \cong \mathbb{Z}_{2}$. Correspondingly, These are the vortices with half-odd integer circulation numbers and the vortices with integer circulation quanta.} 
\label{MappingDiagram}
\end{figure*}

Equation.~(\ref{RelativeGroup}) is the kernel of this chapter, it demonstrates that in the vicinity of the phase transition from the first (polar) phase to the second (PdB) phase, all the topological objects of the first phase described by the group $\pi_n(R_P)$ retain their topological charges in the second phase, and thus convert to the topological objects in the second (PdB) phase. Some of these defects remain free, while the other become the parts of the composite defects -- the string monopole and for the KLS string wall in Sec.~\ref{KLSStringWall} and Sec.~\ref{StringMonopole} respectively. However, Eq.~(\ref{RelativeGroup}) does not resolve between the free and the composite objects of the second (PdB) phase. The full classification of the topological
objects in the second phase depends not only on $\pi_{n}(R_{1},R_{2})$, but also on the details of the mappings in
 the exact sequence of homomorphisms of $\pi_{n}(R_{1},R_{2})$, which is calculated in
Appendix.~\ref{app:derivation}. 
The mapping diagram of the long exact sequence (LES) Eq.~(\ref{sequence}) in Fig.~\ref{MappingDiagram} depicts the relation between different topological objects in $R_{1}$ and $R_{2}$. 

\section{Composite Cosmological Objects and Exact Sequences}
\label{StringMonopoleAndKLSWall}

\subsection{Strings terminated by monopole -- String Monopole}
\label{StringMonopole}
\begin{figure*}
\centerline{\includegraphics[width=0.8\linewidth, height=0.4\linewidth]{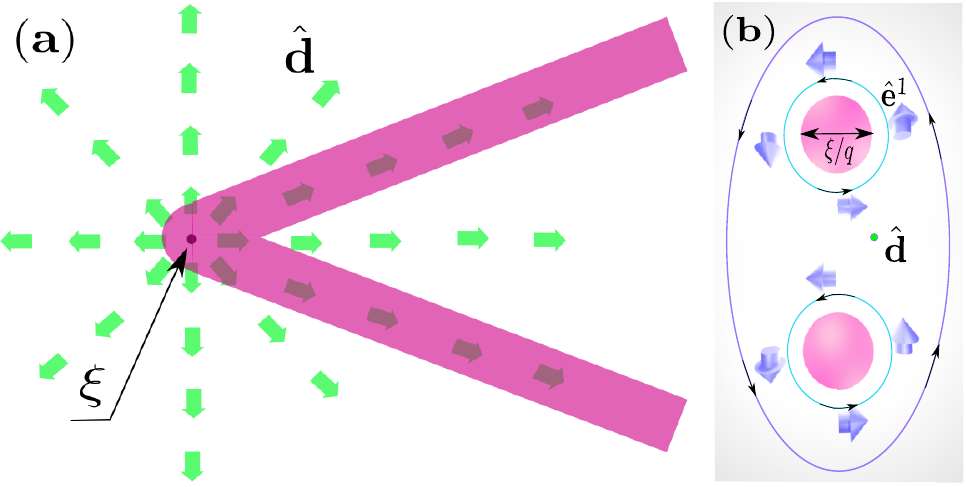}}
\caption{Illustration of string monopole in the PdB phase.  The monopole in the  $\hat{\bf d}$-field has topological charge $n_{2}=1$ and hard core of the coherence length size $\xi$ (black dot). This string monopole terminates two spin vortices  with the soft core size $\xi/q$ and with the total topological charge $n_{1}=1+1=2=2n_{2}$ according to Eq.~(\ref{MonopoleInvariant}). (a) Texture of $\hat{\mathbf{d}}$ vector string monopole with $n_{2}=1$ terminating two spin vortices.  The green arrow corresponds to $\hat{\mathbf{d}}$ vector. The pink region is the soft core of spin vortex of size $\xi/q$. (b) The cross sections of two spin vortices with $n_{1}=1$ each. The blue arrows represent the $\hat{\mathbf{e}}$ vectors. In these cross sections, 
$\hat{\mathbf{e}}_{1}$ and $\hat{\mathbf{e}}_{2}$ vectors rotate around $\hat{\mathbf{d}}$ vector by $2\pi$ for every spin vortex. The core size of every spin vortex is $\xi/q$. The blue line, which surround two cores of spin vortices, is the field line of $\hat{\mathbf{e}}$ vectors rotation, while the cyan lines, which surround core of every spin vortices, illustrate the winding of $\hat{\mathbf{e}}$ vectors around the core of spin vortice.    
} 
\label{MonopoleFig1}
\end{figure*}

The relative homotopy group 
\begin{equation}
 \pi_{2}(R_{1},R_{2})\cong\pi_2(R_P) = \mathbb{Z} 
 \,,
\label{pi2relative}
\end{equation}
describes monopoles (hedgehogs) of $\hat{\mathbf{d}}$ field. They survive in  the vicinity of the second transition as the topological objects which terminate the spin vortices. To demonstrate this, we need the short exact sequence (SES) 
\begin{equation}
\xymatrix@1@R=10pt@C=13pt{
0 \,\, \ar[r] \,\, &  \,\, 0 \,\, \ar[r]^-{i^{*}} \,\, & \,\, \pi_{2}(R_{1},R_{2}) \,\, \ar[r]^-{{\partial}^{*}} \,\, & \,\, 2\mathbb{Z} \,\,  \ar[r] \,\, & \,\, 0\\
}\,, 
\label{SESPi2R1R2}
\end{equation}
of $\pi_{1}(R_{1},R_{2})$, see calculation details in Appendix.~\ref{app:derivation}. The boundary homormophism $\partial^{*}$ describes the possible string defects connecting on the $\hat{\mathbf{d}}$-monopoles. As shown both in Fig.~\ref{MappingDiagram} and in Eq.~(\ref{SESPi2R1R2}), $im{\partial^{*}} = 2\mathbb{Z}$, then there are spin vortices with even winding number connecting on non-trivial $\hat{\mathbf{d}}$-monopole. 
This means the corresponding composite object -- the string monopole -- has two topological charges, $n_{2}$ and $n_{1}$, which are related as: 
\begin{equation}
n_2= \frac{1}{8\pi} e^{ijk} \int_{\rm S^{2}} dS_k \,\hat{\bf d}\cdot 
\left(  \nabla_i \hat{\bf d} \times \nabla_j \hat{\bf d} \right)=\frac{1}{2}n_1
 \,.
\label{MonopoleInvariant}
\end{equation}
Here $S^{2}$ is the surface encircling monopole and the group $\mathbb{Z}$  is the group of integers $n_2$ --  the topological charges of the hedgehog. While $n_{1}$ is the winding number of spin vortices in Eq.~(\ref{SpinVortexInvariant}). The equation $n_1=2n_2$ in Eq.~(\ref{MonopoleInvariant}) shows  that monopole is termination point of spin vortices with the even total charge $n_2$ in vicinity of the second time symmetry breaking. This situation is similar to the monopole  in the chiral A-phase \cite{volovik2000,VolovikBook2009,Blaha1976,Volovik1976}, which either terminates a single vortex with $n_{1}=2$, or forms the nexus with two singly quantized vortices with $n_{1}=1+1=2$, or with four HQVs with $n_{1}=1/2+1/2+1/2+1/2=2$. Those vortices, which connect with monopoles ($n_{2}>0$) or antimonopoles ($n_{2}<0$) allow the existences of complex monopole-antimonopole networks \cite{Kibble2015,Saurabh2019,Shafi2019,Volovik2019d,Lazarides2021}. \\[0.1cm]

Fig. \ref{MonopoleFig1} illustrates the configuration of the string monopole, which consists of the hedgehog with $n_{2}=1$ and two strings -- spin vortices each with $n_1=1$. The spin vortices have a soft core with size $\xi/q$. 

\subsection{Wall bounded by string -- KLS string wall}
\label{KLSStringWall}

\begin{figure*}
\centerline{\includegraphics[width=0.8\linewidth, height=0.4\linewidth]{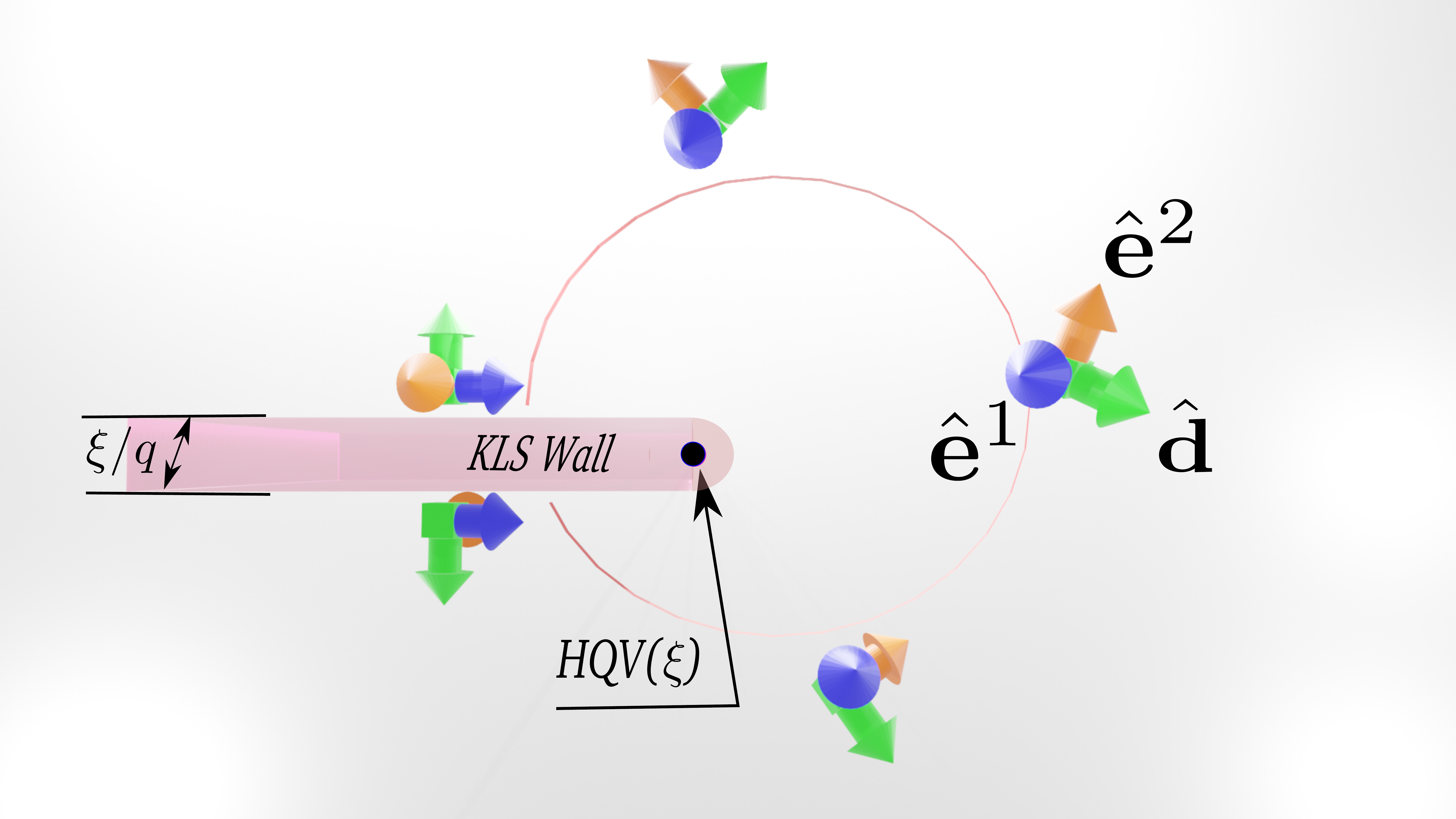}}
\caption{Illustration of the KLS string wall.
The wall itself is topologically protected by $\pi_{0}(R_{2})$ if the order parameter takes values from disconnected parts of $R_{2}$. In the case of two-step sucessive symmetry breaking, it convert to the combined object -- the HQV, which terminates the KLS wall.  The pink region is the topological KLS wall with thickness $\xi/q$, while the black small dot is HQV string, which diameter is $\xi$. The spin tripods show the configurations of order parameter around the HQVs string. The flipping of the tripods on two sides of KLS wall demonstrates the Domain wall feature.  
} 
\label{HoleInWallFig}
\end{figure*}

The relative homotopy group fo KLS string wall is
\begin{equation}
\pi_1( R_1, R_2) \cong \pi_1(R_P) = \tilde{\mathbb{Z}} \,.
\label{pi1relative}
\end{equation}
Even though Eq.~(\ref{pi1relative}) shows that the topological charges of string wall in the second (PdB) phase are the same as those in the first (polar) phase and they form the group of integer and half-odd integers in both phases, the physical realizations of these (composite) objects are different in the two phases. Similar with the case of string monopole, we recall the SES 
\begin{equation}
\xymatrix@1@R=10pt@C=13pt{
0 \,\, \ar[r] \,\, &  \,\, \mathbb{Z} \,\, \ar[r]^-{i^{*}} \,\, & \,\, \pi_{1}(R_{1},R_{2}) \,\, \ar[r]^-{{\partial}^{*}} \,\, & \,\, \mathbb{Z}_{2} \,\,  \ar[r] \,\, & \,\, 0\\
}\, 
\label{SESPi1R1R2}
\end{equation}
of $\pi_{1}(R_{1},R_{2})$ to demonstrate the details of KLS string wall (calculation details see Appendix.~\ref{app:derivation}). Because the image of boundary homomorphism $\partial^{*}$ is $\pi_{0}(R_{2}) = \mathbb{Z}_{2}$, there are two kinds of string defects in the vicinity of the second time symmetry breaking. One is free vortex with integer winding number $n \in \mathbb{Z}$, while the other one is vortex with half-odd integers, $k=n+1/2$, terminate the wall bounded by string --  the KLS string wall. Thus we can see from Fig.~\ref{MappingDiagram} and Eq.~(\ref{SESPi1R1R2}) the $\operatorname{im} i^{*}$ of SES of $\pi_{1}(R_{1},R_{2})$ are vortices of integer winding number because of $\operatorname{im}i^{*} \cong ker{\partial^{*}}$. \\[0.1cm]

Fig. \ref{HoleInWallFig} illustrates the configuration of the composite object. 
This kind of topologically protected KLS string wall induces the cosmological catastrophe in the axion solution of strong \textit{CP} problem \cite{ZeldovichSoviet1974,Zeldovich1974,Vilenkin1982,Sikivie1982,Sikivie1985,Sikivie2008,Kim2010,
Lazarides1982,Lazarides1985,Sato2018,Chatterjee2019,Andrea2019}. \\[0.1cm]


\section{Skyrmions and Nexus in the presence of magnetic field}
\label{VorticesSkyrmion}

\begin{figure*}
\centerline{\includegraphics[width=0.85\linewidth, height=0.5\linewidth]{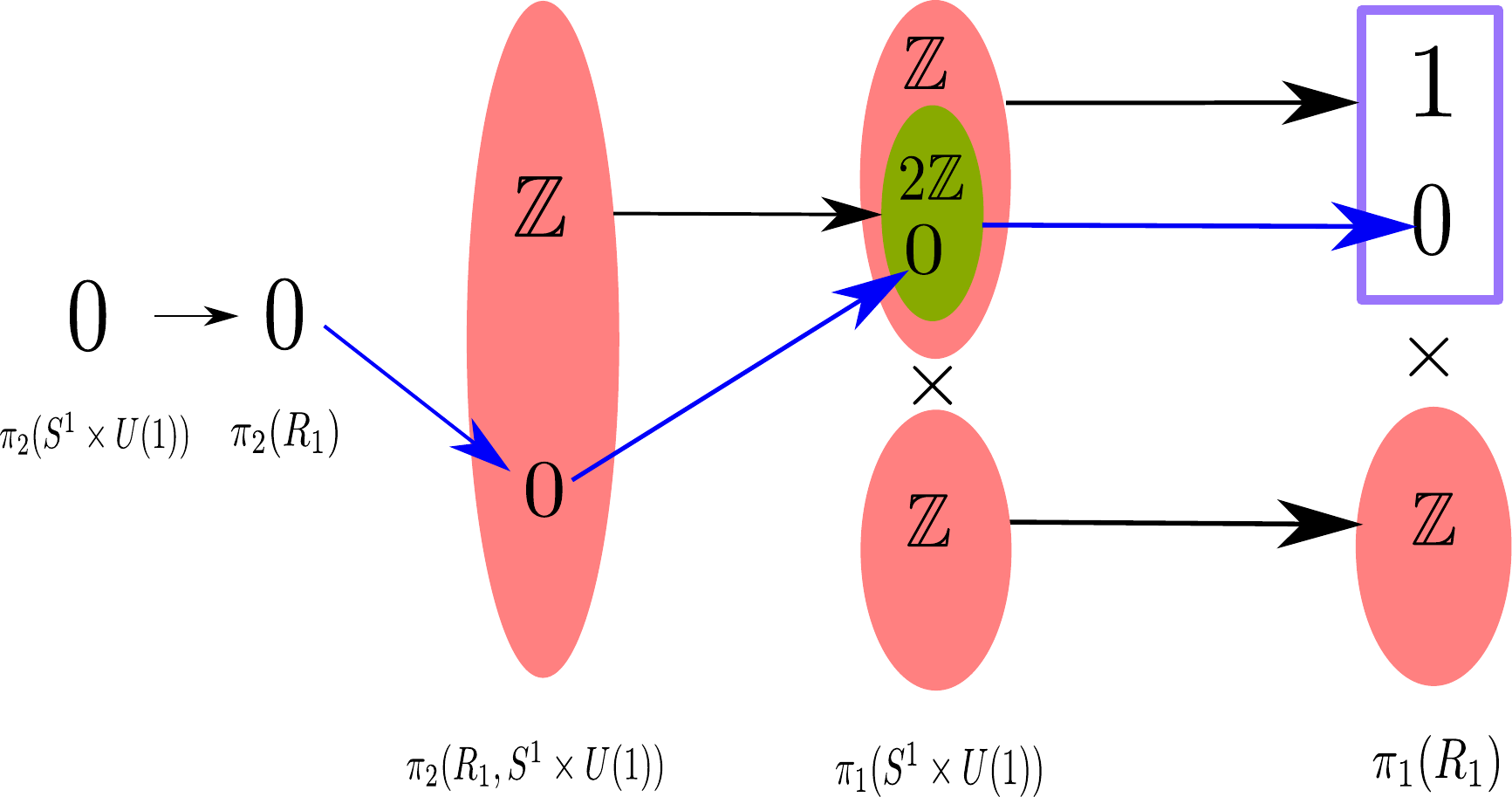}}
\caption{Mapping diagram of exact sequence between $R_{1}$ and $S^{1} \times U(1)$. The black arrows represent the image of homomorphisms, while the blue arrows represent the kernal of every homomorphsim. This diagram shows 
that the skyrmions soften the core of $\mathbb{Z}$ spin vortices with size $\xi/q$ to size $\xi_{H}$ in the presence of magnetic field. In regions larger than $\xi_{\mathbf{H}}$, vortex skyrmions can be connected with spin vortices via string monopole, if their total topological charge is even according to Eq. (\ref{SkyrmInvariant}). This is because $\pi_{2}(R_{1},R_{2}) \cong \pi_{2}(R_{1},S^{1} \times U(1))$. There are also phase vortices described by $\mathbb{Z}$, but here we ignore them because they do not influence the connection between the spin vortices and skyrmions. } 
\label{MappingDiagram_SN}
\end{figure*}

In the presence of magnetic field $\mathbf{H}$, a new length scale appears in the PdB phase -- the magnetic length 
$\xi_{H} \propto |\mathbf{H}|^{-1}$.  The magnetic length  $\xi_{H}$ is the longest length scale if we neglect the spin-orbit coupling. 
As a result, we have the order parameter vacuum manifold $R_1$ at short distances where the magnetic energy can be neglected, and a submanifold $R_{1}^{H}\subset R_1$ at large distances ($> \xi_{H}$), where the space of the order parameter is restricted by magnetic interaction \cite{MineyevVolovik1978,Kondo1992,Seji2019,Liu2020}.
In the region with length scale larger than $\xi_{H}$, the magnetic anisotropy locks the directions of $\hat{\mathbf{d}}$ vector in the plane perpendicular to $\mathbf{H}$ to minimize the magnetic energy, which is proportional to $|\mathbf{H} \cdot \hat{\mathbf{d}}|^{2}$.  The degenerate space of the  order parameter is reduced from $R_1 = SO_{S-L}(3) \times U(1)$ in Eq.~(\ref{RPdB}) to 
$R_1^H =S^{1} \times S^{1} \times U(1)$ in the regions larger than $\xi_{H}$. The first $S^{1}$ is the manifold of in plane $\hat{\mathbf{d}}$ vector, while the second $S^{1}$ is the manifold of rotations of $\hat{\mathbf{e}}^{1}$ and $\hat{\mathbf{e}}^{2}$ about the $\hat{\mathbf{d}}$-axis. 
However, for $q\ll 1$, the gradient energy of the $\hat{\mathbf{d}}$-textures is much larger than that of the textures in $\hat{\mathbf{e}}^{1}$ and $\hat{\mathbf{e}}^{2}$ fields and intensively increase the free energy of system \cite{VollhardtWolfle1990}. That is why we consider only the $S^{1}$ manifold of $\hat{\mathbf{e}}^{1}$ and $\hat{\mathbf{e}}^{2}$, and neglect the $S^{1}$ manifold of $\hat{\mathbf{d}}$. Then the relative second homotopy group which we need in this case is 
\begin{equation}
\pi_{2}(R_{1},S^{1} \times U(1)) = \mathbb{Z}.
\label{RHG_SkyrmionNexus}
\end{equation}
The results for the relative homotopy group have been confirmed by calculations using the exact sequence, see details in appendices Sec. \ref{InThePresentOfMagneticField}.
The mapping diagram of LES of $\pi_{2}(R_{1},S^{1} \times U(1))$ i.e., Eq.~(\ref{sequence_SkyrmionNexus}) is shown in Fig.~\ref{MappingDiagram_SN}. \\[0.1cm]

\begin{figure*}
\centerline{\includegraphics[width=0.9\linewidth, height=0.45\linewidth]{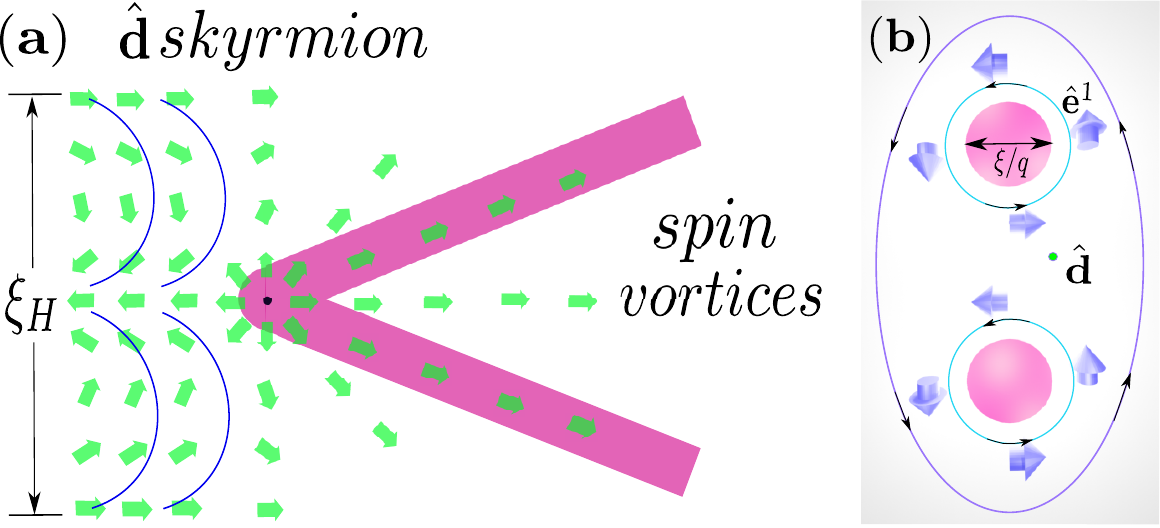}}
\caption{Illustration of nexus object in presence of magnetic field. The nexus connects spin vortices with core size $\xi/q$ and the vortex skyrmion with core size $\xi_H$, where $\xi_{H} \gg \xi/q$. (a) the texture configuration of nexus object with $n_{2}=1$ and $n_{1}=2$. The green arrows are the $\hat{\mathbf{d}}$ vectors and the pink regions are core regions of defects. The black dot is the core of string monopole with size $\xi$, while the pink regions are the cores of spin vortices with size $\xi/q$. The green arrows represent the distribution of $\hat{\mathbf{d}}$ vectors. The vortex skyrmion with core size $\xi_{H}$ transforms to two spin vortices via nexus.  (b) the cross section of two spin vortices. Every spin vortex has the $2\pi$ rotation of $\hat{\mathbf{e}}^{1}$ and  $\hat{\mathbf{e}}^{2}$ vectors around fixed $\hat{\mathbf{d}}$ vector. The blue arrows represent the field of the $\hat{\mathbf{e}}^{1}$ vector. }
\label{NexusFig}
\end{figure*}

Following the convention in Sec.~\ref{StringMonopoleAndKLSWall}, we introduce the SES (for details see Appendix.~\ref{InThePresentOfMagneticField})
\begin{equation}
\xymatrix@1@R=10pt@C=13pt{
0 \,\, \ar[r] \,\, &  \,\, 0 \,\, \ar[r]^-{i^{*}} \,\, & \,\, \pi_{2}(R_{1},S^{1} \times U(1)) \,\, \ar[r]^-{{\partial}^{*}} \,\, & \,\, 2\mathbb{Z} \,\,  \ar[r] \,\, & \,\, 0\\
}\,
\label{SESRi2R1R1H}
\end{equation}
of $\pi_{2}(R_{1},S^{1} \times U(1)$ to discuss the corresponding topological objects. Because the $im{\partial^{*}} \cong 2\mathbb{Z}^{S} \subset \pi_{1}(S^{1} \times U(1))$, this kind of object has spin vortices with even winding number on its boundary. As a result, the object described by $\pi_{2}(R_{1},S^{1} \times U(1))$ is the vortex skyrmion in the presence of magnetic field, which has the soft core of size $\xi_{H}$ represented as skyrmion, see Fig.~\ref{NexusFig}.
  The topological charge of skyrmion is 
\begin{equation}
n_2= \frac{1}{8\pi} e^{ijk} \int_{\rm D_{2}} dS_k \,\hat{\bf d}\cdot 
\left(  \nabla_i \hat{\bf d} \times \nabla_j \hat{\bf d} \right)=\frac{1}{2}n_1
 \,,
\label{SkyrmInvariant}
\end{equation}
where $n_{2} \in \pi_{2}(R_{1},S^{1} \times U(1))$, $D_{2}$ is the cross-section of skyrmion and $n_{1}$ is the winding number of spin vortices in Eq. (\ref{SpinVortexInvariant}). \\[0.1cm]

The Eq. (\ref{SkyrmInvariant}) is the analog of the Mermin-Ho relation in $^3$He-A \cite{MerminHo1976}.
However, there is a more prominent feature of this relation, that is Eq.~(\ref{SkyrmInvariant}) is identical with Eq.~(\ref{MonopoleInvariant}).
In fact, the SESes of $\pi_{2}(R_{1} ,R_{2})$ in Eq.~(\ref{SESPi1R1R2}) and $\pi_{2}(R_{1},S^{1} \times U(1))$ in Eq.~(\ref{SESRi2R1R1H}) are exactly same as well. This means 
\begin{equation}
\pi_{2}(R_{1},S^{1} \times U(1)) \cong \pi_{2}(R_{1},R_{2}).
\label{Pi2Isomorphism}
\end{equation}
As a result, their topological invariants are same. Due to this relation the vortex skyrmion can be connected to $\mathbb{Z}$ spin vortices with core size $\xi/q$ via the string monopole. 
Such composite objects, where the monopole connects several linear objects with different characteristic length scales is called nexus. It demonstrates the interplay between $\pi_1$ and $\pi_2$ topologies. In spite of its novel and complicated structure, nexus actually connect topological objects with different characteristic energies in a topological protected binding in Eq.~(\ref{Pi2Isomorphism}). This property allow researchers to detect topological objects with small length scale via low energy dynamic process. We will see this in next two chapters. \\[0.1cm]

Originally vortex skyrmions formed by orbital and phase degenerate parameters have been suggested in $^3$He-A by Anderson and Toulouse \cite{Anderson1977} 
and by Chechetkin \cite{Chechetkin1976}. The lattice of vortex-skyrmions in rotating $^3$He-A has been discussed in Ref. \cite{Volovik1977}.
 These objects have been identified in different experiments under rotation \cite{Seppala1984,Pekola1990}. The dynamics of the vortex skyrmions  provides an effective electromagnetic fields, which induces the observed effect of chiral anomaly experienced by fermionic excitations (Weyl fermions)  living in the soft core of a vortex skyrmion \cite{Bevan1997}. 


\chapter{Equilibrium Configurations of the Extended Structures of KLS String Wall} 

\label{Chapter4} 

As we have seen in last section of Chapter.~\ref{Chapter3}, the orientation energies with different characteristic lengths reduce the vacuum manifolds and generate nexus, which connects topological objects with different characteristic energies. In this chapter, we discuss how this mechanism runs for one dimensional nexus when we take into account the dipole energy --- the smallest orientation energy in superfluid $^3$He. 
We will see the KLS string wall with length scale $\xi$ and $\xi/q$ connect to spin solitons with dipole length $\xi_{D} \gg \xi/q $ via HQV.
The possible mesoscopic equilibrium state configurations are analyzed and numerically calculated. The resulted data will be directly used for spin dynamic response calculations in Chapter.~\ref{Chapter5} and compare with the experimental observations. 




\begin{figure*}
\centerline{\includegraphics[width=1.0\linewidth]{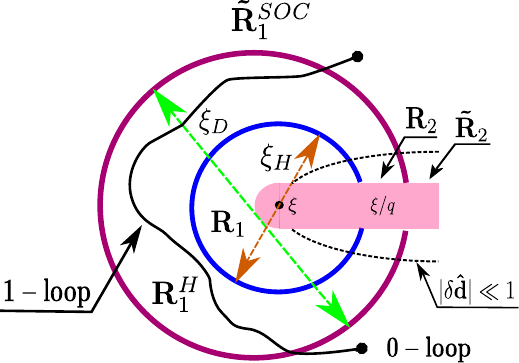}}
\caption{Illustration of vacuum manifolds with length scales $\xi_{H} < r < \xi_{D}$ and $r > \xi_{D}$ in the vicinity of transition from polar phase to PdB phase. As been discussed in Ref.~\cite{Zhang2020}, the vacuum manifolds of PdB in the vicinity of the second symmetry breaking are $R_{1}$ and $R_{2}$ in the region with $r<\xi_{H}$. The hierarchy of length scales extends in the presence of magnetic energy and SOC energy. We have known there is KLS string wall described by $\pi_{1}(R_{1},R_{2}) \cong \tilde{\mathbb{Z}}$ for $r < \xi_{H}$. In larger region with length scale $\xi_{H} < r < \xi_{D}$, $R_{1}$ reduces to $R_{1}^{H} = S^{1}_{S} \times U(1)^{\Phi}$ by magnetic energy. To minimize the magnetic energy, spin vector $\hat{\mathbf{d}}$ is perpendicular to static magnetic field $\mathbf{H}^{(0)}$, while the $R_{2}$ is unchanged. When the SOC energy is taken into account, $R_{1}^{H}$ further reduces to $\tilde{R}_{1}^{SOC} = R_{s}^{SOC} \times U(1)^{\Phi}$ and $R_{2}$ reduces to $\tilde{R}_{2} = \mathbb{Z}_{2}^{S-\Phi}$. As a results, there are linear topological objects described by $\pi_{1}(R_{1}^{H},\tilde{R}_{1}^{SOC})$, in which the relative 1-loop (black solid curve) is mapped to $R_{1}^{H}$ while its end points 0-loop is mapped to $\tilde{R}_{1}^{SOC}$. \label{Degenerate_Spaces}}
\end{figure*}

\section{Vacuum manifolds in the presence of orientation energies}
\label{VacuaManifold}
The PdB phase achieved by two step successive symmetry breaking phase transition, which starts from uniaxial anisotropy normal phase vacuum, has two well separated length scales $\xi$ and $\xi/q$ in the vicinity of transition from polar phase to PdB phase \cite{Zhang2020}. In the Chapter.~\ref{Chapter2}, we discussed the vacua of order parameters of superfluid in the nafen-distorted $^3$He. These vacua have dramatically different characteristic lengths determined by the energy gaps. As a result, the PdB phase in the vicinity of transition from polar phase to PdB phase has several composite topological objects with different dimensions. These novel composite objects are classified by relative homotopy groups $\pi_{n}(R_{1},R_{2})$ between vacua $R_{1}$ and $R_{2}$. Because of fibration in Eq.~(\ref{RFibration}), the stable objects of polar phase are stabilized again in PdB phase by forming composite objects described by relative homotopy groups $\pi_{n}(R_{1},R_{2})$. However, objects with length scales around $\xi$ or $\xi/q$ have higher characteristic energies than the energy scale of typical detecting method i.e., Nuclear Magnetic Resonance (NMR). In the experiment in Ref.~\cite{Makinen2019}, the detection of KLS string wall was done by NMR of spin solitions, which connect with KLS domain wall via HQV. Thus HQV is one dimensional (1D) nexus and KLS string wall extends its structure into mesoscopic length scale via 1D nexus.  In this section, we discuss the orientation energies, their corresponding characteristic lengths and the reduced vacuum manifolds, which make this detection possible. \\[0.1cm]

In nafen-distorted $^3$He system, these length scales are magnetic length $\xi_{H}$ and dipole length $\xi_{D}$ \cite{Yudin2014,Autti2016,VollhardtWolfle1990}. These two length scales are healing lengths and characterize the spatial ranges in which the gradient energy are larger than orientations energies. When the length scale of spatial variations is larger than these characteristic lengths, the vacuum manifolds are reduced to minimize the orientation energies. We discussed the consequence of this kinds of reduction by magnetic energy and magnetic length $\xi_{H}$ i.e., the vortex skyrmions in Chapter.~\ref{Chapter3}. We will see there are more interesting results when dipole length $\xi_{D}$ is introduced in addition to $\xi_{H}$ in rest parts of this review. $\xi_{H}$ is determined by gradient energy density 
\begin{align}
f_{\rm grad} = & \frac{1}{2} K_1 \partial_{i} A_{\alpha j} \partial_{i} A^{*}_{\alpha j} +\frac{1}{2} K_2 \partial_{j} A_{\alpha i} \partial_{i} A^{*}_{\alpha j} \notag \\ & + \frac{1}{2} K_3 \partial_{i} A_{\alpha i} \partial_{j} A^{*}_{\alpha j}
\label{GradientEnergy}
\end{align}    
where 
\begin{equation}
A_{{\alpha}i} \equiv A_{{\alpha}i}^{PdB} = e^{i\Phi} [\Delta_{P}\hat{d}_{\alpha} \hat{z}_{i} + \Delta_{\bot1}\hat{e}^{1}_{\alpha} \hat{x}_{i} + \Delta_{\bot2}\hat{e}^{2}_{\alpha} \hat{y}_{i}]
\label{PdBOrderParameter}
\end{equation}
is the order parameter of PdB phase. $\hat{\mathbf{d}} \equiv \hat{d}_{\alpha}$ and $\hat{\mathbf{e}}^{1(2)} \equiv \hat{e}_{\alpha}^{1(2)}$ are the spin degenerate parameters and they form the triad in spin space. $\Phi$ and $\hat{x}_{i}\equiv\hat{x}$,$\hat{y}_{i}\equiv\hat{y}$,$\hat{z}_{i}\equiv\hat{z}$ are phase and orbital degenerate parameters respectively. Here $|\Delta_{\bot1}|=|\Delta_{\bot2}|=|q|\Delta_{P}$ with $|q| \leq 1$, and $K_{1}=K_{2}=K_{3}$ \cite{VollhardtWolfle1990}. The magnetic energy density is
\begin{equation}
f_{\rm H} = -\frac{1}{2} \chi_{\alpha \beta} H_{\alpha} H_{\beta}=\frac{1}{2}\gamma^{2} S_{a}S_{b}(\chi^{-1})_{ab}- \gamma H_{a}S_{a},
\label{MagneticEnergy}
\end{equation}
here the $\chi_{\alpha \beta}$ is uniaxial tensor of magnetic susceptibility of PdB phase, $H_{\alpha}$ are magnetic field strengths with $\alpha=1,2,3$, $S_{a}$ are spin densities with $a=1,2,3$ and $\gamma$ is gyromagnetic ratio \cite{VollhardtWolfle1990}. With the help of Eq.~(\ref{GradientEnergy}) and Eq.~(\ref{MagneticEnergy}), the magnetic length is given as 
\begin{equation}
\xi_{H}=[\frac{K_{1}\Delta_{P}^{2}}{(\chi_{\bot}-\chi_{\|})H^{2}}]^{\frac{1}{2}},
\label{MagneticLength}
\end{equation}
where $\chi_{\bot}$ and $\chi_{\|}$ are transverse and longitude spin magnetic susceptibilities of PdB phase. In the experiment, a static magnetic field $\mathbf{H}^{(0)}$ with fixed direction is turned on \cite{Makinen2019}. Then the degenerate space of PdB order parameter reduces to 
\begin{equation}
R^{H}_{1} = S^{1}_{S} \times U(1)^{\Phi}
\end{equation}
from $R_{1}$ in the region where length scale of spatial variation is larger than $\xi_{H}$, as we discussed in Sec.~\ref{VorticesSkyrmion}. Because the magnetic energy locks the $\hat{\mathbf{d}}$ vector into the plane perpendicular to $\mathbf{H}^{(0)}$, $R_{2}$ keeps the same form as it is inside the region with length scale $\xi_{H}$. Then we still have $R_{2} = SO(2)_{S-L} \times \mathbb{Z}^{S-\Phi}_{2}$ in the region where condition $|\delta \hat{\mathbf{d}}| \ll 1$ is satisfied. In Fig.~\ref{Degenerate_Spaces}, we illustrate the $R_{1}^{H}$ and $\xi_{H}$ in the presence of KLS string wall. \\[0.1cm]

Following the same idea, the dipole length $\xi_{D}$ is determined by gradient energy density $f_{grad}$ and SOC energy density
\begin{equation}
f_{\rm soc} = \frac{3}{5}g_D(A^*_{ii}A_{jj}+A^*_{ij}A_{ji} - \frac{2}{3}A^*_{ij}A_{ij}), 
\label{EnergyDensityOfSOC}
\end{equation}
where $g_{D}$ is strength of spin orbital coupling. Then we have
\begin{equation}
\xi_{D}=(\frac{5K_{1}}{6g_{D}})^{\frac{1}{2}}.
\label{DipoleLength}
\end{equation}
When the Spin-Orbit coupling (SOC) is taken into account, vacuum manifolds of order parameters are  further reduced from $R^{H}_{1}$ and $R_{2}$. In general consideration, the requirement of minimizing SOC energy in region with length scale larger than $\xi_{D}$ fixes the relative directions between spin vectors and orbital vectors. The resulted vacuum manifold always could be represented by spin degree of freedom because the broken symmetry is relative symmetry \cite{VollhardtWolfle1990}. Thus $R_{1}^{H}$ reduces to
\begin{equation}
\tilde{R}_{1}^{SOC} = R_{S}^{SOC}\times U(1)^{\Phi}
\end{equation}
in the region with length scale larger than $\xi_{D}$, where $R_{S}^{SOC}$ is the reduced vacuum manifold of spin degree of freedom. In general case, $R_{S}^{SOC}$ is a complicated space. However $R_{S}^{SOC}$ may be simplified by using parametrization of $\hat{\mathbf{d}}$ and $\hat{\mathbf{e}}^{1(2)}$ vectors of $A_{{\alpha}i}^{PdB}$. To facilitate comparison between experimental observations and our theoretical analysis, the paramentrizations 
\begin{equation}
\mathbf{\hat{d}} =\hat{x}cos\theta-\hat{z}sin\theta, \,\,
\mathbf{\hat{e}}^{1} =-\hat{x}sin\theta-\hat{z}cos\theta, \,\, 
\mathbf{\hat{e}}^{2} =\hat{y}, \,\, \mathbf{H}^{(0)} = H\hat{y}
\label{PARA1}  
\end{equation}
would be used in this work, where $\theta$ is the angle between $\hat{\mathbf{d}}$ and local orbital-coordinate frame \cite{Makinen2019}. In this case, we find 
\begin{equation}
R_{S}^{SOC} = \{\theta_{0}, \pi-\theta_{0}, -\theta_{0}, \pi+\theta_{0} \},
\end{equation}
where $\theta_{0}=arcsin[q/(1-|q|)]$. There is a discrete symmetry for free energy of system and this discrete symmetry turns out to be the mirror symmetry between parametrization in Eq.~(\ref{PARA1}) and the alternative in the presence of KLS domain wall. We discuss the details of this discrete symmetry and its violation in Sec.~\ref{DiscreteSymmetry}. Before that, we mainly use the parametrization in Eq.~(\ref{PARA1}). In the region where condition $|\delta \hat{\mathbf{d}}| \ll 1$ is satisfied, SOC energy fixes the relative rotation of $SO(2)_{S-L}$, thus $R_{2}$ reduces to 
\begin{equation}
\tilde{R}_{2} = \mathbb{Z}_{2}^{S-\Phi}
\end{equation}
in the region with length scale larger than $\xi_{D}$. \\[0.1cm]

From illustrtion of $R_{1}^{H}$, $\tilde{R}_{1}^{SOC}$ and $\tilde{R}_{2}$ in Fig.~\ref{Degenerate_Spaces}, we find again the possibility of utilizing the relative homotopy group to investigate the novel topological objects because of the presence of multiple characteristic length scales \cite{NashBook1988}. This multilength-scales system belongs to type (i) of the classifications in Ref.~\cite{Zhang2020}. Recently. other example of this class is solitons terminated by HQVs observed in spinor Bose condensate with quadratic Zeeman energy \cite{Seji2019,Liu2020}. Both of these systems can be described by the first relative homopoty group. In next section, we discuss this topic.  

\section{1D nexus objects and spin solitons}
\label{SpinSolitonsAndRelativeHomotopyGroup}


\subsection{Spin configuration of KLS string wall -- half spin vortices}
In the region with length scale $\xi_{H} \leq r \leq \xi_{D}$, we have the long exact sequence (LES) of homomorphism of $\pi_{1}(R_{1}^{H},R_{2})$ (for the details of LES, see Appendix.~\ref{LongAndShortES})
\begin{equation}
\xymatrix@1@R=10pt@C=13pt{
\pi_{1}(R_{2}) \ar@{-}[d] \ar[r]^{i^{*}} & \pi_{1}(R_{1}^{H}) \ar@{-}[d] \ar[r]^{j^{*}} &
\pi_{1}( R_{1}^{H}, R_{2}) \ar@{-}[d] \ar[r]^-{\partial^{*}} & \pi_{0}(R_{2}) \ar@{-}[d] \ar[r]^{k^{*}} & \pi_{0}(R_{1}^{H}) \ar@{-}[d]\\
\mathbb{Z}^{S} \ar[r]^{i^{*}} & \mathbb{Z}^{S} \times \mathbb{Z}^{\Phi} \ar[r]^{j^{*}} & \pi_{1}( R_{1}^{H}, R_{2}) \ar[r]^-{\partial^{*}} & \mathbb{Z}_{2} \ar[r]^{k^{*}} & 0
}\,,
\label{LES1a}
\end{equation}
where $i^{*}$ projects spin vortices of $\pi_{1}(R_{2})$ to the spin vortices of $\pi_{1}(R_{1}^{H})$ \cite{NashBook1988,suzuki1982}. And boundary homomorphism $\partial^{*}$ maps all relative $1$-loops of $\pi_{1}(R_{1}^{H},R_{2})$ to their $0$-loops of $\pi_{0}(R_{2})$. Because $\pi_{0}(R_{2}) = \mathbb{Z}_{2}$, the end-points of relative $1$-loop may take values from connected or disconnected subsets of $R_{2}$. This LES can be split to the short exact sequence (SES) 
\begin{equation}
\xymatrix@1@R=10pt@C=13pt{
0 \ar[r] & \mathbb{Z}^{\Phi} \ar[r]^-{\iota} &
\pi_{1}( R_{1}^{H},R_{2}) \ar[r]^-{\pi} & \mathbb{Z}_{2} \ar[r] & 0\\
}\,,
\label{SES1}
\end{equation}
where $\iota$ and $\pi$ are inclusion and surjection respectively. Eq.~(\ref{SES1}) suggests 
\begin{equation}
\pi_{1}( R_{1}^{H},R_{2}) \cong \tilde{\mathbb{Z}},
\end{equation}
which is isomorphic to $\pi_{1}(R_{1},R_{2})$ in the region smaller than $\xi_{H}$ \cite{Zhang2020}. This means KLS string wall, which determined by two length scales $\xi$ and $\xi/q$ in two-step phase transition, extends into the region with length scale $\xi_{H} \leq r \leq \xi_{D}$.  However Eq.~(\ref{SES1}) only contains degree of freedom (DOF) of phase factor $\Phi$, all information about spin degree of freedom lose because they are trivial elements of $\pi_{1}(R_{1}^{H}, R_{2})$. To understand the spin part of KLS string wall, we should take account the continuity of order parameter. The continuity of order parameter $A_{\alpha i}^{PdB}$ requires spin vectors simultaneously change by $(2n+1)\pi$ in the present of KLS string wall \cite{Volovik1990}. This consideration suggests that the spin textures of KLS string wall in the spatial region with length scale $\xi_{H} \leq r \leq \xi_{D}$ are classified by group
\begin{equation}
M \equiv \{ n^{s}/2 | n^{s} \in \mathbb{Z} \},
\end{equation}
such that $M/\pi_{1}(S^{1}_{S}) \cong \mathbb{Z}_{2} = \{ [0], [1/2]\}$. The cosets $[1/2]$ and $[0]$ correspond to the presence or absence of the KLS string wall in the region $\xi_{H} < r \leq \xi_{D}$ respectively. Coset $[0] \cong 2\mathbb{Z}$ contains all free spin vortices. While  Coset $[1/2] \cong \{n + 1/2 | n \in \mathbb{Z}\}$ contains all spin vortices with half-odd winding number i.e., it is set of half spin vortices. 

\begin{figure*}
\centerline{\includegraphics[width=0.95\linewidth]{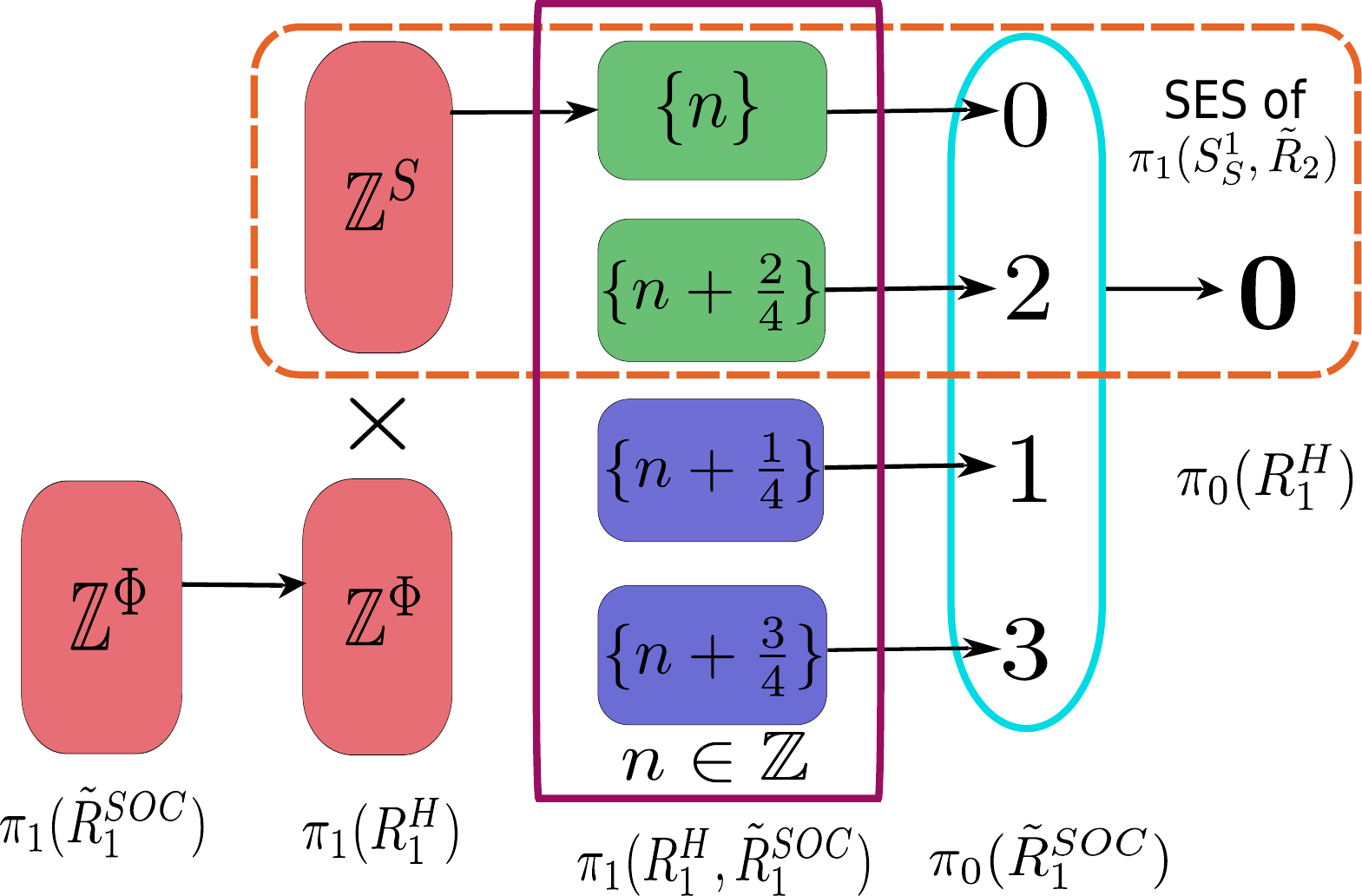}}
\caption{Illustrations of LES of homomorphism for $\pi_{1}(R_{1}^{H},\tilde{R}_{1}^{SOC})$ and short exact sequence of homomorphism for $\pi_{1}(S^{1}_{S},\tilde{R}_{2})$. The black arrows represent the image of homomorphisms between homotopy groups. This mapping diagram demonstrates the linear objects of $\pi_{1}(R_{1}^{H},\tilde{R}_{1}^{SOC})$ are spin solitons. This is because the mapping between $\pi_{1}(\tilde{R}_{1}^{SOC})$ and $\pi_{1}(R_{1}^{H})$ is projection, the image of homomorphism $i^{*} : \pi_{1}(\tilde{R}_{1}^{SOC}) \rightarrow \pi_{1}(R_{1}^{H}) = \mathbb{Z}^{\Phi}$ i.e., topological invariant of all phase vortices.  As a result, the trivial linear objects of $\pi_{1}(R_{1}^{H},\tilde{R}_{1}^{SOC})$ are all phase vortices because of $\operatorname{im}i^{*} \cong \ker j^{*}$. We found there are one kind of spin vortices and three kinds of spin solitons because $\ker k^{*} \cong \operatorname{im} \partial^{*} = \mathbb{Z}_{4}$ and $j^{*}$ is projection. Moreover, we found form this illustration that the subgroup $G=\{[n],[n+2/4]\}$ of $\pi_{1}(R_{1}^{H},\tilde{R}_{1}^{SOC})$ is extension of $\mathbb{Z}^{S}$ by $\pi_{0}(\tilde{R}_{2}^{SOC}) = \mathbb{Z}_{2}$ and then isomorphic to $M$. In the orange dash line panel, we shows the corresponding SES of $G$. As a result, HQV is 1D nexus between spin soliton of coset $[2/4]$ and KLS domain wall in PdB phase. \label{MappingDiagramAndSES}} 
\end{figure*}
\subsection{Spin solition described by $\pi_{1}(R^{H}_{1},\tilde{R}^{SOC}_{1})$}
\label{FourSpinSolitons}
When taking into account SOC, $R_{1}^{H}$ reduces to $\tilde{R}_{1}^{SOC} = R_{S}^{SOC} \times U(1)$ as mentioned in Sec.~\ref{VacuaManifold}. As a result, there are linear objects which classified by $\pi_{1}(R_{1}^{H}, \tilde{R}_{1}^{SOC})$. $\pi_{1}(R_{1}^{H}, \tilde{R}_{1}^{SOC})$ has LES 
\begin{equation}
\xymatrix@1@R=10pt@C=13pt{
\pi_{1}(\tilde{R}_{1}^{SOC}) \ar@{-}[d] \ar[r]^-{i^{*}} & \pi_{1}(R_{1}^{H}) \ar@{-}[d] \ar[r]^-{j^{*}} &
\pi_{1}( R_{1}^{H}, \tilde{R}_{1}^{SOC}) \ar@{-}[d] \ar[r]^-{\partial^{*}} & \pi_{0}(\tilde{R}_{1}^{SOC}) \ar@{-}[d] \ar[r]^-{k^{*}} & \pi_{0}(R_{1}^{H}) \ar@{-}[d]\\
\mathbb{Z}^{\Phi} \ar[r]^-{i^{*}} & \mathbb{Z}^{S} \times \mathbb{Z}^{\Phi} \ar[r]^-{j^{*}} & \pi_{1}( R_{1}^{H}, \tilde{R}_{1}^{SOC}) \ar[r]^-{\partial^{*}} & \mathbb{Z}_{4} \ar[r]^-{k^{*}} & 0
}\,,
\label{LES1}
\end{equation}
where $i^{*}$ is projection and $\partial^{*}$ is boundary homomorphism \cite{Zhang2020b,NashBook1988,suzuki1982}. Figure~\ref{MappingDiagramAndSES} depicts the mapping relation of Eq.~(\ref{LES1}). The relative 1-loop of $\pi_{1}(R_{1}^{H}, \tilde{R}_{1}^{SOC})$ and the boundary 0-loop are shown in Fig.~\ref{Degenerate_Spaces}. Because $\operatorname{im} {\partial}^{*} \cong {\ker} k^{*} = \mathbb{Z}_{4}$, the boundary 0-loop (two end points) of 1-loop takes values from four disconnected subsets of $\tilde{R}_{1}^{SOC}$. For every element of $\tilde{R}_{1}^{SOC}$, there are four possible combinations of elements of $\tilde{R}_{1}^{SOC}$ for 0-loop because of $\pi_{0}(\tilde{R}_{1}^{SOC}) = \mathbb{Z}_{4}$. As a result, we found there are four kinds of linear objects in general, which might be distinguished by four boundary homotopy classes of $\pi_{0}(\tilde{R}_{1}^{SOC})$.  \\[0.1cm]

Now we split the LES in Eq.~(\ref{LES1}) into SES (see details in Sec.~\ref{LongAndShortES})
\begin{equation}
\xymatrix@1@R=10pt@C=13pt{
0 \ar[r] & \mathbb{Z}^{S} \ar[r]^-{\iota} &
\pi_{1}( R_{1}^{H},\tilde{R}_{1}^{SOC}) \ar[r]^-{{\partial}^{*}} & \mathbb{Z}_{4} \ar[r] & 0\\
}\,.
\label{SES2}
\end{equation}
Then we find 
\begin{equation}
\pi_{1}( R_{1}^{H},\tilde{R}_{1}^{SOC}) = \{n^{S}/4 | n^{S} \in \mathbb{Z}\} \cong \mathbb{Z},
\end{equation}
such that 
\begin{equation}
\pi_{1}( R_{1}^{H},\tilde{R}_{1}^{SOC})/\mathbb{Z}^{S} \cong \mathbb{Z}_{4}.
\end{equation}
Because Eq.~(\ref{SES2}) is merely determined by $ \mathbb{Z}^{S} = \pi_{1}(S^{1}_{S})$ and $\mathbb{Z}_{4} = \pi_{0}(R_{S}^{SOC})$, $\pi_{1}(R_{1}^{H}, \tilde{R}_{1}^{SOC})$  actually is isomorphic to $\pi_{1}(S^{1}_{S}, R_{S}^{SOC})$ i.e.,
\begin{equation}
\pi_{1}(R_{1}^{H}, \tilde{R}_{1}^{SOC}) \cong \pi_{1}(S^{1}_{S}, R_{S}^{SOC}).
\label{isomorphism1}
\end{equation}
\begin{figure*}
\centerline{\includegraphics[width=1.0\linewidth]{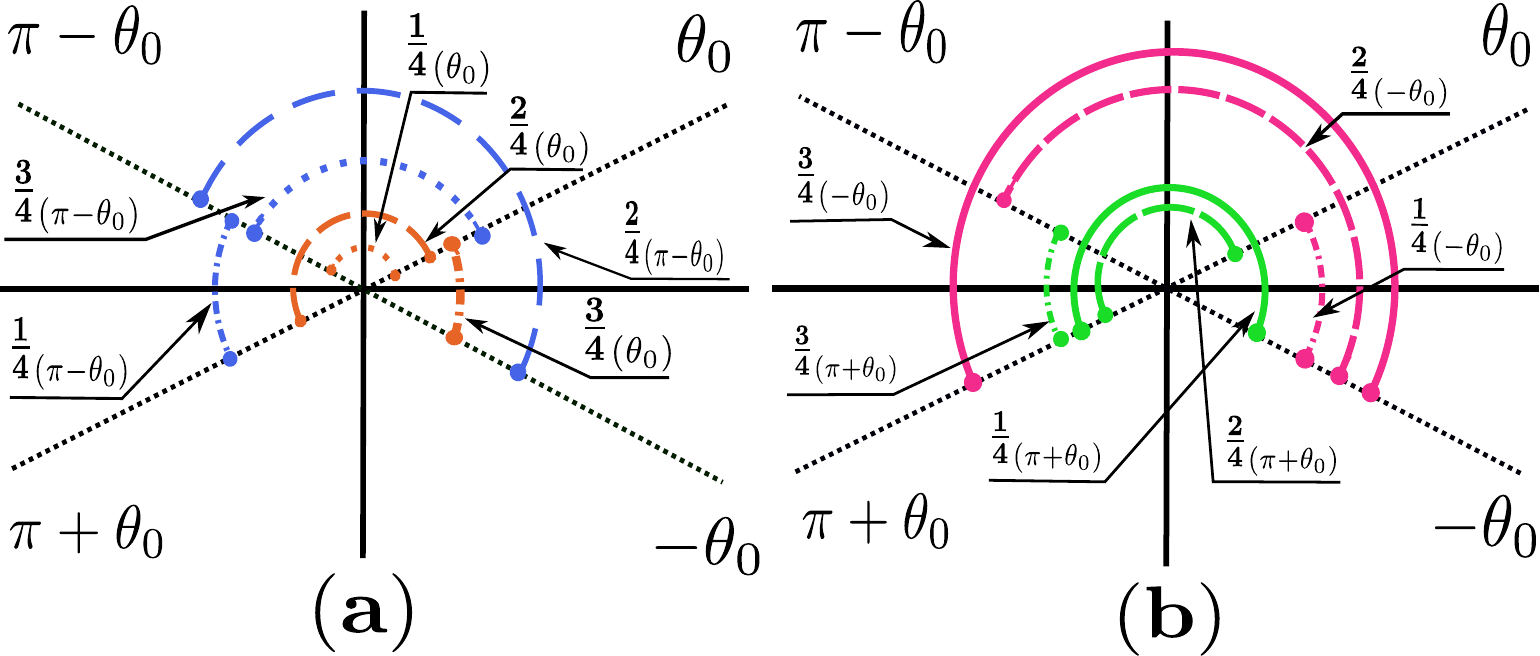}}
\caption{Illustrations of three kinds of spin solitons of $\pi_{1}(S^{1}_{S},R_{S}^{SOC})$. The black dot lines represent the four elements of $R_{S}^{SOC}$ i.e., $\pm \theta_{0}$ and $\pi \pm \theta_{0}$. The dash line, dot line, dash-dot line and solid line correspond to $\pi$-soliton ($|\Delta{\theta}| = \pi$), solition ($|\Delta{\theta}| = \pi -2\theta_{0}$), KLS-Solition ($|\Delta{\theta}| = 2\theta_{0}$) and big-solition ($|\Delta{\theta}| = \pi+2\theta_{0}$) respectively. (a) Spin solitons with topological invariants $1/4$, $2/4$ and $3/4$ for $\theta_{0}$ (orange) and $\pi-\theta_{0}$ (blue) respective. (b) Spin solitons with topological invariants $1/4$, $2/4$ and $3/4$ for $-\theta_{0}$ (pink) and $\pi+\theta_{0}$ (green) respective. \label{ClassesOfSolitons}}
\end{figure*}
This means the linear objects classified by $\pi_{1}(R_{1}^{H}, \tilde{R}_{1}^{SOC})$ are spin solitons and spin vortices \cite{mineyev1979,Zhang2020b}. And the four cosets of $\pi_{1}(S^{1}_{S}, R_{S}^{SOC})$ are 
\begin{equation}
[0] = \{n^{S} \} \,\,, [\frac{1}{4}] = \{n^{S}+\frac{1}{4} \}, \,\, [\frac{2}{4}]  =  \{n^{S}+\frac{2}{4} \} \,\,, [\frac{3}{4}] = \{n^{S}+\frac{3}{4} \}.
\label{cosets}
\end{equation} 
These cosets give out the topological invariants of the four different kinds of linear objects distinguished by four classes of boundary 0-loop of $\pi_{0}(R_{S}^{SOC})$. They correspond to free spin vortices and three kinds of spin solitons respectively. Figure.~\ref{ClassesOfSolitons} shows the representatives of the three classes of spin solitons for every element of $R_{S}^{SOC}$. We omit the spin vortices of $[0]$ from now because it is not energy-favored. From  Fig.~\ref{ClassesOfSolitons}, we found there are four types of spin solitons distinguished by $|\Delta \theta|$. Following the terminologies in Ref.~\cite{Makinen2019}, they are big-solition ($|\Delta \theta| = \pi +2\theta_{0}$), solition ($|\Delta \theta_{0}|=\pi-2\theta_{0}$), KLS-soliton ($|\Delta \theta_{0}|=2\theta_{0}$) and $\pi$-soliton ($|\Delta \theta|=\pi$). To avoid terminological confusion, we claim here that we use phrase "spin soliton" to denote spin textures of $\pi_{1}(S^{1}_{S},R_{S}^{SOC})$ in rest of this review, while use phrases "solitons", "big-solitons", "KLS-solitons" and "$\pi$-solitons" to denote particular spin textures with different $|\Delta \theta|$.  

\subsection{Short exact sequence of $\pi_{1}(S^{1}_{S},\tilde{R}_{2})$ and 1D nexus}
\label{RelativeHomotopyGroupof1DNexus}
A significant property of $\pi_{1}(S^{1}_{S},R_{S}^{SOC})$ is that it has a subgroup $G \equiv \{[0],[2/4]\}$ such that $G/\mathbb{Z}^{S} \cong \mathbb{Z}_{2}$. The SES of $G$ is given as  
\begin{equation}
\xymatrix@1@R=10pt@C=13pt{
0 \ar[r] & \mathbb{Z}^{S} \ar[r] &
G \ar[r]^{\partial^{*}} & \mathbb{Z}_{2} \ar[r] & 0\\
}\,
\label{SES3}
\end{equation}
by Eq.~(\ref{SES2}). The mapping diagram of Eq.~(\ref{SES3}) is shown in the dash panel of Fig.~\ref{MappingDiagramAndSES}. Because $\pi_{0}(\tilde{R}_{2}) \cong \mathbb{Z}_{2}$, Eq.~(\ref{SES3}) can be written as
\begin{equation}
\xymatrix@1@R=10pt@C=13pt{
\pi_{1}(\tilde{R}_{2}) \ar[r] & \pi_{1}(S_{S}^{1}) \ar[r] &
G  \ar[r]^{\partial^{*}} & \pi_{0}(\tilde{R}_{2})  \ar[r] & 0\\
}\,.
\label{SES4}
\end{equation}
This LES suggests 
\begin{equation}
G = \pi_{1}(S^{1}_{S}, \tilde{R}_{2}) \cong \hat{\mathbb{Z}} = M,
\label{isomorphism2}
\end{equation}
here $\hat{\mathbb{Z}} \equiv \{n^{S}/2 | n^{S} \in \mathbb{Z}\}$. \\[0.1cm]

Eq.~(\ref{isomorphism2}) is one of main results in this chapter. This relation means spin solitons, which are classified by coset $[2/4]$ of $\pi_{1}(S^{1}_{S}, \tilde{R}_{2})$ can continuously transform to half spin vortices of $M$. In  other word, KLS domain wall smoothly connects to $[2/4]$ spin soliton via HQV. Similar with 2D nexus which connects string monopole and vortex skyrmion, the HQV is 1D nexus which connects KLS domain wall and $[2/4]$ spin soliton \cite{Zhang2020}. The composite object formed by $[2/4]$ spin soliton and KLS domain wall is then named as 1D nexus object. 

\begin{figure*}
\centerline{\includegraphics[height=0.35\linewidth, width=0.9\linewidth]{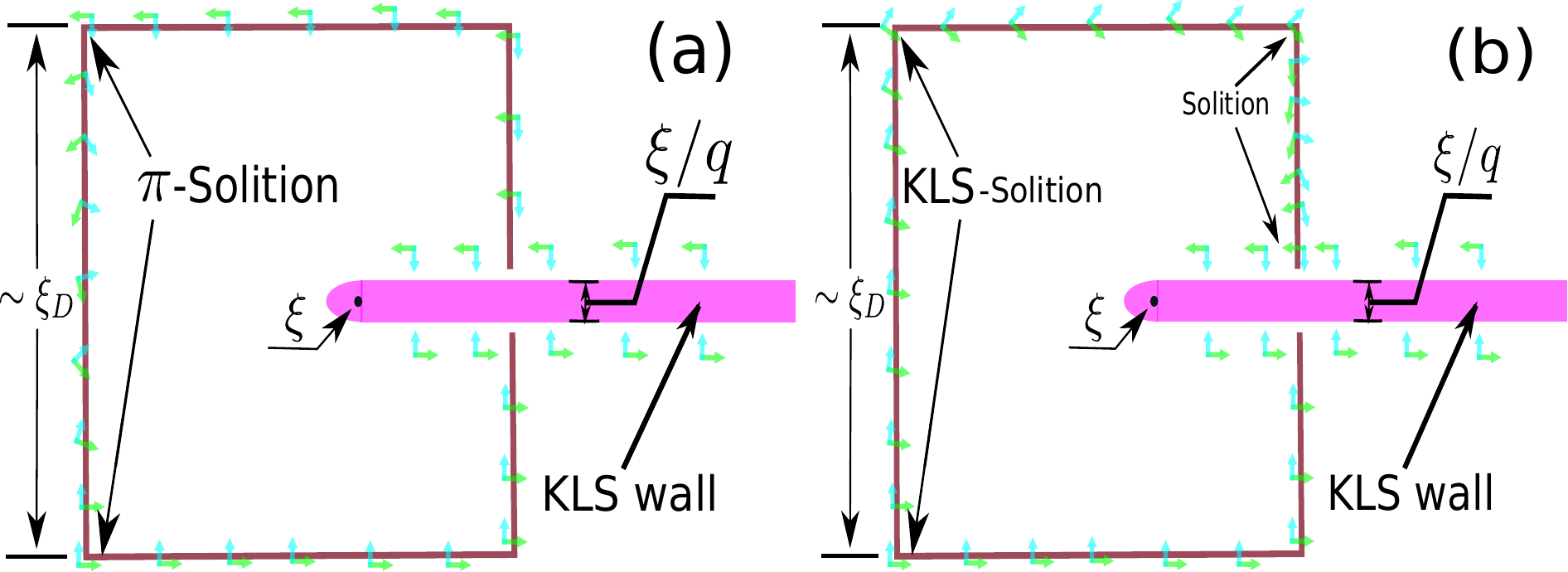}}
\caption{Illustrations of two dramatically different spin configurations of 1D nexus object, which consists of spin solitons, HQV and KLS domain wall. The green and cyan arrows represent the $\hat{\mathbf{d}}$ and $\hat{\mathbf{e}}$ vectors respectively. (a) Configuration of inseparable spin soltion. In this configuration, the topological invariant of spin soliton is literally $1/2$, which corresponds to $\pi$-soliton. In contrast, there are (b) two spin solitons with topological invarant $1/4$ when the group $\pi_{1}(S^{1}_{S}, \tilde{R}_{2})$ is implemented in alternative way. Following the requirement of continuity of order parameter, these two spin solitons are KLS-soliton ($\Delta{\theta}=2\theta_{0}$) and soliton ($\Delta{\theta}=\pi-2\theta_{0}$). 
\label{IllustrationsOfInseparableAndSeparableSolitons}}
\end{figure*}

\subsection{Two different configurations of 1D nexus object  -- separable and inseparable }
\label{TwoDifferentSolitonConfiguretions}
Because $\pi_{1}(S^{1}_{S},R^{SOC}_{S})/\mathbb{Z}^{S} \cong \mathbb{Z}_{4}$, we have $[2/4] = [1/4] + [1/4]$. Thus $\pi_{1}(S^{1}_{S},\tilde{R}_{2})$ could also be represented as $\{[0],[1/4]+[1/4]\}$ besides $\pi_{1}(S^{1}_{S},\tilde{R}_{2}) \cong \{[0],[2/4]\}$. This means there are two kinds of spin soliton configurations connecting with KLS domain wall via HQV for a given element of $\pi_{1}(S^{1}_{S},\tilde{R}_{2})$. When the topological invariant is literally $2/4$, the spin soliton is spatially inseparable $\pi$-soliton in Fig.~\ref{ClassesOfSolitons}. The illustration of configuration of this kind of 1D nexus object is shown in Fig.~\ref{IllustrationsOfInseparableAndSeparableSolitons}(a). \\[0.1cm] 

When the topological invariant is $1/4+1/4$, the spin soliton is combination of two spatially separable spin solitons with topological invariant $1/4$. 
In order to identify these two spatially separable spin solitons, we take in account the requirement of continuity of the order parameters. This requirement is equivalent to the requirement of single-value and continuity of $\theta$. Then the accumulation of $|\Delta\theta|$ of those two spin solitons must equal to $\pi$ because of the KLS domain wall. Based on the discussions of Sec.~\ref{FourSpinSolitons} and Fig.~\ref{ClassesOfSolitons}, These two spatially separated spin solitons are KLS-soliton ($|\Delta{\theta}|= 2\theta_{0}$) and soliton ($|\Delta{\theta}| = \pi - 2\theta_{0}$). As illustrated in Fig.~\ref{IllustrationsOfInseparableAndSeparableSolitons}(b), the 1D nexus object contain two spin solitons in this case. \\[0.1cm]

We will see these two dramatically different spin textures of 1D nexus objects have different equilibrium free energies, different spin dynamic response properties and different NMR frequency shifts in Sec. \ref{EqulibriumTextures} and Sec. \ref{NMRShifts}. These properties help us to identify the objects which be observed in experiment. 

\section{Equilibrium textures of pseudo-random lattices}
\label{EqulibriumTextures}
For the PdB phase generated from symmetry breaking of nonuniform polar phase, we can use the Ginzburg-Landau model to describe the system when $|q|$ is small enough. The Ginzburg-Landau (GL) free energy consists of gradient energy and orientation energies \cite{VollhardtWolfle1990}. In order to quantitatively analyze the equilibrium configurations of 1D nexus objects containing spin solitons with length scale around $\xi_{D}$, we must find out the extreme point of GL free energy under given external parameters. \\[0.1cm]

Because $\xi_{D} \gg \xi$ and the strongly uniaxial anisotropy in the presence of nafen strands, we actually did this procedure under London limit \cite{Ikeda2019,Zhang2020b,VollhardtWolfle1990,volovik1992}. In London limit, all gap parameters attain equilibrium structures and then their magnitudes are constants over whole calculations. This means the HQV and KLS domain wall both achieve their equilibrium structure and the contributions to free energy from them are identical  constant for both spin configurations of 1D nexus objects. When the static magnetic field $\mathbf{H}^{(0)}$ is big enough, the magnetic length $\xi_{H}$ is far smaller than the dipole length $\xi_{D}$, then the magnetic energy has achieved equilibrium over the PdB superfluid as well. In this situation the Ginzburg-Landau free energy in London limit is 
\begin{equation}
F_{London} =\int\nolimits_{\Sigma} (f_{\rm soc} + f_{\rm grad} ) d\Sigma,
\label{GinzburgLandauFreeEnergyTheta}
\end{equation} 
where $\Sigma$ is the volume of the PdB phase sample. \\[0.1cm]

Plunging $A_{\alpha i}^{PdB}$ into Eq.~(\ref{GinzburgLandauFreeEnergyTheta}) and substituting $\hat{\mathbf{d}}$, $\hat{\mathbf{e}}^{1}$ and $\hat{\mathbf{e}}^{2}$ with their parametrizations in Eq.~(\ref{PARA1}), we get the gradient energy density and SOC energy density in term of $\theta$ and $\Phi$
\begin{align}
f_{\rm grad}(\Phi,\theta) = 
& \frac{K_{1}}{2}(\Delta_{P}^{2}+\Delta_{\bot1}^{2}+\Delta_{\bot2}^{2})\partial_{i}\Phi \partial_{i}\Phi + \frac{K_{1}}{2}(\Delta_{P}^{2}+\Delta_{\bot1}^{2})\partial_{i} \theta \partial_{i} \theta  \notag \\ 
& + \frac{1}{2} (K_{2}+ K_{3}) (\Delta_{P}^{2} \partial_{z} \Phi \partial_{z} \Phi + \Delta_{\bot1}^{2} \partial_{x} \Phi \partial_{x} \Phi \notag  \\
& + \Delta_{\bot2}^{2} \partial_{y} \Phi \partial_{y} \Phi + \Delta_{P}^{2}\partial_{z}\theta \partial_{z}\theta + \Delta_{\bot1}^{2} \partial_{x}\theta \partial_{x}\theta),\\
f_{\rm soc}(\theta) = & \frac{g_{D}}{5}(\Delta_{P}^{2} + \Delta_{\bot1}^{2} + \Delta_{\bot2}^{2}) \notag \\ 
& -\frac{3g_{D}}{5}(\Delta_{P}+\Delta_{\bot1})^{2}cos2\theta-\frac{6g_{D}}{5}(\Delta_{P}+\Delta_{\bot1})\Delta_{\bot2}sin\theta, \notag
\end{align}
where $i=1,2,3$ are the summation indexes of spatial coordinates. In London limit, the term ($g_{D}/5)(\Delta_{P}^{2} + \Delta_{\bot1}^{2} + \Delta_{\bot2}^{2})$ is constant over the sample, thus we drop it in the rest. Because spin degree of freedom does not couple with phase degree of freedom, $f_{grad}(\Phi,\theta)$ is simply the summation of $f_{grad}(\Phi)$ and $f_{grad}(\theta)$, where $f_{grad}(\Phi)$ and $f_{grad}(\theta)$ are the gradient energy densities of phase and spin vectors respectively. Then $f_{grad}(\Phi)$ achieves equilibrium independently and can be safely dropped. Moreover, because the HQVs are pinned by nafen strands, the system is translation invariant along the direction of nafen strands, thus all $\partial_{z}\theta$ terms vanish. Finally the free energy, which determines the equilibrium textures in London limit is
\begin{equation}
F(\theta)_{London} =\int\nolimits_{\Sigma} [f_{soc}(\theta) + f_{grad}(\theta)] d\Sigma ,
\label{FreeEnergyTheta}
\end{equation}
where $f_{\rm grad}(\theta)$ and $f_{\rm soc}(\theta)$ are given as
\begin{align}
f_{\rm grad}(\theta) = & \frac{K_{1}}{2}(\Delta_{P}^{2}+\Delta_{\bot1}^{2})(\partial_{x} \theta \partial_{x}\theta + \partial_{y} \theta \partial_{y} \theta) + \frac{1}{2} (K_{2}+K_{3}) \Delta_{\bot1}^{2} \partial_{x}\theta \partial_{x}\theta,  \\
f_{\rm soc}(\theta)= & -\frac{3g_{D}}{5}(\Delta_{P}+\Delta_{\bot1})^{2}cos2\theta -\frac{6g_{D}}{5}(\Delta_{P}+\Delta_{\bot1})\Delta_{\bot2}sin\theta.\notag 
\end{align} 
In this section, we utilize the nonlinear optimization BFGS algorithm to minimize the free energy functional Eq.~(\ref{FreeEnergyTheta}) \cite{JorgeBook2006}. (We discuss the details of BFGS algorithm and its implements in Appendix.~\ref{RitzStrategyAndFiniteElementsPartition} and Appendix.~\ref{BFGSAlgorithmAndImplements}.)  The saddle points $\theta$ of free energy under different parameters are the equilibrium textures of spin solitons of 1D nexus objects. To facilitate minimization of free energy with nonlinear optimization algorithm, we reduce Eq.~(\ref{FreeEnergyTheta}) to 
\begin{align}
\tilde{F}(\theta)_{London} 
 = & {\frac{1}{\xi_{D}}}\int\nolimits_{\Sigma} [\frac{1}{2} (\gamma_{1}+2\gamma_{2}) \partial_{x}\theta \partial_{x}\theta \notag  + \frac{1}{2}\gamma_{1}\partial_{y}\theta \partial_{y}\theta + \frac{1}{\xi_{D}^{2}} 
(-\frac{1}{2}{\gamma_{4}}cos2\theta-\gamma_{3}sin\theta)] d\Sigma  \\
 = & \frac{1}{\xi_{D}} \int_{\Sigma} (\tilde{f}_{grad} + \tilde{f}_{soc}) d\Sigma
\label{FreeEnergyThetaDimensionless}
\end{align}
by multiplying $(\xi_{D}K_{1}\Delta_{P}^{2})^{-1}$, where 
\begin{align}
q = \frac{\Delta_{\bot2}}{\Delta_{P}},\,\,
 \gamma_{1}=1+|q|^{2},\,\, 
\gamma_{2}=|q|^{2},\,\, 
\gamma_{3}=q(1+|q|),\,\,
\gamma_{4}=(1+|q|)^{2},
\end{align}
and 
\begin{align}
\tilde{f}_{grad} & = \frac{1}{2} (\gamma_{1}+2\gamma_{2}) \partial_{x}\theta \partial_{x}\theta + \frac{1}{2}\gamma_{1}\partial_{y}\theta \partial_{y}\theta \notag \\ 
\tilde{f}_{soc} & = \frac{1}{\xi_{D}^{2}} (-\frac{1}{2}{\gamma_{4}}cos2\theta-\gamma_{3}sin\theta).
\label{ReducedEnergyDensity}
\end{align}
$\xi_{D}K_{1}\Delta_{P}^{2}$ is the characteristic unit of London limit free energy in this review. \\[0.1cm]

Before talking about those numeric results and analyzing the corresponding physics, we discuss the random lattice of HQVs and $2/4$ spin solitons formed by the random pinning effect of nafen strands \cite{Makinen2019,Volovik2008,Dmitriev2008}. We analyze the condition under which the coupling between spin solitons induced by random distributions of HQVs can be neglected. The random lattice of spin solitons is pseudo-random lattices as long as this condition is satisfied. This allows us to understand the network of 1D nexus objects consisting of $2/4$ spin solitons and KLS string walls by calculating and analyzing unit cell of pseudo-random lattices consisting of spin solitons.      
\begin{figure*}
\centerline{\includegraphics[width=1.0\linewidth]{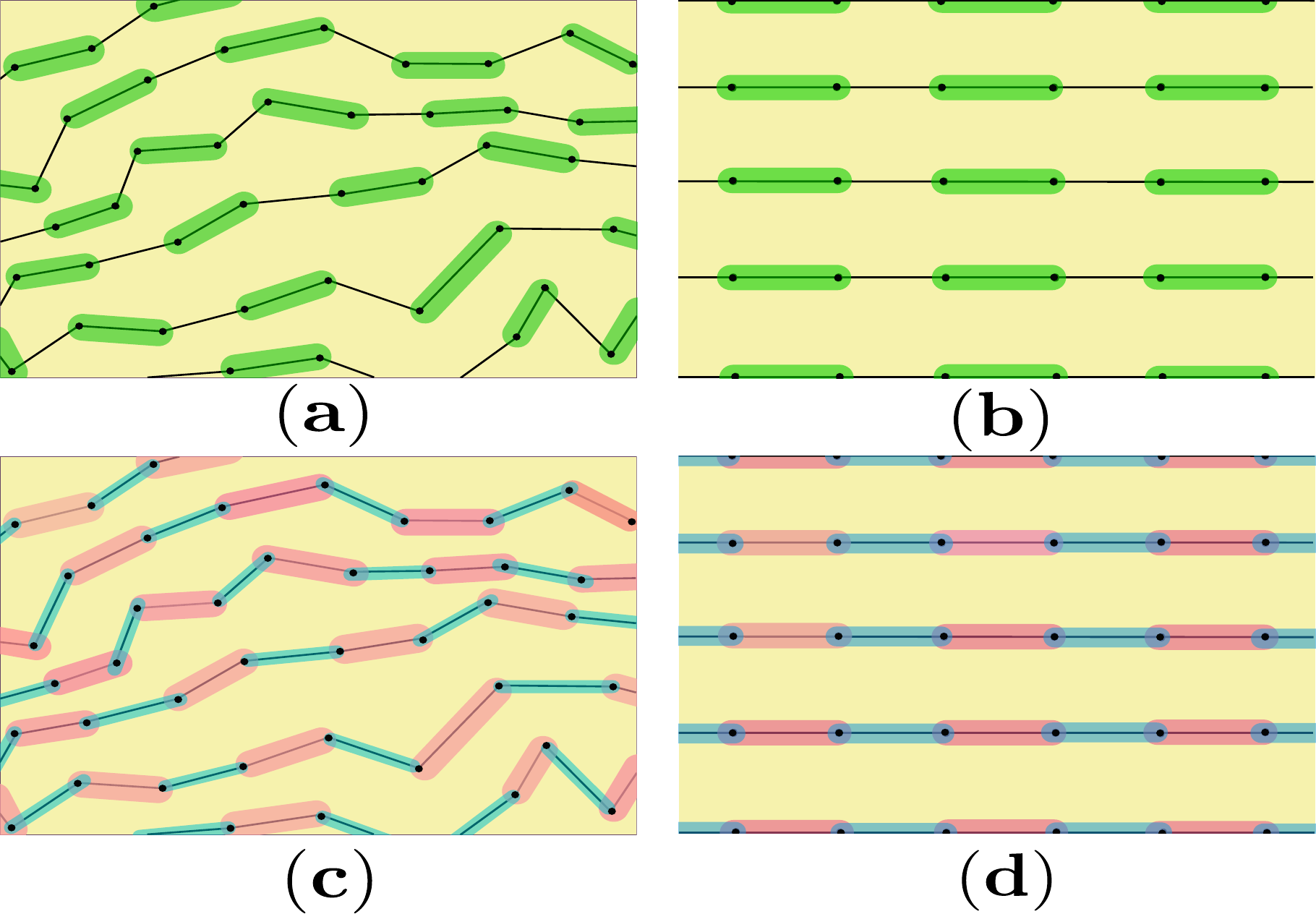}}
\caption{Illustrations of pseudo-random lattices of 1D nexus objects and their equivalent regular lattices. The black dots represent the HQVs and the black solid lines represent the KLS domain walls. Because every HQV is 1D nexus, two HQVs connect with each others via separable or inseparable $2/4$ spin solitons. These topological objects with different characteristic lengths and spatial dimensions give rise to the network of 1D nexus objects with complex hierarchy of length scales. Thus this network is an instance of the interplay between different homotopy groups of topological objects with dramatically different length scales.
The small green filleted rectangles represent the inseparable spin solitons, while pink and blue filleted rectangles represent separable spin solitons. (a) pseudo-random lattice of inseparable spin solitons ($\pi$-solitons) when $\Omega \ll \Omega_{c}$. The spin solitons are almost identical and well spatially separated with each others. The spin dynamic response properties of pseudo-random lattice is equivalent to (b) 2D regular lattice of $\pi$-solitons. Similarly, (c) pseudo-random lattice of separable spin solitons (KLS-solitons and solitons) has same spin dynamic response with (d) 2D regular lattice consisting of KLS-solitons ($|\Delta{\theta}| = 2\theta_{0}$) and solitons ($|\Delta{\theta}| = \pi - 2\theta_{0}$). \label{PsudoRamdomLatticeOfSolitons}} 
\end{figure*}

\subsection{Pseudo-random lattices}
\label{PseudoRandomLattice}
In the experiment of PdB phase, the HQVs are pinned by nafen strands when they appear during cooling down. Hence the 
HQVs and KLS string walls randomly distribute in the PdB sample and form network.
The statistic distribution of HQVs is uniform because there is no reason provides preferable location for HQV. This means the number of HQVs in unit area is constant for rotating PdB superfluid with angular velocity $\Omega$. Then the average area occupied by one HQV is constant as well. We denote the average area occupied by HQV as $A=D(\Omega)^{2}$, where $D(\Omega)$ is the average distance between two HQVs. $D(\Omega)$ depends on the angular velocity as 
\begin{equation}
D(\Omega)=\sqrt{A}=\sqrt{\frac{\kappa_{0}}{4\Omega}},
\label{DOmega}
\end{equation}
where $\kappa_{0}=h/2m$ is the circulation of HQV and $m$ is mass of $^3$He atom \cite{Salomaa1985,Autti2016}. \\[0.1cm]

In Fig.~\ref{PsudoRamdomLatticeOfSolitons} (a) and (c), we illustrate the uniformly distributed HQVs under given $\Omega$. These HQVs, as we have known at Sec. \ref{RelativeHomotopyGroupof1DNexus} and \ref{TwoDifferentSolitonConfiguretions}, are 1D nexuses which connect $2/4$ spin solitons and KLS domain walls. Because the random distribution of HQVs, the $2/4$ spin solitons are also randomly distributed over the PdB superfluid. Therefore the HQVs and spin solitons form a 2D random lattice \cite{volovik2019}. These spin solitons have almost identical spin configuration and geometric size determined by gradient energy and SOC energy. Their spin dynamic response under weak magnetic drive are almost identical as well. As a result, the spin dynamic response of these spin solitons under weak drive is independent to the distribution of HQVs and spin solitons. The NMR frequency shift under weak magnetic drive is merely determined by the configuration of one spin soliton, and the total ratio intensity of system is the summation of ratio intensities of all spin solitons. We call this kind of random lattice of HQVs and spin solitons as pseudo-random lattice. This means the spin dynamic response properties of pseudo-random lattice of $2/4$ spin solitons are equivalent to the spin dynamic properties of regular lattice of $2/4$ spin solitons. There are two types of regular lattices as shown in Fig.~\ref{PsudoRamdomLatticeOfSolitons} (b) and (d), which correspond to inseparable and separable $2/4$ spin solitons respectively. \\[0.1cm]  
\begin{figure*}
\centerline{\includegraphics[width=0.8\linewidth]{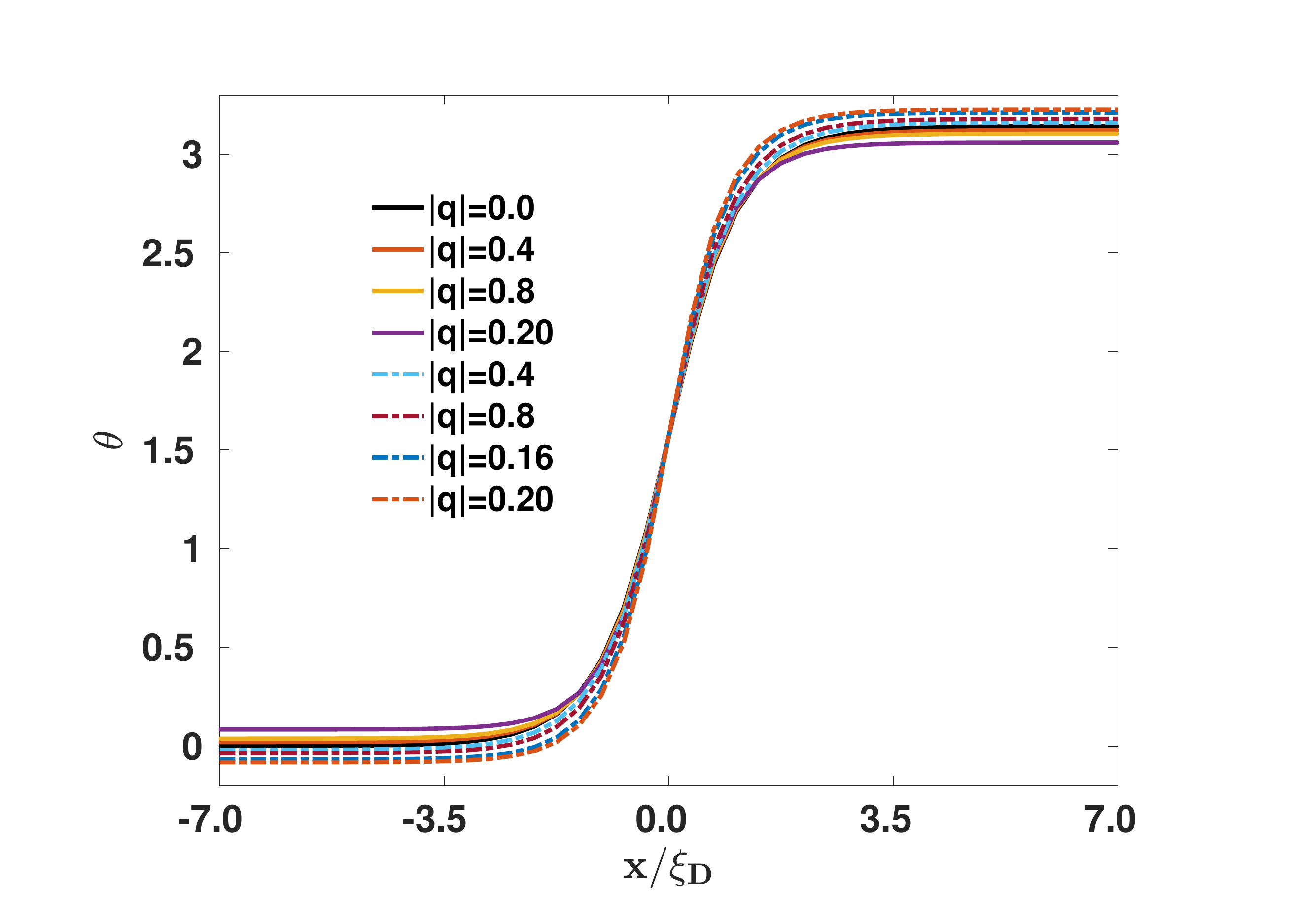}}
\caption{Equlibrium spin configurations of solitons ($|\Delta{\theta}|=2\pi-\theta_{0}$) and big-solitons ($|\Delta{\theta}|=2\pi+\theta_{0}$) in uniform domains. Black solid line represents spin soliton of polar phase ($|q|=0$). The colored dash lines represent big-solitons of PdB phase with $q<0$ from $|q|=0.04$ until $|q|=0.2$. The colored solid lines represent solitons of PdB phase with $q>0$ from $q=0.04$ until $q=0.2$. The spin vectors of all solitons and big-solitons have same relative direction respect to orbital frame because $\theta=\pi/2$ is the stationary point of $\partial_{x} {\theta}$. \label{SolitonAndBigSoliton}} 
\end{figure*}

However, the pseudo-random lattice model doesn't work any more when the angular velocity of PdB system increase up to critical value $\Omega_{c}$. Recalling that the average distance $D(\Omega)$ between two HQVs is proportional to $1/\sqrt{\Omega}$ as shown in Eq.~(\ref{DOmega}). This means spin solitons overlap and couple with each others when $\Omega$ is big. This is because the characteristic thickness of spin soliton i.e., $\sim2\xi_{D}$ is constant under given external parameters. The independence of the spin solitons between two 1D nexus objects loses when $D(\omega) \sim 2\xi_{D}$ and the static textures of spin solitons strongly depend on the distribution of HQVs. As a result, the spin dynamic response of the random lattice of spin solitons under weak magnetic drive strongly depends on the distribution of HQVs as well. Thus the upper limit of $\Omega$, under which pseudo-random lattice model works, is determined by $\sqrt{\kappa_{0}/4\Omega_{c}} \sim 2 \xi_{D}$ and then
\begin{equation}
\Omega_{c} \sim \frac{\kappa_{0}}{16 \xi_{D}^{2}}.
\label{OmegaC}
\end{equation}   
For PdB system with $\kappa_{0}=6.62 \times 10^{-8} m^{2}/s$ and $ \xi_{D} \sim 10^{-6}m$ to $\sim 10^{-5}m$, Eq.~(\ref{OmegaC}) suggests $\Omega_{c} \sim 10^{1} rad/s$ to $ \sim 10^{3} rad/s$. These values is larger enough than the angular velocity of PdB system in the experiment of Ref.~\cite{Makinen2019}, then pseudo-random lattice model is good and we keep working with it in the rest parts of this review. 

\subsection{Spin solitons in the absence of KLS string walls -- solitons and big-solitons}
\label{AbsenceOfHQVs}
In order to understand the 1D nexus object consisting of $2/4$ spin soliton and KLS string wall, we start from the simpler situation in which there is absence of KLS string wall. In this case $\Delta_{\bot2}$ is single valued over the sample of superfluid, then only solitons ($|\Delta \theta|=\pi-2\theta_{0}$) with topological invariant $1/4_{(\theta_{0})}$, $3/4_{(\pi-\theta_{0})}$ and big-solitons ($|\Delta \theta|=\pi+2\theta_{0}$) with topological invariant $1/4_{(\pi+\theta_{0})}$, $3/4_{(-\theta_{0})}$ are possible in the system. These two different cases correspond to spin solitons in uniform domain with $\Delta_{\bot2}=+|q|\Delta_{P}$ or $\Delta_{\bot2}=-|q|\Delta_{P}$ respectively. Moreover, the spin textures have translation symmetry along transverse direction of spin solitons, then the question reduces to one dimensional question. As mentioned before, we use the BFGS non-linear optimization algorithm on Eq.~(\ref{FreeEnergyThetaDimensionless}) to get the equilibrium configuration of spin solitons \cite{JorgeBook2006}. (We discuss the details of BFGS algorithm and its implements in Appendix.~\ref{RitzStrategyAndFiniteElementsPartition} and Appendix.~\ref{BFGSAlgorithmAndImplements}.)\\[0.1cm]

In Fig.~\ref{SolitonAndBigSoliton}, we show the equilibrium configuration of solitons and big-solitons from $|q|=0$ to $|q|=0.2$. The spin textures with $q>0$ are solitons, while the spin textures with $q<0$ are big-solitons. We find that the spin vectors of all solitons and big-solitons have common direction $\theta=\pi/2$. This is because $\theta=\pi/2$ is stationary point of $\partial_{x} \theta$, then $\partial_{x}\partial_{x} \theta|_{\theta=\pi/2}=0$ for all solitons and big-solitons. We will soon see this important feature helps us to set appropriate boundary condition for searching equilibrium textures of pseudo-random lattices consisting of $\pi$-solitons.

\subsection{Spin solitons in the presence of KLS string walls -- Boundary conditions on wall}
\label{EquilibriumTexturesOfInseparableAndSeparableSolitons}
As we have discussed in Sec. \ref{RelativeHomotopyGroupof1DNexus} and Sec. \ref{TwoDifferentSolitonConfiguretions}, the HQV is 1D nexus connecting KLS domain wall and $2/4$ spin solitons. In London limit, the free energy of network of 1D nexus objects is free energy of pseudo-random lattices consisting of $2/4$ spin solitons. The equilibrium configuration of pseudo-random lattices is the saddle point of Eq.~(\ref{FreeEnergyThetaDimensionless}). The only complexity here is the topological invariant $2/4$ has two different representations i.e., literal $2/4$ or $1/4 + 1/4$. Based on the topological analysis, we have known these two cases correspond to inseparable $\pi$-soliton configuration and separable configurations of KLS-soliton ($|\Delta{\theta}|= 2\theta_{0}$) and soliton ($|\Delta{\theta}| = \pi - 2\theta_{0}$). \\[0.1cm]

In order to quantitatively get the equilibrium spin textures for both configurations, we minimize the London limit free energy Eq.~(\ref{FreeEnergyThetaDimensionless}) in the presence of KLS string wall. For parametriztion Eq.~(\ref{PARA1}), KLS string wall separates two domains with oppsite $\Delta_{\bot2}$ in an unit cell of pseudo-random lattice of spin solitons. \\[0.1cm]

However, different from the situation with uniform domain for soliton and big-soliton in Sec.~\ref{AbsenceOfHQVs}, the existence of KLS domain wall induces a singularity of the London limit free energy $\tilde{F}(\theta)$. That is because the order parameter $A_{\alpha i}^{PdB}$ in the London limit is ill-defined on the KLS domain wall. As a result, the free energy Eq.~(\ref{FreeEnergyThetaDimensionless}) and corresponding Lagrangian equation of $\theta$ are also ill-defined on the KLS domain wall. On the other hand, we know $\theta$ is a continuous function everywhere for $2/4$ spin soliton because the relative 1-loop of $\pi_{1}(S_{S}^{1},\tilde{R}_{2})$ is continuous mapping. Then $\theta$ keeps single-valued and continuous. These facts require us to set a proper boundary condition of $\theta$ on the KLS domain wall. The London limit free energy Eq.~(\ref{FreeEnergyThetaDimensionless}) can be minimized with this boundary condition. \\[0.1cm]

\begin{figure*}
\centerline{\includegraphics[width=1.0\linewidth]{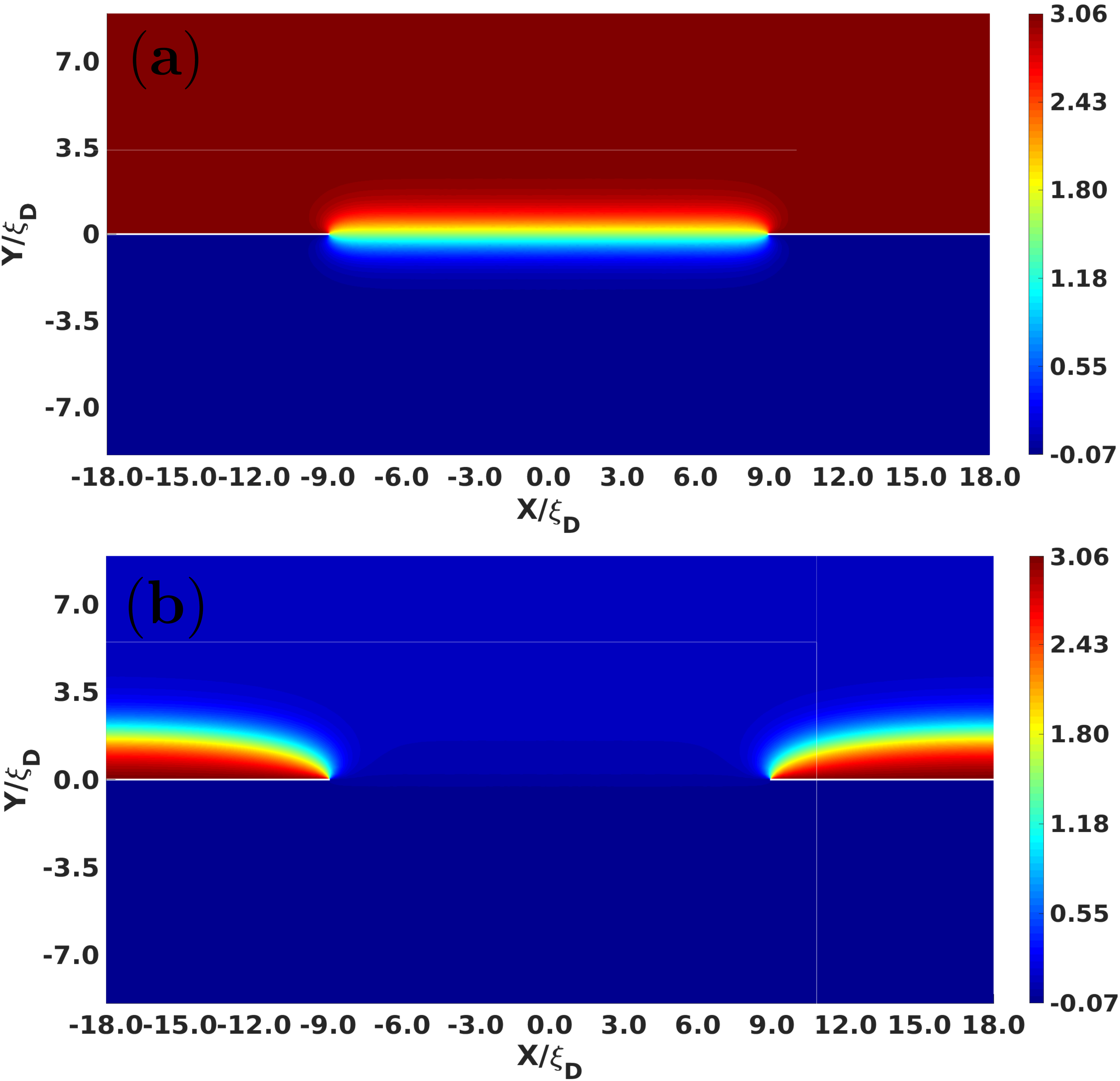}}
\caption{Two kinds of equilibrium configurations of unit cell of pseudo-random lattices \cite{Zhang2020b}. These equilibrium spin textures are gotten by minimizing the reduced London limit free energy $\tilde{F}(\theta)_{London}$ in Eq.~(\ref{FreeEnergyThetaDimensionless}) by BFGS algorithm. The resulted equilibrium distributions of $\theta$ depict the equilibrium textures of spin vectors in a unit cell of pseudo-random lattices.
(a) depicts the unit cell of pseudo-random lattice consisting of inseparable $2/4$ spin solitons ($\pi$-solitons) for $|q|=0.18$ and $D=18\xi_{D}$. (b) depicts the unit cell of pseudo-random lattice consisting of separable spin solitons (KLS-Solitons and solitons) with same parameters of (a) but its topological invariant is $1/4+1/4$. \label{SeparableAndInSeparableSolitons}} 
\end{figure*}

\begin{figure*}
\centerline{\includegraphics[width=1.0\linewidth]{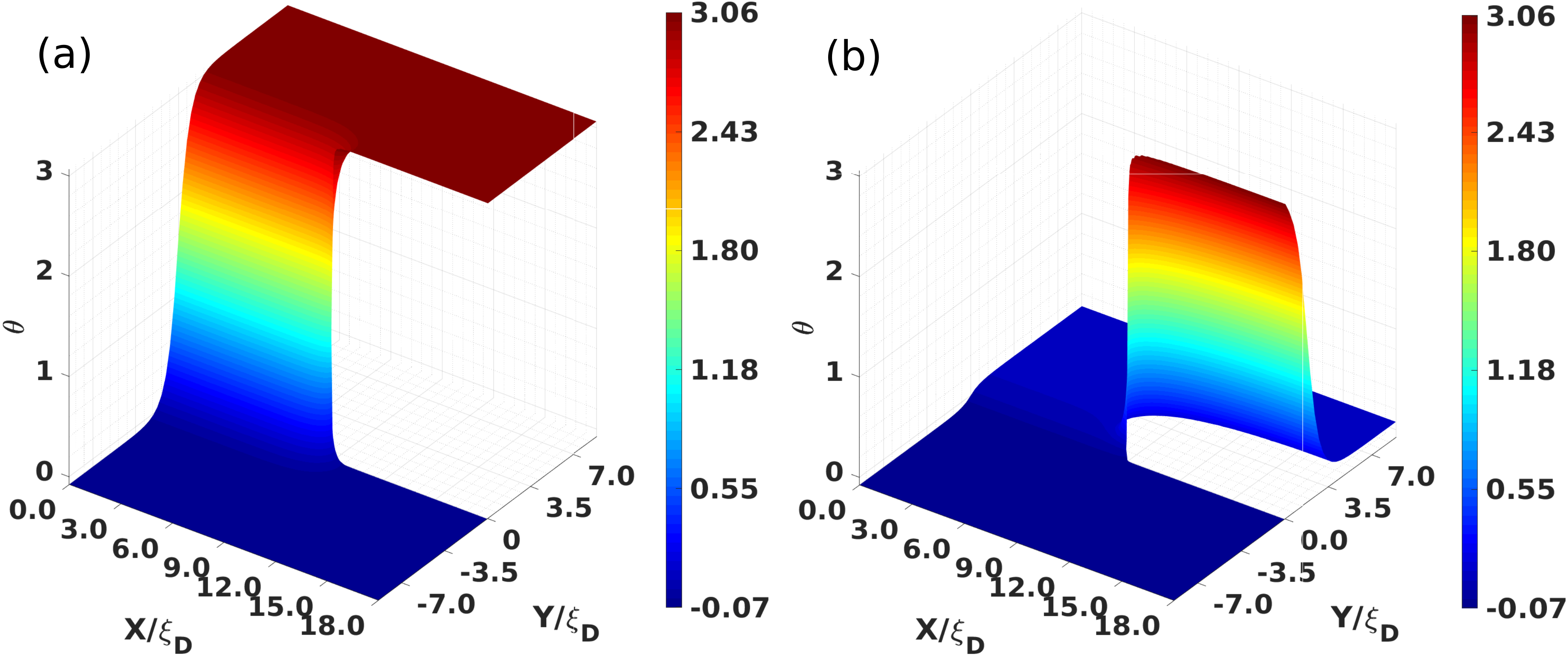}}
\caption{The equilibrium configurations of inseparable and separable spin solitions in one-half unit cell with $|q|=0.18$ and $D=18\xi_{D}$. (a) Inseparable ($\pi$-soliton) configuration; (b) Separable (KLS-soliton and soliton) configuration.  
\label{OneHalf}   
} 
\end{figure*}

Because the free energy and Lagrangian equation of $\theta$ is ill-defined on the KLS domain wall in London limit, $\theta$ of different domains in the vicinity of the KLS domain wall does not relate to each other by Lagrangian equation. Then $\theta$ in two different domains, which are separated by KLS domain wall, are determined independently in two uniform domains with opposite $\Delta_{\bot2}$. In this situation, to keep the continuity of $\theta$ on the KLS domain wall, the boundary condition of $\theta$ must be a common value of spin solitons in both two domains with opposite $\Delta_{\bot2}$. For the inseparable spin soliton with literally topological invariant $2/4$, the natural choice is the stationary point of big-soliton and soliton i.e., $\theta_{KLS}=\pi/2$. This boundary condition indicates the $\pi$-soliton may be understood as a hybrid of big-soliton and soliton in London limit. As for the separable spin soliton with topological invariant $1/4+1/4$, because all KLS-solitons have common values $\theta=0$ or $\theta=\pi$ on the KLS domain wall, there are two options of boundary condition \cite{Zhang2020b}. However, these two options are identical, they give rise to same spin textures of pseudo-random lattices consisting of separable spin solitons, see details in Appendix.~\ref{SeparableSolitionTwoBoundryConditions}. Thus in the rest of this review, we exclusively use $\theta_{KLS} = 0$ for all calculations about separable spin solitons in main text.

\begin{figure*}
\centerline{\includegraphics[width=0.87\linewidth]{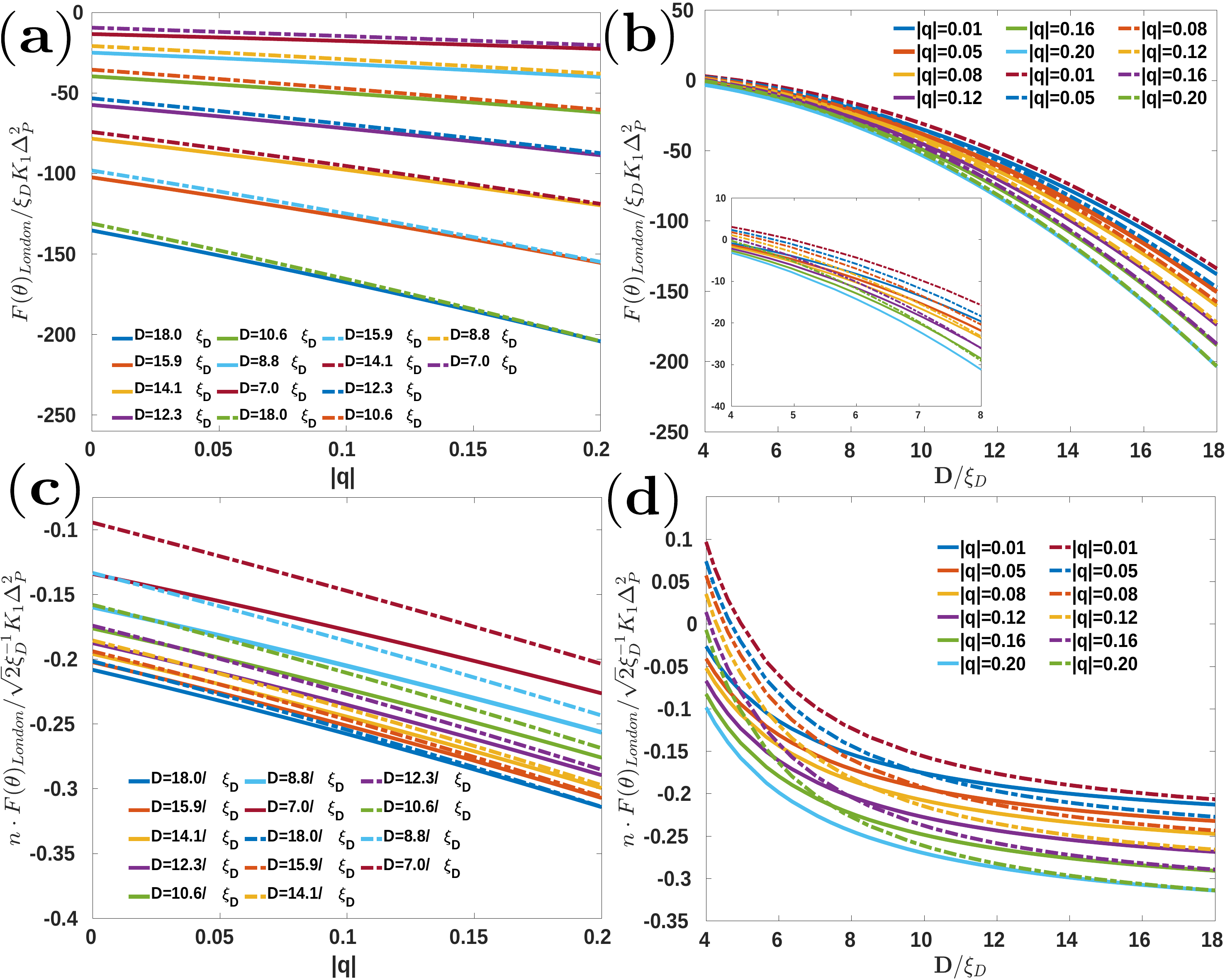}}
\caption{Free energies $F(\theta)_{London}$ of one-half unit cell and surface densities of free energies $n \cdot F(\theta)_{London}$ of pseudo-random lattices consisting of $2/4$ spin solitons. The external parameters $|q|$ are from $0.01$ to $0.20$ and $D$ are from $4\xi_{D}$ to $18\xi_{D}$.
The reduced free energies $\tilde{F}(\theta)_{London}$ and reduced densities of free energies $n \cdot \tilde{F}(\theta)_{London}$ are calculated based on the equilibrium spin textures of one-half unit cell of the lattices with the pseudo-random lattice model. These results depict the equilibrium free energies and energy densities of spin degree of freedom of 1D nexus objects network in PdB system. The solid lines represent the inseparable spin solitons, and the dash-dot lines represent the separable spin solitons. (a) shows the free energies $F(\theta)_{London}$ as functions of $|q|$ with different average distances $D$. When SOC energy dominates the system in big unit cell, $F(\theta)_{London}$ is monotonically decreasing respect to $|q|$. While $F(\theta)_{London}$ does not show remarkably change as $|q|$ changes when gradient energy is competitive to SOC energy in a small unit cell. (b) shows the free energies $F(\theta)_{London}$ are monotonically decreasing  functions of $D$. The zooming plot between $4\xi_{D}$ and $8\xi_{D}$ in (b) demonstrates this monotonicity is held even when the gradient energy is competitive with the SOC energy. Similarly, (c) depicts the London limit free energy density $n \cdot \tilde{F}(\theta)_{London}$ is monotonically decreasing function of $|q|$. However, (d) demonstrates the London limit free energy densities $n \cdot \tilde{F}(\theta)_{London}$ asymptotically trend to constants determined by SOC energy when SOC energy is the dominating energy in big unit cell. When $D$ is small enough ($D<6\xi_{D}$) and gradient energy becomes to the dominating energy, the free energy densities increase rapidly as $D$ decrease because $n \cdot \tilde{F}(\theta)_{London}|_{D<6\xi_{D}}\propto{1/D}$. All these results show the equilibrium free energies of pseudo-random lattice consisting of inseparable spin solitons ($\pi$-solitons) are lower than those of separable spin solitons (KLS solitons and solitons). \label{FreeEnergyBoxAndDensity}   
} 
\end{figure*}
\subsection{Equilibrium spin textures and free energies of pseudo-random lattices consisting of inseparable and separable $2/4$ spin solitons}
\label{EquilibriumSpinTexturesAndFreeEnergy}
In Fig.~\ref{SeparableAndInSeparableSolitons}, we show the equilibrium textures of pseudo-random lattices consisting of inseparable and separable $2/4$ spin solitons with $|q|=0.18$ and $D=18\xi_{D}$. These two equilibrium configurations of a pair of 1D nexus objects are two kinds of unit cells of pseudo-random lattices. Both of them are axially symmetric about $x=0$ as suggested in Sec.~\ref{PseudoRandomLattice}. This property allows me to increase the precision of calculation by just calculating one-half of unit cell. As an example, the equilibrium structures of spin solitons, which were gotten from BFGS optimization in one-half unit cell, are shown in Fig.~\ref{OneHalf}. To collect enough data which could be used to calculate spin dynamic response and compare with experiment, I calculated spin textures with parameters $|q|$ from $0$ to $0.2$ and $D$ from $4\xi_{D}$ to $18\xi_{D}$. Based on these data, we further calculated the reduced London limit free energy Eq.~(\ref{FreeEnergyThetaDimensionless}) of these two types of pseudo-random lattices, the results are shown in Fig.~\ref{FreeEnergyBoxAndDensity}. Before we discussing these numeric results, we first evaluate the Eq.~(\ref{FreeEnergyThetaDimensionless}) for one-half of unit cell when $D \geq 10\xi_{D}$. In this case, 
\begin{align}
\int_{\Sigma} \tilde{f}_{grad} \, d\Sigma\,\, & \sim \frac{1}{2}(1+|q|^{2})\frac{\pi^{2}}{4\xi_{D}^{2}}\int_{\Sigma'}d\Sigma, \notag \\ \int_{\Sigma} \tilde{f}_{soc} \, d\Sigma\,\, & \sim \int_{\Sigma'} \tilde{f}_{soc} \, d\Sigma\,\, + \int_{\Sigma-\Sigma'} \tilde{f}_{soc} \, d\Sigma,\,\,
\label{EvaluationOfFreeEnergy1} 
\end{align} 
where $\Sigma'$ is the region which spin solitons occupy. Its area in $x$-$y$ plane is around $D\xi_{D}$. Then the integral of $\tilde{f}_{soc}$ in Eq.~(\ref{EvaluationOfFreeEnergy1}) can be evaluated as
\begin{align}
\int_{\Sigma'} \tilde{f}_{soc} \, d\Sigma\,\, & \sim 0, \notag \\ 
\int_{\Sigma-\Sigma'} \tilde{f}_{soc} \, d\Sigma\,\, & \sim \int_{\Sigma-\Sigma'} \tilde{f}_{soc}|_{y>0} \, d\Sigma\,\, + \int_{\Sigma-\Sigma'} \tilde{f}_{soc}|_{y<0} \, d\Sigma.\,\,
\label{EvaluationOfFreeEnergy2} 
\end{align} 
The first integral in Eq.~(\ref{EvaluationOfFreeEnergy2}) vanishes because $\tilde{f}_{soc}$ is not negative-definite function in $\Sigma'$. In contrary, $\tilde{f}_{soc}$ has negative-definite equilibrium values in regions $(\Sigma-\Sigma')_{y>0}$ and  $(\Sigma-\Sigma')_{y<0}$. Hence 
\begin{equation}
\int_{\Sigma-\Sigma'} \tilde{f}_{soc} \, d\Sigma\,\, \sim  \frac{{\xi_{D}}D(D-\xi_{D})}{2} [\tilde{f}_{soc}|_{y>0} \,\, + \tilde{f}_{soc}|_{y<0}]_{q} .\,\,
\label{EvaluationOfFreeEnergy3}
\end{equation}
As a result, the reduced London limit free energy is evaluated as
\begin{align}
\tilde{F}(\theta)_{London} & \, \sim 
 \frac{1}{8}(1+|q|^{2})\frac{\pi^{2}D}{\xi_{D}} \, + \, \frac{D(D-\xi_{D})}{2} [\tilde{f}_{soc}|_{y>0} \,\, + \tilde{f}_{soc}|_{y<0}]_{q} \notag \\
& \sim \frac{1}{8}(1+|q|^{2})\frac{\pi^{2}D}{\xi_{D}} \, + \, \frac{D(D-\xi_{D})}{2{\xi_{D}^{2}}} \notag \\ 
& \times [-(1+|q|)^{2}cos2|\theta_{0}| -2(1+|q|)|q|sin|\theta_{0}|]. \label{EvaluationOfFreeEnergy4}
\end{align}
Eq.~(\ref{EvaluationOfFreeEnergy4}) immediately suggests SOC energy is dominating energy of London limit free energy when the average distance $D$ is big and the $\tilde{F}(\theta)_{London}<0$ because $(\tilde{f}_{soc}|_{y>0} \,\, + \tilde{f}_{soc}|_{y<0})<0$ over $\Sigma-\Sigma'$. For $|q|\in[0,0.2]$, $\tilde{F}(\theta)_{London}$ in Eq.~(\ref{EvaluationOfFreeEnergy4}) is around $-130$ to $-200$ with $D=18\xi_{D}$. This is exactly what the numeric results show in Fig.~\ref{FreeEnergyBoxAndDensity}(b).  When $D$ decreases during the angular velocity $\Omega$ of PdB system increases, Eq.~(\ref{EvaluationOfFreeEnergy4}) increases monotonically as shown in Fig.~\ref{FreeEnergyBoxAndDensity} (a) and (b). Other information which Eq.~(\ref{EvaluationOfFreeEnergy4}) indicates is the London limit free energy of unit cell of pseudo-random lattice is decreasing function for $|q|$ as long as SOC energy is dominating energy. This is because $\tilde{f}_{soc}|_{y>0} \,\, + \tilde{f}_{soc}|_{y<0}$ is decreasing function of $|q|$. However, this is not true any more when $D$ is small. Because SOC energy is not dominating energy in this case, the positive-definite gradient energy is competitive with SOC energy. As a result, we can find from Fig.~\ref{FreeEnergyBoxAndDensity} (a) and (b) that the $\tilde{F}(\theta)_{London}$ of one-half unit cell does not change remarkably for different $|q|$ in small unit cell with $D \sim [4\xi_{D},8\xi_{D}]$. The free energy density of per unit area of equlibrium pseudo-random lattices can be evaluated by multiplying the surface density of 1D nexues $n=D^{-2}$ to the Eq.~(\ref{EvaluationOfFreeEnergy4}),
\begin{align}
n \cdot & \tilde{F}(\theta)_{London}\, \sim \, \frac{1}{8}(1+|q|^{2})\frac{\pi^{2}}{\xi_{D}D} \notag \\ 
& + \, \frac{1}{2{\xi_{D}^{2}}} [-(1+|q|)^{2}cos2|\theta_{0}| -2(1+|q|)|q|sin|\theta_{0}|].
\label{FreeEnergyDensity}
\end{align}
Then we find the London limit free energy density of pseudo-random lattices trends to be a constant determined by $q$ when SOC energy is dominating with large $D$. We can clearly see this form Fig.~\ref{FreeEnergyBoxAndDensity}(d) when $D$ is larger than $10\xi_{D}$. From Eq.~(\ref{FreeEnergyDensity}), we find the magnitude of $n \cdot F(\theta)_{London}$ is around $10^{-1}{\sqrt{2}}{\xi_{D}^{-1}}K_{1}{\Delta_{P}^{2}}$ for $|q|\in[0,0.2]$ when $D>10\xi_{D}$. This coincides with the numerical results in Fig.~\ref{FreeEnergyBoxAndDensity} (c) and (d). When the system is dominated by gradient energy if $D$ is small enough, the free energy density increases rapidly as shown in Fig.~\ref{FreeEnergyBoxAndDensity} (d). If the angular velocity increase successively, the system will go into a parameters region in which pseudo-random lattice model violates. \\[0.1cm]

In all cases, we find the equilibrium free energies of one-half unit cell of separable spin solitons (KLS soliotns and solitons) are slightly higher than those of inseparable spin solitons ($\pi$-solitons). As a result, the equilibrium free energy densities of pseudo-random lattices consisting of separable spin solitons (KLS-soliotns and solitons) are also slightly higher than those of inseparable spin solitons ($\pi$-solitons). This significant fact suggests that the equilibrium states which was observed in experiment of rotating PdB system is the pseudo-random lattice of inseparable $2/4$ spin solitons ($\pi$-solitons) of 1D nexus objects. We will see this is true in next chapter by calculating the spin dynamic response under weak magnetic drive.

\section{The mirror symmetry of 1D nexus objects and its breaking} 
\label{DiscreteSymmetry}
\begin{figure*}
\centerline{\includegraphics[width=1.0\linewidth]{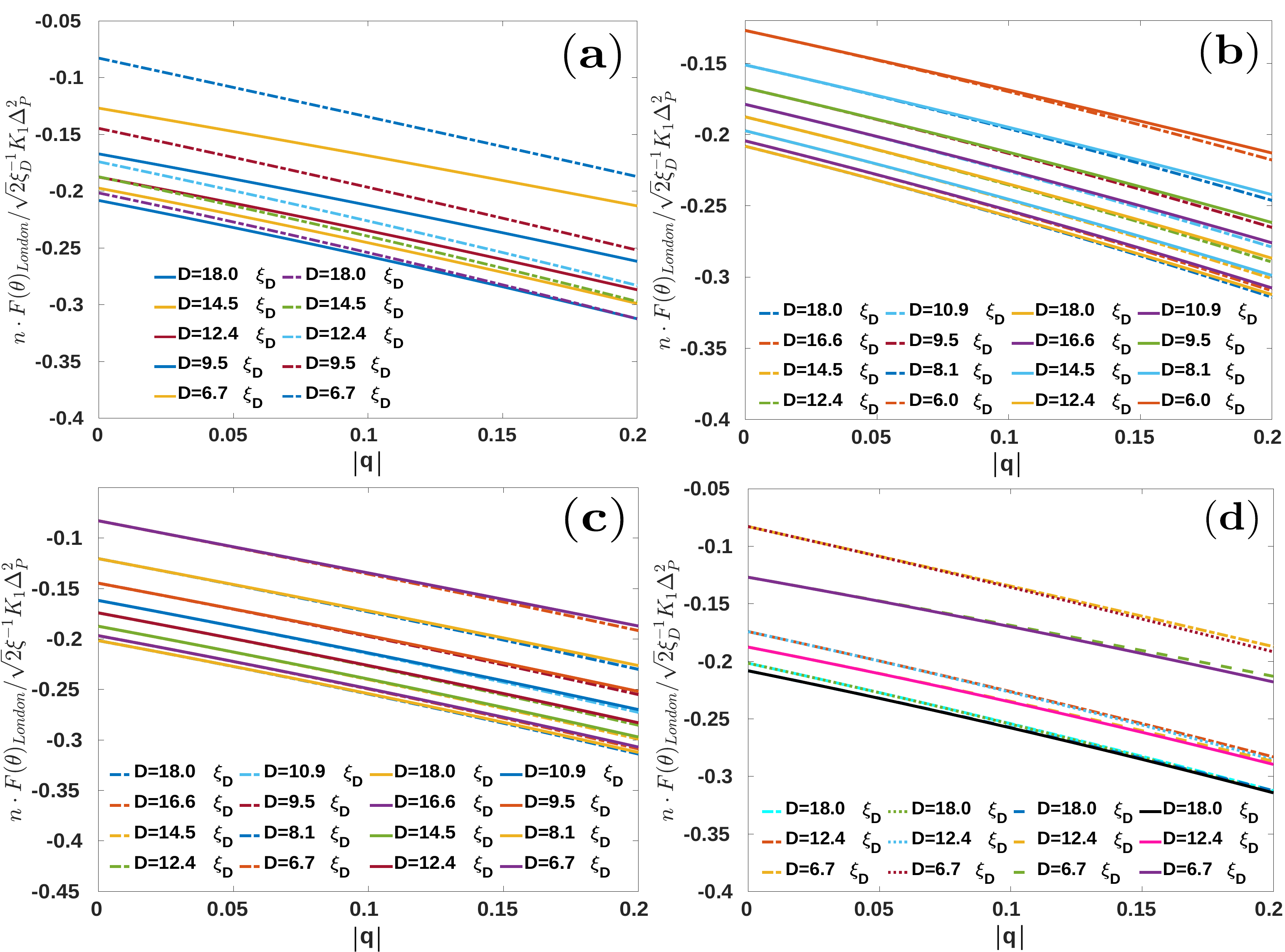}}
\caption{Surface densities of equilibrium London limit free energies of pseudo-random lattices with vacuum state (i) and vacuum state (ii) respective. These two vacuum states are not degenerated with same free energy any more when the direction of KLS domain wall is fixed. (a) The surface densities of London limit free energies of pseudo-random lattices consisting of inseparable spin solitons ($\pi$-solitons) and separable spin solitons (KLS-solitons and solitons) in the vacuum state (ii). This figure shows similar features with Fig.~\ref{FreeEnergyBoxAndDensity}(c). (b) depicts the surface densities of London limit free energies of pseudo-random lattices consisting of inseparable $2/4$ spin solitons of vacuum states (i) and (ii) respectively. The dash-dot lines represent the vacuum state (i) while the solid lines represent the vacuum state (ii). (c) depicts the surface densities of London limit free energies of pseudo-random lattices consisting of separable $2/4$ spin solitons of vacuum states (i) and (ii) respectively. The dash-dot lines represent the vacuum state (i) while the solid lines represent the vacuum state (ii). (d) depicts the surface densities of London limit free energies of pseudo-random lattices in vacuum states (i) and (ii). The solid lines and dash lines represent the pseudo-random lattices consisting of inseparable spin solitons ($\pi$-solitons) within vacuum states (i) and (ii) respectively. The dot lines and dash-dot lines represent the pseudo-random lattices consisting of separable spin solitons (KLS-solitons and solitons) within vacuum states (i) and (ii) respectively. We found the pseudo-random lattices consisting of $2/4$ inseparable spin solitons within vacuum state (i) have lowest equilibrium free energies. \label{DensityOfLondonLimitFreeEnergyPARA2}}
\end{figure*}

As we mentioned in Sec.~\ref{VacuaManifold}, The London limit free energy $F(\theta)_{London}$ has a mirror symmetry when the coordinates are permuted to each other i.e., $F[\theta(x',y')] = F[\theta(x,y)]$ with $x' = y$ and $y' = x$. This mirror symmetry does not vanish even in the presence of 1D nexus object. As a result, the spin textures of $2/4$ spin solitons have this mirror symmetry as well. \\[0.1cm]

This discrete symmetry originates from the reduction of degenerate space of order parameter by requirement of continuity of order parameter in the presence of KLS domain wall. In order to understand this, we start from the degenerate manifold of PdB which generates from symmetry breaking transition of polar phase vacuum. In this case, $R_{PdB} \cong SO(2)_{S-L} \times \mathbb{Z}_{2}^{S-\Phi}$, in which the nontrivial element of $\mathbb{Z}_{2}^{S-\Phi}$ corresponds to the presence of KLS domain wall \cite{Zhang2020}. The requirement of continuity of order parameter reduces the degenerate space of $\hat{\mathbf{e}}^{1}$ and $\hat{\mathbf{e}}^{2}$ on both sides of domian wall from $SO(2)_{S-L}$ to (i) $\hat{\mathbf{e}}^{1}$ $\longrightarrow$ $-\hat{\mathbf{e}}^{1}$, while $\hat{\mathbf{e}}^{2}$ keeps its direction and (ii) $\hat{\mathbf{e}}^{2}$ $\longrightarrow$ $-\hat{\mathbf{e}}^{2}$, while $\hat{\mathbf{e}}^{1}$ keeps its direction. The parametrization in Eq.~(\ref{PARA1}), which we used in previous calculations and discussions, corresponds to the vacuum state (i) and the direction of static magnetic field $\mathbf{H}^{(0)}$ is set to parallel with the $\hat{\mathbf{e}}^{2}$. Because the vacuum state (ii) is another possible vacuum state with same free energy of case (i) in the presence of KLS domain wall, the London limit free energy $F(\theta)_{London}$ is invariant when we transform from vacuum state (i) to vacuum state (ii). In our case, the parametrization of vacuum state (ii) is
\begin{align}
\mathbf{\hat{d}} = \hat{y}cos\theta-\hat{z}sin\theta, \, \mathbf{\hat{e}}^{2} = -\hat{y}sin\theta-\hat{z}cos\theta,  
\mathbf{\hat{e}}^{1}  = \hat{x}, \, \mathbf{H}^{(0)} = H\hat{x}, 
\label{PARA2} 
\end{align}
and the corresponding dimensionless London limit free energy is 
\begin{align}
\tilde{F}(\theta)_{London}  = 
& {\frac{1}{\xi_{D}}}\int\nolimits_{\Sigma} [\frac{1}{2} (\gamma_{1}+2\gamma_{2}) \partial_{y}\theta \partial_{y}\theta \notag  + \frac{1}{2}\gamma_{1}\partial_{x}\theta \partial_{x}\theta \notag \\ 
& + \frac{1}{\xi_{D}^{2}} 
(-\frac{1}{2}{\gamma_{4}}cos2\theta-\gamma_{3}sin\theta)] d\Sigma \label{FreeEnergyThetaDimensionlessPARA2}
\end{align}
where 
\begin{align}
q  = \frac{\Delta_{\bot1}}{\Delta_{P}},\,\,
 \gamma_{1}=1+|q|^{2},\,\, 
\gamma_{2}=|q|^{2},\,\, 
 \gamma_{3}=q(1+|q|),\,\,
\gamma_{4}=(1+|q|)^{2}.
\end{align}
Comparing Eq.~(\ref{FreeEnergyThetaDimensionlessPARA2}) and Eq.~(\ref{FreeEnergyThetaDimensionless}), we can see the mirror symmetry. \\[0.1cm]

However, this discrete symmetry may be destroyed if the direction of domain wall is fixed in both vacuum states (i) and (ii). In this case, the term containing $\gamma_{3}$ in Eq.~(\ref{FreeEnergyThetaDimensionlessPARA2}) is invariant for both parametrizations, and thus violates this mirror symmetry. As a result, the equilibrium sates of Eq.~(\ref{FreeEnergyThetaDimensionlessPARA2}) and Eq.~(\ref{FreeEnergyThetaDimensionless}) are not identical any more. Then we need to check the equilibrium London limit free energy of these two different equilibrium states. We did the same numeric minimizations of London limit free energy with parameteization Eq.~(\ref{PARA2}) and calculated the surface densities of equilibrium free energies of pseudo-random lattices in vacuum state (ii). The latter can be evaluated as
\begin{align}
n \cdot \tilde{F}(\theta)_{London}|_{(ii)}\, & \sim \,\frac{1}{8}(1+|q|^{2})\frac{\pi^{2}}{\xi_{D}D}  + \frac{\pi^{2}}{4\xi_{D}D}|q|^{2}
 + \, \frac{1}{2} [\tilde{f}_{soc}|_{y>0} \,\, + \tilde{f}_{soc}|_{y<0}]_{q}  \notag \\ 
 & \sim n \cdot \tilde{F}(\theta)_{London}|_{(i)}\, + \frac{\pi^{2}}{4\xi_{D}D}|q|^{2}. \label{FreeEnergyDensityPARA2}
\end{align}
Then we can expect the surface densities of equilibrium London limit free energy of vacuum state (ii) are slightly higher than those of vacuum state (i) when $|q|\leq0.2$. In Fig.~\ref{DensityOfLondonLimitFreeEnergyPARA2}, we show this  
for pseudo-random lattices consisting of inseparable and separable $2/4$ spin solitons respectively. In all cases, the surface densities of London limit free energy of vacuum state (ii) are indeed higher than those of vacuum state (i).  

\chapter{Spin Dynamical Response and NMR} 

\label{Chapter5} 
   
In this chapter, we discuss how the network of mesoscopic extended structure of KLS string wall --- pseudo random lattice of $[1/2]$ spin solitions response the weak magnetic drive in the NMR experiment in Ref.~\cite{Makinen2019}. We firstly demonstrate the spin dynamic equation and discuss how the static spin textures modify the NMR spectrum by generating a satellite peak related to the NMR main peak. Then we use the data of static spin textures, which we got in Chpater.~\ref{Chapter4}, to solve spin response equation and get the NMR frequency shift of satellite peak.  The  experimentally observed scaling rule of the NMR ratio intensity against the rotation angular velocity $\Omega$ also be checked. \\[0.1cm]    





We have talked the topological origin of network of 1D nexus objects with topological invariant $2/4$ as well as their equilibrium free energies in previous chapters. Because there are two kinds of spin solitons connecting with KLS string wall, the pseudo-random lattices consisting of them have different equilibrium free energy densities. To compare with the experiments and check  the theories, we must calculate the spin dynamic response. Under weak enough magnetic drive, the nuclear spin magnetization of PdB superfluid responds a NMR signal when the frequency of magnetic drive matches the transverse spin dynamic mode. Because the spin dynamics of symmetry breaking states of $^3$He is strongly influenced by SOC energy which is determined by the relative orientations between spin and orbital degenerate parameters, the NMR of continuous wave drive is a perfect tool, which can be used to detect the pseudo-random lattice of 1D nexus objects network via dynamics of spin solitons  \cite{VollhardtWolfle1990}.  \\[0.1cm]  

When the PdB superfluid is equilibrium, the spin density has equilibrium value $\mathbf{S}^{(0)}$ over the system. If the weak homogeneous magnetic drive is turned on, the spin density gets a tiny variation ${\delta}\mathbf{S}(\mathbf{r},t)$, where $\mathbf{r}$ and $t$ are spatial and time coordinates respectively. In this perturbed system, the transverse spin density ${\delta}S_{+}$ may be expanded as 
\begin{equation}
{\delta}S_{+}({\mathbf{r}},t)={\int}{d{\sigma}'}{\int}{dt'}{\frac{{\delta}{S_{+}}}{{\delta}{H_{a}}}}(\mathbf{r},t,\mathbf{r}',t'){\delta}{H_{a}}({\mathbf{r}}',t')+O({\delta}{H_{a}}^{2}),
\label{LinearResponse}
\end{equation}
where ${\delta}H_{a} \equiv {\delta}\mathbf{H}$ is the homogeneous weak magnetic drive and $a=1,2,3$ are spatial coordinate indexes. Thus the PdB superfluid under magnetic drive is a linear response system if $|{\delta}\mathbf{H}| \ll |\mathbf{H}^{(0)}|$ \cite{Alxander2010}. The poles of the transverse spin dynamic response function ${\delta}S_{+}/{\delta}H_{a}$ correspond to eigen-modes of the NMR. The most classic example of this response is the detection of Larmor frequency in homogeneous system, and we will see soon that spin textures also put their fingerprints in the response mode. We calculate these eigen-modes for two kinds of pseudo-random lattices of 1D nexus objects with topological invariant $2/4$ in this chapter.  
  
\begin{figure*}
\centerline{\includegraphics[width=1.0\linewidth]{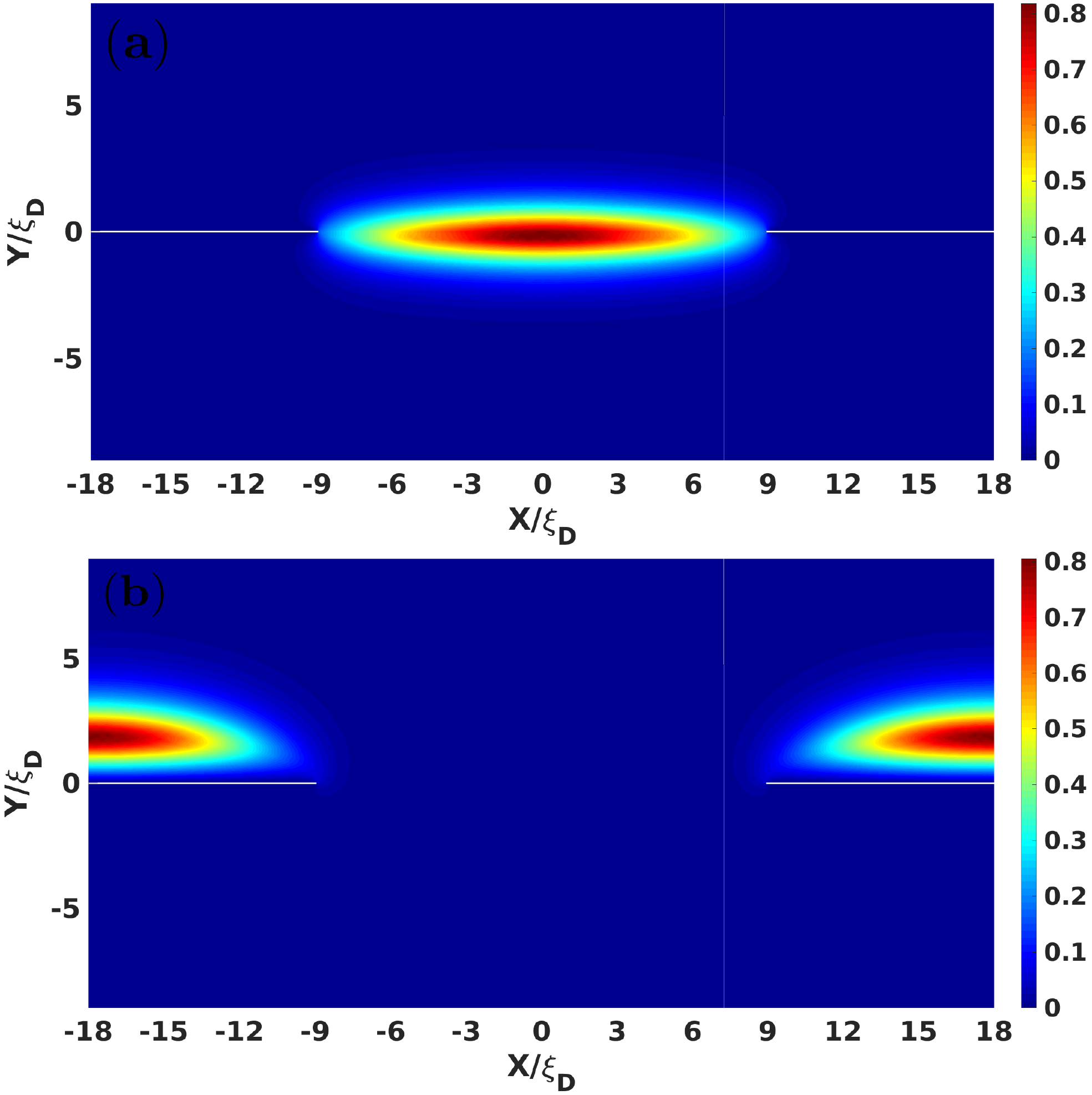}}
\caption{The modulus $|\delta{S_{+}}(\omega)|$ of the lowest transverse spin dynamic response modes located in the unit cells of pseudo-random lattices of inseparable and separable $2/4$ spin solitons. In both cases, we depict the results with parameters $|q|=0.18$ and $D=18\xi_{D}$. (a) $|\delta{S_{+}}(\omega)|$ in unit cell of inseparable spin solitons ($\pi$-solitons); (b) $|\delta{S_{+}}(\omega)|$ in unit cell of separable spin solitons (KLS-solitons and Solitons). \label{SpinDynamicResponseModes}}
\end{figure*} 
\section{Spin dynamic response equations}
Spin-orbit coupling plays an important role in the NMR measurements of significant properties of different superfluid phases in $^3$He system. This is because the coherence of superfluid states, which breaks relative symmetry between spin and orbital degree of freedom , strengthens the SOC energy \cite{Anderson1973,AndersonVarma1973}. This gives rise to the observable NMR frequency shift of nuclear spin magnetization. In our case, the SOC energy takes into account all the information and effects of spin vectors in spin solitons, which connect to the KLS domain wall via 1D nexus. Thus, we must calculate the spin dynamic response function $\delta{S_{+}}/\delta{H_{a}}$ dominated by SOC energy. \\[0.1cm]

In this section, we utilize the spin dynamic equations dominated by SOC energy to get $\delta{S_{+}}/\delta{H_{a}}$ and corresponding eigenequations of poles \cite{chaikin1995}. Because the SOC energy is much smaller than the microscopic energy scales of PdB superfluid i.e., $\Delta_{P}$,  the characteristic time scales of spin dynamic response function $\delta{S_{+}}/\delta{H_{a}}$ is much longer than the time scales of microscopic processes which are proportional to $\Delta_{P}^{-1}$. All the microscopic processes with time scales $\Delta_{P}^{-1}$ are equilibrium in the spin dynamic processes under weak magnetic drive. This means the spin dynamic equations are a system of hydrodynamic equations of spin densities $\delta{S_{a}}$ and spin vectors of order parameter \cite{VollhardtWolfle1990, chaikin1995}. \\[0.1cm]    

In the limit of hydrodynamics, the system of dynamic equations of spin densities $S_{\alpha}$ and spin vectors are system of Liouville equations 
\begin{align}
{\frac{\partial{S_{\alpha}}}{\partial{t}}} =\{F_{hydrodynamics},S_{\alpha}\}, \,\,
 {\frac{\partial{V_{\alpha}^{a}}}{\partial{t}}}  =\{F_{hydrodynamics},V_{\alpha}^{a}\},\,\,
V_{\alpha}^{a}={\hat{e}}_{\alpha}^{1}, {\hat{e}}_{\alpha}^{2}, \hat{d}_{\alpha},
\label{LiouvilleEquations}
\end{align}
where $\alpha=1,2,3$ are the indexes of spatial coordinates. And $V_{\alpha}^{a}$ denote the three spin vectors of order parameter i.e.,  $V_{\alpha}^{1}={\hat{e}}_{\alpha}^{1},$ $V_{\alpha}^{2}={\hat{e}}_{\alpha}^{2}$, $V_{\alpha}^{3}=\hat{d}_{\alpha}$. The hydrodynamic free energy of PdB superfluid dominated by SOC energy is   
\begin{equation}
F_{hydrodynamics}=\int\nolimits_{\Sigma} (f_{\rm H} + f_{\rm soc} + f_{\rm grad})d\Sigma.
\label{FreeEnergyOfHydrodynamics}
\end{equation} 
Thus Eq.~(\ref{LiouvilleEquations}) can be further written as
\begin{align}
{\frac{\partial{S_{\alpha}}}{\partial{t}}} & = {\int_{\Sigma}}{{d^{3}}r'}\frac{{\delta}F_{hydrodynamics}}{{\delta}{S_{\beta}}}(r')\{S_{\beta}(r'),{S_{\alpha}}(r)\} \notag \\ & + {\int_{\Sigma}}{{d^{3}}r'}\frac{{\delta}F_{hydrodynamics}}{{\delta}{V_{\beta}^{a}}}(r')\{V_{\beta}^{a}(r'),{S_{\alpha}}(r)\},
\label{LiouvilleEquations1}
\end{align}
and
\begin{equation}
{\frac{\partial{V_{\alpha}^{a}}}{\partial{t}}}={\int_{\Sigma}}{{d^{3}}r'}\frac{{\delta}F_{hydrodynamics}}{{\delta}{S_{\beta}}}(r')\{S_{\beta}(r'),{V_{\alpha}^{a}}(r)\},
\label{LiouvilleEquations2}
\end{equation}
where $\beta=1,2,3$ are indexes of spatial components of hydrodynamic variables. The Poisson brackets between $S_{\alpha}$ and $V_{\alpha}^{a}$ can be gotten by the commutators-based methods in Ref.~\cite{Dzyaloshinskii1980} as 
\begin{align}
\{{S_{\alpha}}(r_{1}),{{S_{\beta}}(r_{2})}\} = {\epsilon}_{{\alpha}{\beta}{\gamma}}{S_{\gamma}}{\delta}(r_{1}-r_{2}), \,\,
\{{S_{\alpha}}(r_{1}),{{V_{\beta}^{a}}(r_{2})}\} = {\epsilon}_{{\alpha}{\beta}{\gamma}}{V_{\gamma}^{a}}{\delta}(r_{1}-r_{2}), \label{PoissonBrackets1}
\end{align}
where $r_{1}$ and $r_{2}$ are the spatial coordinates and ${\epsilon}_{{\alpha}{\beta}{\gamma}}$ is the Levi-Civita symbol.
After plugging Eq.~(\ref{PoissonBrackets1}) into Eq.~(\ref{LiouvilleEquations1}) and Eq.~(\ref{LiouvilleEquations2}), the coupled first order dynamic equations of spin densities $S_{\alpha}$ and $V_{\alpha}^{a}$ are given as
\begin{align}
\frac{{\partial}{S_{\alpha}}}{{\partial}{t}} & ={\gamma}{H_{\beta}}{\epsilon_{{\alpha}{\beta}{\gamma}}}{S_{\gamma}} -{\frac{6}{5}}{g_{D}}{V_{j}^{d}}{V_{\gamma}^{b}}{{\epsilon}_{{\alpha}{\beta}{\gamma}}}{Q_{{\beta}j}^{bd}}+({{\partial}_{i}}{{\partial}_{j}}{V_{\beta}^{b}}){V_{\gamma}^{a}}{\epsilon_{{\alpha}{\beta}{\gamma}}}{K_{ij}^{ba}}, 
\label{1stOderEquationsA}
\end{align}
\begin{align}
\frac{{\partial}{V_{\alpha}^{a}}}{{\partial}t} & = {\gamma}{H_{\beta}}{\epsilon_{{\alpha}{\beta}{\gamma}}}{V_{\gamma}^{a}} -{\delta}{\gamma^{2}}{\chi_{\perp}^{-1}}{S_{\eta}V_{\eta}^{3}}{V_{\beta}^{3}}{\epsilon_{{\alpha}{\beta}{\gamma}}}{V_{\gamma}^{a}}-{\gamma^{2}}{\chi_{\perp}^{-1}}{S_{\beta}}{\epsilon_{{\alpha}{\beta}{\gamma}}}{V_{\gamma}^{a}},
\label{1stOderEquationsB}
\end{align}
where $\delta  = (\chi_{\bot} -\chi_{\|})/\chi_{\|}$ in which $\chi_{\bot}$ and $\chi_{\|}$ are the transverse magnetic susceptibility and the longitude magnetic susceptibility of PdB phase respectively.
\begin{align}
K_{ij}^{ba} =K_{1}{\delta}_{ij}{X_{m}^{b}}{X_{m}^{a}}+K_{2}{X_{j}^{a}}{X_{i}^{b}}+K_{3}{X_{j}^{b}}{X_{i}^{a}},
Q_{{\beta}{j}}^{bd} & ={X_{\beta}^{b}}{X_{j}^{d}}+{X_{\beta}^{d}}{X_{j}^{b}}
\end{align}
with
\begin{align}
{X_{i}^{1}}={{\Delta}_{{\perp}1}}{\hat{x}_{i}},\,\, {X_{i}^{2}}={{\Delta}_{{\perp}2}}{\hat{y}_{i}},\,\, {X_{i}^{3}}={{\Delta}_{{\parallel}}}{\hat{z}_{i}}.
\label{2ndEQ}
\end{align}
The details of calculation from Eq.~(\ref{LiouvilleEquations1}) to Eq.~(\ref{1stOderEquationsB}) are shown in Appendix.~\ref{For1stOrderEqations}. \\[0.1cm]

Starting from the first order equations of spin densities and degenerate parameters in Eq.~(\ref{1stOderEquationsA}) and  Eq.~(\ref{1stOderEquationsB}), we can further derive the second order spin dynamic response equations of $\delta{S_{\alpha}}$ under weak magnetic drive $\delta{H_{\alpha}}$. This was done by plugging 
\begin{equation}
S_{\alpha}=S_{\alpha}^{(0)}+{\delta}{S_{\alpha}(\mathbf{r},t)}, V_{\alpha}^{a}=V_{\alpha}^{a(0)}+{\delta}{V_{\alpha}^{a}(\mathbf{r},t)}
\end{equation}
and 
\begin{equation}
H_{\alpha}=H_{\alpha}^{(0)}+{\delta}{H_{\alpha}(t)}
\end{equation}
into Eq.~(\ref{1stOderEquationsA}) and Eq.~(\ref{1stOderEquationsB}). Here the $S_{\alpha}^{(0)}$ and $V_{\alpha}^{a(0)}$ are the equilibrium spin densities and equilibrium degenerate parameters respectively. While the ${\delta}{S_{\alpha}(\mathbf{r},t)}$ and ${\delta}{V_{\alpha}^{a}(\mathbf{r},t)}$ are the dynamic parts of the perturbed spin densities and degenerate parameters. The $H_{\alpha}^{(0)}$ is the static magnetic field and ${\delta}{H_{\alpha}(t)} = |\delta\mathbf{H}|\hat{x}e^{-i{\omega}t}$ is the homogeneous RF continuous-wave drive. We put the details of calculations in Appendix.~\ref{For2stOrderEqations} and the derived spin dynamic response equations within frequency form is 
\begin{align}
i{\omega}{{\delta}{S_{\alpha}}({\omega})} ={\gamma}{\epsilon_{{\alpha}{\beta}{\gamma}}}{H_{\beta}^{(0)}}{{\delta}S_{\gamma}(\omega)} +{\gamma}{\epsilon_{{\alpha}{\beta}{\gamma}}}S_{\gamma}^{(0)}{{\delta}{H_{\beta}}}(\omega) 
+{\frac{\Xi_{{\alpha}{\lambda}}}{i\omega}}{\delta}{S_{\lambda}}(\omega) +{\frac{C_{{\alpha}{\eta}}}{i{\omega}}}{\delta}{H_{{\eta}}}(\omega)
\label{2ndOderEquations}
\end{align}
and
\begin{align}
{\Xi_{{\alpha}{\lambda}}} & ={\frac{\gamma^{2}}{\chi_{\perp}}}{K_{ij}^{ba}}{\Lambda_{ij{\alpha}{\lambda}}^{ba}}+{\frac{6{g_{D}}{\gamma^{2}}}{5{\chi_{\perp}}}}{R_{j{\lambda}{\alpha}{\beta}}^{db}}{Q_{{\beta}j}^{bd}}+{\frac{6{g_{D}}{\gamma}^{2}}{5\chi_{\perp}}}{V_{\zeta}^{d{(0)}}}{V_{\gamma}^{b(0)}}{\epsilon_{j{\lambda}{\zeta}}}{\epsilon_{{\alpha}{\beta}{\gamma}}}Q_{{\beta}{j}}^{bd}, \notag \\
C_{{\alpha}{\eta}} & = {\gamma}G_{{i}{j}{\alpha}{\eta}}^{ba}{K_{{i}{j}}^{ba}} -{\frac{6{g_{D}}{\gamma}}{5}}{R_{j{\eta}{\alpha}{\beta}}^{db}}{Q_{{\beta}j}^{bd}}-{\frac{6{g_{D}}{\gamma}}{5}}{V_{\zeta}^{d(0)}}{V_{\gamma}^{b(0)}}{\epsilon_{j{\eta}{\zeta}}}{\epsilon_{{\alpha}{\beta}{\gamma}}}Q_{{\beta}{j}}^{bd}, 
\label{XiAndC}
\end{align}
where
\begin{align}
R_{j{\eta}{\alpha}{\beta}}^{db} & = {V_{j}^{d(0)}}{V_{{\beta}}^{b(0)}}{\delta_{{\eta}{\alpha}}}-{V_{j}^{d(0)}}{V_{{\alpha}}^{b(0)}}{\delta_{{\eta}{\beta}}}, \,\,
G_{ij{\alpha}{\gamma}}^{ba} = ({\partial_{i}}{\partial_{j}}{V_{{\alpha}}^{b(0)}}){V_{{\gamma}}^{a(0)}}-({\partial_{i}}{\partial_{j}}{V_{{\beta}}^{b(0)}}){\delta_{{\beta}{\gamma}}}{V_{{\alpha}}^{a(0)}}, \notag \\
{\Lambda_{ij{\alpha}{\lambda}}^{ba}} & = ({\partial_{i}}{\partial_{j}}{V_{{\beta}}^{b(0)}}){{\delta}_{{\beta}{\lambda}}}{V_{{\alpha}}^{a(0)}} + ({V_{{\gamma}}^{b(0)}}{V_{{\gamma}}^{a(0)}}{\delta_{{\alpha}{\lambda}}}-{\delta_{{\gamma}{\lambda}}}{V_{{\alpha}}^{b(0)}}{V_{{\gamma}}^{a(0)}}){\partial_{i}}{\partial_{j}} \notag \\ & + [({\partial_{i}}{V_{{\gamma}}^{b(0)}}){V_{{\gamma}}^{a(0)}}{\delta_{{\alpha}{\lambda}}}-({\partial_{i}}{V_{{\alpha}}^{b(0)}}){V_{{\gamma}}^{a(0)}}{\delta_{{\gamma}{\lambda}}}]{\partial_{j}} \notag \\ 
& + [({\partial_{j}}{V_{{\gamma}}^{b(0)}}){V_{{\gamma}}^{a(0)}}{\delta_{{\alpha}{\lambda}}}-({\partial_{j}}{V_{{\alpha}}^{b(0)}}){V_{{\gamma}}^{a(0)}}{\delta_{{\gamma}{\lambda}}}]{\partial}_{i} \notag \\ 
& - {\delta_{{\gamma}{\lambda}}({\partial_{i}}{\partial_{j}}{V_{{\alpha}}^{b(0)}}){V_{{\gamma}}^{a(0)}}}. 
\label{RGandLambda}
\end{align}
The first two terms of Eq.~(\ref{2ndOderEquations}) correspond to the NMR response of Larmor precession of $\delta{S_{\alpha}}$ with frequency $\omega_{L}=\gamma H^{(0)}$. While the last two terms of Eq.~(\ref{2ndOderEquations}) induce the NMR frequency shift, and they conventionally are called torque terms. \\[0.1cm]

From Eq.~(\ref{XiAndC}) and Eq.~(\ref{RGandLambda}), we found that torque terms are fully determined by the equilibrium textures of spin vectors. In our case with pseudo-random lattices of $2/4$ spin solitons, this means the NMR frequency shifts are totally induced by equilibrium textures of spin solitons in 1D nexus objects. That's why the transverse NMR spectrum is perfect tool to observe the network of 1D nexus objects and network of KLS string wall.
Taking into account the static magnetic field $\mathbf{H}^{(0)}=|\mathbf{H}^{(0)}|{\hat{\mathbf{y}}}$ and the parametrization Eq.~(\ref{PARA1}), we can derive the dynamic response equations of transverse spin density 
\begin{equation}
{\delta}S_{+}=\frac{1}{\sqrt{2}}[{\delta}S_{1}(\omega)+i{\delta}S_{3}(\omega)]
\end{equation}
under weak magnetic drive $\delta{\mathbf{H}(t)}$, see the detail of calculation in Appendix.~\ref{TransverseNMRResponseEquation}. This calculation gives
\begin{align}
({\omega}^{2} -{\omega}_{L}^{2}) {\delta}S_{+}(\omega)  & =  ({\Xi}_{11}+{\Xi}_{33}){\delta}S_{+}(\omega)+i({\Xi}_{13}-{\Xi}_{31}){\delta}S_{+}(\omega) \notag \\ 
& -[{\frac{1}{2}}({C_{11}}+{C_{31}})-{\frac{{\chi}_{\perp}}{\sqrt{2}{\gamma}}}({\Xi_{33}}+i{\Xi_{13}}-i{\Xi_{31}})]{\delta}{H_{1}}(\omega).
\label{TransverseResponseEquation}
\end{align}
Thus 
\begin{equation}
\frac{{\delta}S_{+}(\omega)}{{\delta}H_{1}(\omega)} \propto \frac{1}{{\omega}^{2}-{\omega}_{L}^{2}-({\Xi}_{11}+{\Xi}_{33})-i({\Xi}_{13}-{\Xi}_{31})}.
\label{ResponsFunction}
\end{equation}
The poles of spin dynamic response function ${\delta}S_{+}/{\delta}H_{1}$, which are determined by eigenequation
\begin{equation}
({\omega}^{2}-{\omega}_{L}^{2}){\delta}S_{+}(\omega) =({\Xi}_{11}+{\Xi}_{33})+i({\Xi}_{13}-{\Xi}_{31}){\delta}S_{+}(\omega),
\label{NMREigenEquation}
\end{equation} 
correspond to the eigen-modes of transverse NMR frequency shift related to Larmor frequency induced by pseudo-random lattices of 1D nexus objects. We numerically solve this eigen-equation in next section with different $D$ and $|q|$.

\begin{figure*}
\centerline{\includegraphics[width=1.0\linewidth]{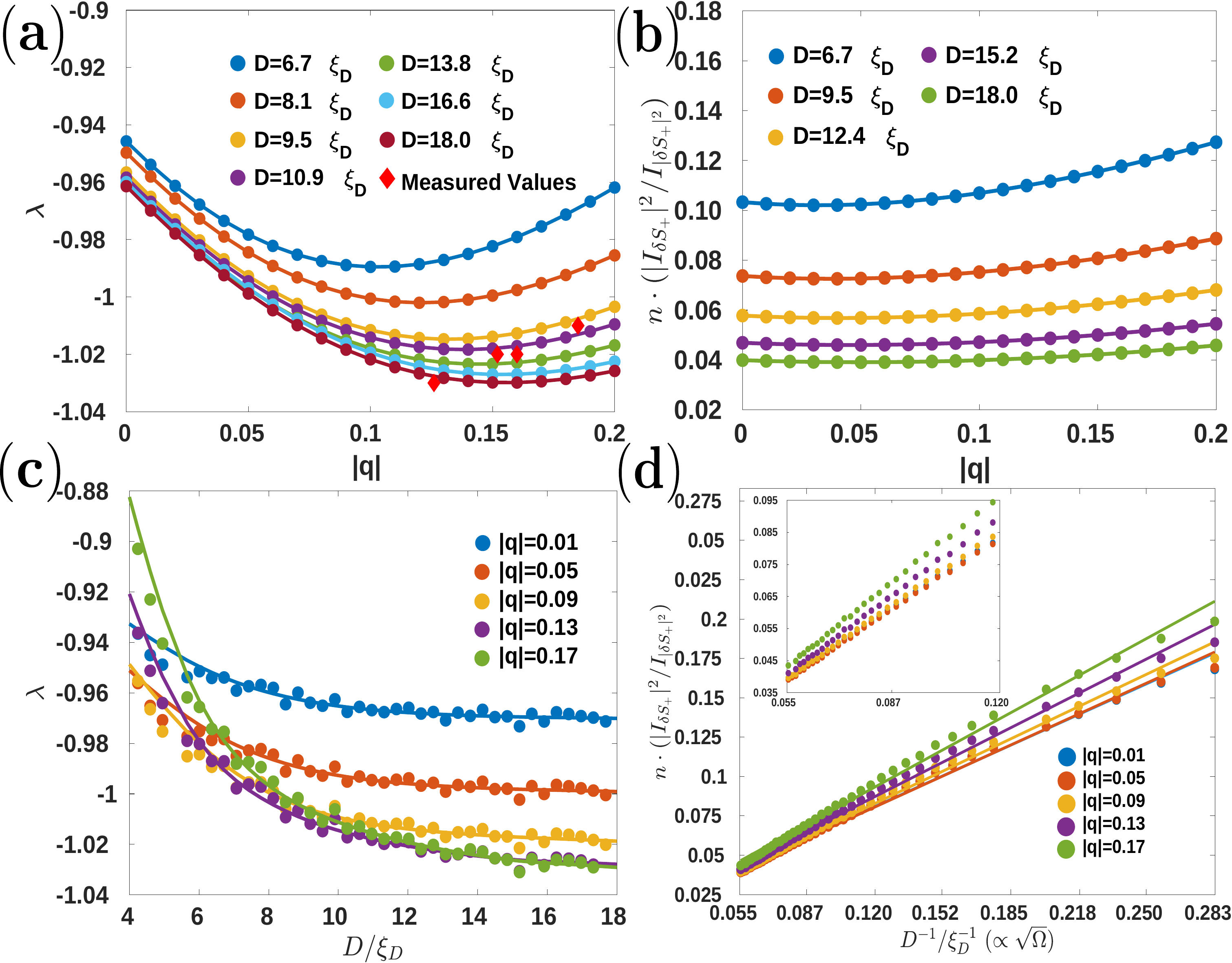}}
\caption{Transverse NMR frequency shifts $\lambda$ and surface densities of ratio intensity $n \cdot (|I_{\delta{S_{+}}}|^{2}/I_{|\delta{S_{+}}|^{2}})$ of pseudo-random lattices consisting of inseparable spin solitons ($\pi$-solitons). The frequency shifts $\lambda$ are eigenvalues of Eq.~(\ref{NMREigenEquationDimensonless}) with equilibrium textures of $\pi$-solitons in London limit. The surface densities of NMR ratio intensity are calculated by using Eq.~(\ref{DensityOfRatioIntensity}). All scattering dots represent the original numeric results, while colored lines are smoothing spline fittings of these original numeric results. (a) Transverse NMR frequency shifts $\lambda$ as functions of $|q|$ with different $D$. For large enough unit cells ($D>10\xi_{D}$), we found $\lambda$ decreases when $|q|$ increases as long as $|q|\leq0.16$. The typical values of $\lambda$ are around $-1.015$ to $-1.03$ when pseudo-random lattices model is good enough. This exactly coincides with the region of $\lambda$ which was observed in experiment of Ref.~\cite{Makinen2019}, as shown via red diamonds. (b) depicts the surface densities of ratio intensity $n \cdot (|I_{\delta{S_{+}}}|^{2}/I_{|\delta{S_{+}}|^{2}})$ of the eigenmodes. (c) depicts the transverse frequency shifts $\lambda$ as function of $D$. (d) The surface densities of ratio intensity $n \cdot (|I_{\delta{S_{+}}}|^{2}/I_{|\delta{S_{+}}|^{2}})$ as function of $1/D \propto \sqrt{\Omega}$. We found $n \cdot (|I_{\delta{S_{+}}}|^{2}/I_{|\delta{S_{+}}|^{2}})$ increases linearly if $\sqrt{\Omega}$ increases. This coincides with the results of experimental observation in Ref.~\cite{Makinen2019}. The inset is the magnified plot between $1/D=0.055$ till $1/D=0.120$. \label{NMRLambdaAndRatioIntensityInseparableSoliton}}   
\end{figure*}
\section{NMR of 1D nexus objects}
\label{NMRShifts}
For facilitating the numeric calculation, we firstly transform the spin dynamic eigen-equation Eq.~(\ref{NMREigenEquation}) into dimensionless form. All $\Xi_{{\alpha}{\lambda}}$ operators in Eq.~(\ref{NMREigenEquation}) must be simplified with prarametrization Eq.~(\ref{PARA1}), See the details in Appendix.~\ref{SimplifyTheNMREigenEquation}. This gives
\begin{align}
\lambda {\delta}S_{+}(\omega) = & \xi^{2}_{D}[(6\rho_{2}^{2}+\rho_{1}^{2} + 1)\partial_{y}\partial_{y} + (3\rho_{1}^{2}+2 \rho_{2}^{2} + 1)\partial_{x}\partial_{x}-2iV]\delta{S_{+}}(\omega) + U\delta{S_{+}}(\omega)
\label{NMREigenEquationDimensonless}
\end{align}
with
\begin{align}
V =  (1+3{\rho_{1}^{2}}cos2\theta) & {\partial_{x}}\theta{\partial_{x}} + (1+\rho_{1}^{2}){\partial_{y}}\theta{\partial_{y}} +\frac{1}{2 \xi_{D}^{2}} [(1 + \rho_{1})^{2}sin2{\theta} - (1 + \rho_{1})\rho_{2}cos{\theta}], \notag \\
U = & (1+\rho_{1})[-(1+\rho_{1})cos2\theta-5\rho_{2}sin\theta]+1+\rho_{1}^{2}+4\rho_{2}^{2},
\end{align}
where $\rho_{1}=\Delta_{\bot1}/\Delta_{P}$ and  $\rho_{2}=\Delta_{\bot2}/\Delta_{P}$. Here the dimensionless eigenvalue 
\begin{equation}
\lambda=\frac{(\omega^{2}-\omega_{L}^{2})}{\tilde{\Omega}^{2}}
\end{equation}
is the transverse NMR frequency shift under weak magnetic drive and
\begin{equation}
\tilde{\Omega}^{2}=(\frac{5\chi_{\bot}}{6\gamma^{2}{\Delta_{P}^{2}}g_{D}})^{-1}.
\end{equation}
We use the Galerkin strategy under finite-element partition to solve Eq.~(\ref{NMREigenEquationDimensonless}) \cite{ciarlet1978}, see details of algorithm see Appendix.~\ref{GarlkinStretargy}. The solving regions are the unit cells of pseudo-random lattices of 1D nexus objects. The equilibrium spin textures of pseudo-random lattices, which we got in Sec. \ref{EquilibriumTexturesOfInseparableAndSeparableSolitons}, are directly used to solve Eq.~(\ref{NMREigenEquationDimensonless}). Because $\delta{H_{\alpha}}$ is low energy drive, we merely consider the spin dynamic response mode with the lowest $\lambda$ of Eq.~(\ref{NMREigenEquationDimensonless}). \\[0.1cm]

In addition, the ratio intensity of NMR signal is another observable besides the frequency shift $\lambda$. the scaling rule between ratio intensity and angular velocity $\Omega$ is a significant feature of the system. The surface density of ratio intensity, which is generated by unit area of pseudo-random lattices of 1D nexuses, is given as
\begin{equation}
n \cdot \frac{|I_{\delta{S_{+}}}|^{2}}{I_{|\delta{S_{+}}|^{2}}}=n \cdot \frac{|\int_{\sigma}\delta{S_{+}}d{\sigma}|^{2}}{\int_{\sigma}|\delta{S_{+}}|^{2}d{\sigma}},
\label{DensityOfRatioIntensity}
\end{equation} 
where $n=D^{-2}$ is the density of 1D nexuses and $\sigma$ is area of  one-half of unit cell of pseudo-random lattice. \\[0.1cm]

\begin{figure*}
\centerline{\includegraphics[width=1.0\linewidth]{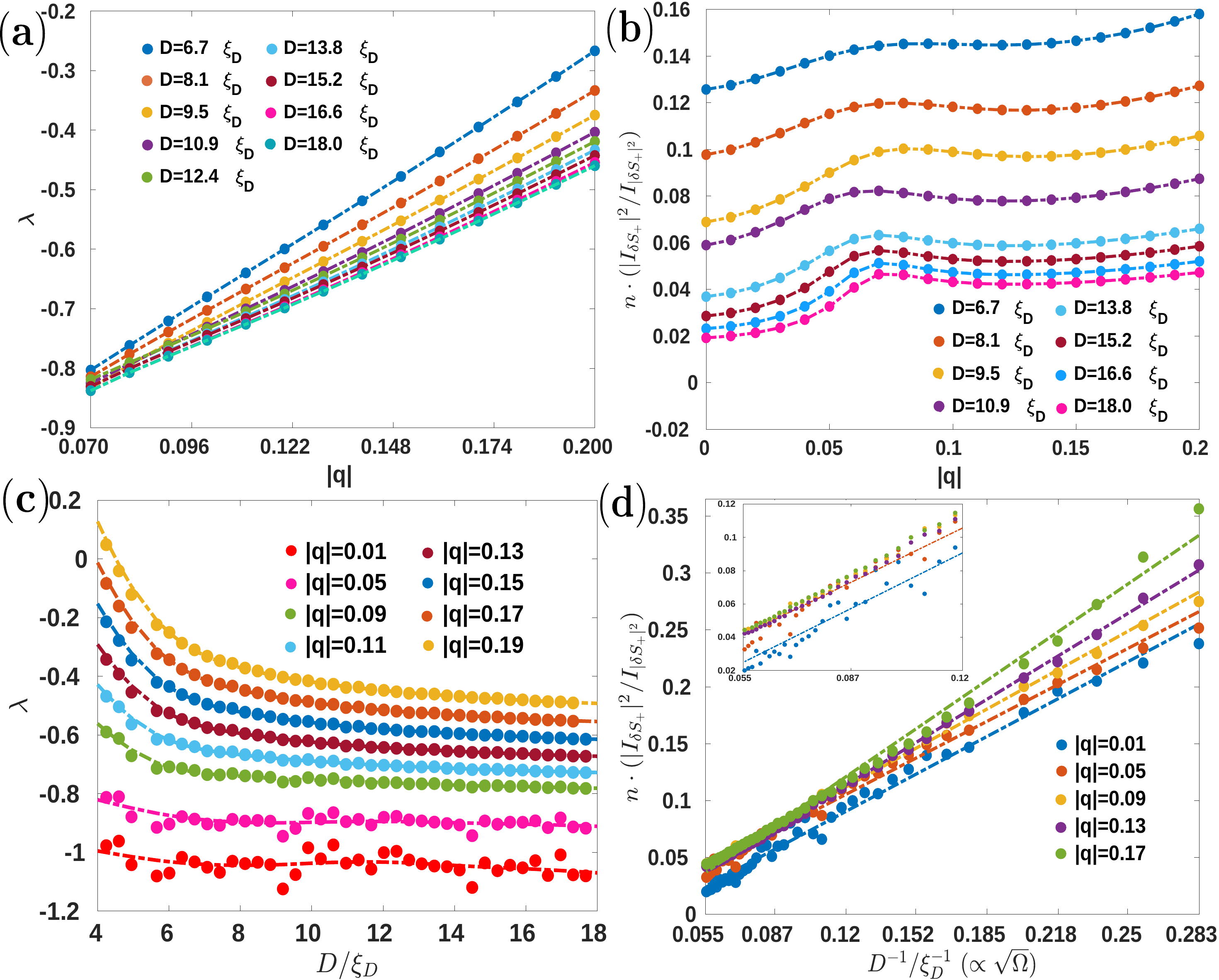}}
\caption{Transverse NMR frequencies shifts $\lambda$ and surface densities of ratio intensity $n \cdot (|I_{\delta{S_{+}}}|^{2}/I_{|\delta{S_{+}}|^{2}})$ of pseudo-random lattices of separable spin solitons (KLS-solitons and solitons). The frequency shifts $\lambda$ are eigenvalues of Eq.~(\ref{NMREigenEquationDimensonless}) with equilibrium textures of $1/4+1/4$ spin solitons in London limit. The surface densities of NMR ratio intensity are calculated by using Eq.~(\ref{DensityOfRatioIntensity}). All scattering dots represent the original numeric results, while colored lines are smoothing spline fittings of these original numeric results. (a) Transverse NMR frequency shifts $\lambda$  increases when $|q|$ increases. This is because only solitons ($|\Delta{\theta}|=\pi-2\theta_{0}$) contribute to the lowest transverse spin dynamic response mode. When $|q|$ increases, $\lambda$ generated by solitons increases, see the details in appendices Sec. \ref{SpinDynamicsSolionAndBigSoliton}. The typical values of $\lambda$ are larger than $-0.9$ when $D\geq10\xi_{D}$. (b) depicts the surface densities of ratio intensity $n \cdot (|I_{\delta{S_{+}}}|^{2}/I_{|\delta{S_{+}}|^{2}})$ of the eigenmodes. (c) depicts the transverse frequency shifts $\lambda$ as function of $D$. (d) The surface densities of ratio intensity $n \cdot (|I_{\delta{S_{+}}}|^{2}/I_{|\delta{S_{+}}|^{2}})$ as function of $1/D \propto \sqrt{\Omega}$. The inset is the magnified plot between $1/D=0.055$ till $1/D=0.120$. \label{NMRLambdaAndRatioIntensitySeparableSoliton}}
\end{figure*}
We demonstrate the moduli of the lowest transverse spin dynamic response modes $|\delta{S_{+}}(\omega)|$ located in the unit cells of pseudo-random lattices of 1D nexus objects with inseparable and separable $2/4$ spin solitons in Fig.~\ref{SpinDynamicResponseModes}. In the unit cell of pseudo-random lattice consisting of inseparable $2/4$ spin soliton, the lowest spin dynamic response mode locates on the region which is occupied by $\pi$-soliton. While, in the unit cell of pseudo-random lattice consisting of separable $2/4$ spin soliton, the lowest spin dynamic response mode locates on the region which is occupied by soliton ($|\Delta{\theta}|=\pi-2\theta_{0}$). This means the KLS-soliton in the separable spin soliton does not respond the continuous-wave magnetic drive. The transverse NMR frequency shifts $\lambda$ and surface densities of ratio intensity of pseudo-random lattices for inseparable and separable $2/4$ spin solitons are shown in Fig.~\ref{NMRLambdaAndRatioIntensityInseparableSoliton} and Fig.~\ref{NMRLambdaAndRatioIntensitySeparableSoliton} respectively. Let's discuss them separately.   

\subsection{Transverse NMR frequency shifts and surface densities of ratio intensity --- inseparable spin solitons}
\label{EigenModeOfInSeparableSoliton}
The transverse NMR frequency shifts $\lambda$ of pseudo-random lattices of inseparable spin solitons ($\pi$-solitons) exactly coincide with the observed values in the experiment of Ref.~\cite{Makinen2019}. As been shown in Fig.~\ref{NMRLambdaAndRatioIntensityInseparableSoliton} (a), the numeric values of $\lambda$ generated by network of $\pi$-solitons is around $-1.01$ to $-1.03$ when pseudo-random lattice model is good enough i.e., $D\geq10\xi_{D}$. In this case, the transverse NMR frequency shifts $\lambda$ slightly increase as $|q|$ increasing when $|q|>0.16$. This phenomenon has also been observed in experiment of Ref.~\cite{Makinen2019}. The ratio intensities generated by unit area of pseudo-random lattice consisting of $\pi$-solitons linearly increase when the square root of angular velocity $\sqrt{\Omega} \propto 1/D$ increases, as shown in Fig.~\ref{NMRLambdaAndRatioIntensityInseparableSoliton} (d). This coincides with the $\sqrt{\Omega}$-scaling of satellite intensity observed in the experiment when $T=0.38T_{c}$ ($|q|\approx0.152$) \cite{Makinen2019}. \\[0.1cm]

In Sec. \ref{EquilibriumSpinTexturesAndFreeEnergy}, based on the topological analysis and free energy calculations, we suggested the possible equilibrium state which was observed in experiment is the pseudo-random lattices of $2/4$ inseparable spin solitons of 1D nexus objects network. Here we see the results of numeric simulations of transverse NMR spin dynamic response  of this kind of pseudo-random lattices indeed coincide with the experimental observations.   

\subsection{Transverse NMR frequency shifts and surface densities of ratio intensity --- separable spin solitons}
\label{EigenModeOfSeparableSoliton}

In contrast with pseudo-random lattices consisting of inseparable $2/4$ spin solitons, the transverse NMR frequency shifts of pseudo-random lattices consisting of separable spin solitons strongly deviate from the results of experimental observations, see Fig.~\ref{NMRLambdaAndRatioIntensitySeparableSoliton} (a). $\lambda$ increase when $|q|$ increases. This is because only the solitons ($|\Delta{\theta}|=\pi-2\theta_{0}$) of separable spin solitons contribute to the transverse NMR frequency shift, and the frequency shifts $\lambda$ of the soliton ($|\Delta{\theta}|=\pi-2\theta_{0}$) increase as $|q|$ increases, see the details in Appendix.~\ref{SpinDynamicsSolionAndBigSoliton}. Moreover, the magnitudes of the surface densities of ratio intensity $n \cdot (|I_{\delta{S_{+}}}|^{2}/I_{\delta{S_{+}}^{2}})$ generated by pseudo-random lattices of separable spin solitons are larger than those generated by pseudo-random lattices consisting of inseparable spin solitons, as shown in Fig.~\ref{NMRLambdaAndRatioIntensitySeparableSoliton} (d).


\chapter{Conclusion and Outlook} 

\label{Chapter6} 




In this review, we discussed a series of significant results originated from fibration between vacuum manifolds of nafen-distorted superfluid $^3$He. This fibration occurs in the vicinity of the second time symmetry breaking of two-step successive symmetry breaking phase transition from normal phase vacuum to PdB phase via polar phase. In this symmetry breaking pattern, any topological objects of the polar phase, which are described by $\pi_{n}(R_{P})$, covert to the composite topological defects described by $\pi_{n}(R_{1},R_{2})$ in PdB phase because of the fibration. For the superfluid $^3$He system, the possible composite topological defects are string monopole (Nambu monopole) and cosmological KLS string wall. \\[0.1cm]
 
The ROTA group has experimentally observed the KLS string wall 
in nafen-disorted superfluid $^3$He \cite{Makinen2019}. In chapter.~\ref{Chapter3}, we demonstrated that in the vicinity of the second transition, such composite object is described by the relative homotopy groups $\pi_{1}(R_{1},R_{2})$ with length scale $r \leq \xi_{H}$. The reason for that is the existence of the two well separated length scales. The coherence length $\xi$, which relates to the symmetry breaking phase transition form the normal liquid to the polar phase, determines the core size of the half-quantum vortex (Alice cosmic string). The larger length scale $\xi/q\gg \xi$, which relates to the second symmetry breaking phase transition form the polar phase to the PdB phase, determines the soft core size of the KLS wall terminated by this string. \\[0.1cm]

In the first time symmetry breaking from normal phase to polar phase, the vacuum manifold is $R_{P}$. After the second time symmetry breaking, the vacuum manifold of system turn to be $R_{1}$. Because the fibration between $R_{1}$ and $R_{P}$, the topological objects of polar phase  convert to the composite cosmological objects in PdB phase. While the submanifold $R_{2}$ of $R_{1}$ becomes to the fiber, which describes the vacuum manifold of PdB generated from a given element of $R_{P}$. In physics, this is equivalent to symmetry breaking of polar phase with fixed value of the order parameter vacuum to PdB phase. Thus the observed  KLS domain wall terminated by the HQV is determined 
by the nontrivial element of $\pi_1(R_1,R_2)$, which isomorphic to $\pi_{1}(R_{P})$. The other composite object, which is still waiting for  observation, is the string monopole (hedgehog), which terminates the string (the spin vortex). Its topology is determined 
by the nontrivial element of $\pi_2(R_1,R_2) \cong \pi_{2}(R_{P})$. The core of the monopole is of coherence length size $\xi$, while the spin vortices have the soft core of size $\xi/q\gg \xi$. \\[0.1cm]

In the work from 1970s and later, the relative homotopy groups have been applied for classification of topological defects on the surface of the ordered system  i.e., the boojum \cite{Volovik1978}, and for classification of topological solitons terminated by point or linear defects \cite{MineyevVolovik1978}. The topology of these combined objects demonstrates new application of the relative homotopy groups. Moreover, this applications for the successive symmetry breaking phase transition reveals the footage of deep algebraic topological concepts in the novel physical system.\\[0.1cm]

Following the idea of group theory application, We also considered the more complicated object -- the nexus, which combines the monopole, the string terminated by monopole, and skyrmion (topological soliton) terminated by the same monopole. Such object in the PdB phase arises in the presence of orientation energies i.e., magnetic energy, which provides magnetic length $\xi_{H}$. 
In other condensed system, such as superconductor-ferromagnet heterostructures, the objects combining vortices and skyrmions were recently considered. These objects were suggested contain Majorana bound states \cite{Samme2019,Stefan2019}. \\[0.1cm]

As discussions in Chapter.~\ref{Chapter4} show, the nexus is powerful and useful tool, which allows the composite cosmological objects with high energies can be directly detected through low energy spin dynamics.
We discussed in details that the topological origin of the novel 1D nexus objects in PdB phase of nafen-distorted $^3$He superfluid. In contrast to the topological objects named 2D nexus objects which are similar but live in higher spatial dimension in PdB superfluid \cite{Anderson1977,Chechetkin1976,Volovik1977,Seppala1984,Pekola1990,Zhang2020}, the network of 1D nexus objects have been directly detected in ROTA's NMR measurement \cite{Makinen2019}. \\[0.1cm]

The observation is supported not only by topological analysis but also by the calculations of equilibrium free energies and the spin dynamic response. For the superfluid $^3$He distorted by nafen-strands, the locations of the HQVs are fixed once they appear during cooling down with a given angular velocity. In the limit of low angular velocity i.e., $\Omega \ll \Omega_{c}$, the average distance between pinned HQVs is around hundred microns. As a result, the KLS domain walls attached on the HQVs have very large geometric sizes when the symmetry breaking transition from polar phase to PdB phase occurs. In the spatial regions with length scales $\xi_{D}$, the SOC energy reduces the vacuum manifolds to discrete sets. The reduced vacuum manifolds have spin solitons, which are described by relative homotopy group $\pi_{1}(R_{1}^{H},\tilde{R}^{SOC}_{1})$. Similar process also happens in bulk Helium-3 superfluid and spinor Bose condensate \cite{Kondo1992,Seji2019,Liu2020}. 
The textures of spin soltions with length scales $\xi_{D}$ strongly influence the SOC energy and then modifies the low frequency NMR signals under continuous wave drive.  \\[0.1cm]

In Chapter.~\ref{Chapter4} and Chapter.~\ref{Chapter5}, we proved the $2/4$ spin solitons smoothly connected on every KLS string wall and form pseudo-random lattices in small angular velocity. The equilibrium configurations and the surface densities of equilibrium free energies of two different pseudo-random lattices with topological invariant $2/4$ are numerically evaluated with self-developed non-linear optimization library. These two types of pseudo-random lattices correspond to two representations of group $G = \pi_{1}(S_{S}^{1},\tilde{R}_{2})$, the relative homotopy group of $2/4$ spin solitons of 1D nexus objects. Our analysis shows the pseudo-random lattices of inseparable spin solitons are energy favorable. And we further calculated the transverse spin dynamic response under NMR continuous wave drive to compare with the experimental observations. The results of pseudo-random lattices consisting of inseparable spin solitons exactly coincide with the experimental measurements.   \\[0.1cm]

In the limit of low angular velocity, the pseudo-random lattices models work very well because the randomness of the network of 1D nexus objects doesn't influence the spin textures of spin solitons. This means the randomness and disorder introduced by nafen-aerogel are lost in the low angular velocity.
However, when the angular velocity approaches the critic value $\Omega_{c}$, the coupling between spin solitons may dramatically modify the equilibrium spin textures of random lattices of 1D nexus objects. In this case, the random distributions of KLS string wall lead to spin solitons glasses \cite{volovik2019}. Thus we can expect the observable effects of this randomness on the NMR spectrums under high enough angular velocity. Moreover, PdB phase could be a good platform to observe the monopole-antimonople networks because the string monopole is topologically protected by $\pi_{2}$ relative homotopy group \cite{Zhang2020}. These kinds of complex networks are predicted in condensed matter system and also in the Grand Unified Theories \cite{Kibble2015,Saurabh2019,Lazarides1980,Shafi2019,Volovik2019d,Lazarides2021,Lazarides2021b,Chakrabortty2021}. The Grand Unified Theories may have a huge variety of networks consisting of monopoles and strings because of their complex symmetry breaking chains \cite{Chakrabortty2018,Chakrabortty2019}. In the absence of magnetic field, the string monopoles in PdB phase may connect to planar solitons with geometric size around $\xi_{D}$ because of the reduction of vacuum manifold by SOC energy. Similar with pseudo-random lattices of spin solitons, these planar solitons may result in observable influence on NMR spectrum. \\[0.1cm]

In summary, we demonstrated and reveled the significant influences of fibration between vacuum manifolds for a given symmetry breaking system. Similar phenomena may also occurs in other systems, in which the similar fibration of vacuum manifolds happens. For example, the ferroelectric nematic liquid crystal, which was observed recently \cite{Chen2020} and the spinor Bose-Einstein condensate \cite{Takeuchi2020}. In fact the string-domain wall structure has been observed in the spin-1 Bose-Einstein condensate system. Another systems, in which the fibration mechanism could happens, are GUT models. In such kinds of models, the symmetry breaking patterns may be very complicated and may generate abundant novel results \cite{Chakrabortty2018,Chakrabortty2019}.   \\[0.1cm]

Typically the state of the system  with topological defects represents the excited state of the system. However, the topological defects can form the ground state. Earlier it was suggested that the suppression of the B-phase on the boundary of superfluid $^3$He may lead to formation of stripe phase in $^3$He-B under nanoscale confinement in a slab geometry \cite{Vorontsov2007}. On a microscopic level, this inhomogeneous phase is thought as the periodic array of the KLS domain walls between the degenerate states of the B-phase, see Refs. \cite{Volovik1990,SalomaaVolovik1988}. 
Recently the experiments, which are suggested as the possible observation of such spatially modulated phase, has been reported \cite{Shook2019,Levitin2019}. 
For nafen-distorted $^3$He, similar situation may take place as well. The strands of nafen could play the same role as the boundaries in the slab confinement. The suppression of the order parameter near the strands may result in the spontaneous proliferation of the composite defects leading to the stripe phases or stripe glasses. \\[0.1cm]
 


\appendix 



\chapter{Fibration, Relative Homotopy Groups and Exact Sequences} 
\label{AppendixA} 

\section{Long and short exact sequences}
\label{LongAndShortES}
The homotopy groups and relative homotopy groups of vacuum manifolds $R_{1}$ and $R_{2}$ form a long exact sequence (LES)
\begin{equation}
\xymatrix@1@R=10pt@C=12pt{
... \,\, \pi_{k}(R_2) \ar[r]^{i_{*}} & \pi_{k}(R_1) \ar[r]^{j_{*}} & \pi_{k}( R_1, R_2) \ar[r]^{{\partial}^{*}} & \,\, \pi_{k-1}(R_{2}) \,\,... 
}
\label{LongExactSequenceDefination}
\end{equation}
by the meaning of their definitions \cite{NashBook1988}.
The exact sequence of (relative) homotopy groups means that the image of any homomorphism $x_{*}^{n}:A \rightarrow B$ of Eq.~\ref{LongExactSequenceDefination} (the sets of the elements of the group $B$ into which the elements of $A$ are mapped) is the kernel of the next homomorphism $x_{*}^{n+1}:B \rightarrow C$ (the sets of the elements of $B$ which are mapped to the zero element of $C$) i.e. $\operatorname{im}x^{n}_{*} \cong {\ker}x^{n+1}_{*}$, with $n\in \mathbb{Z}$ \cite{NashBook1988}. \\[0.1cm]

The relative homotopy classes of $\pi_{k+1}(R_{1},R_{2})$ are mapped to the homotopy classes of $\pi_{k}(R_{2})$ by mapping the $k$-dimension subset of $k+1$ sphere, which surrounds the defects, into $R_{2}$. This mapping between two homotopy classes with different dimensions is called boundary homomorphism $\partial_{*}$ \cite{NashBook1988}. Boundary homormophism shows how topological objects with different dimensions connect to each other. \\[0.1cm]

In principle, Eq.~\ref{LongExactSequenceDefination} has infinite terms, thus Eq.~\ref{LongExactSequenceDefination} is called LES. This make the calculation and analysis difficult. Then we need to split LES in to short exact sequences (SES) \cite{suzuki1982}. For every relative homotopy group $\pi_{1}(R_{1},R_{2})$, the LES can be split as
\begin{equation}
\xymatrix@1@R=10pt@C=13pt{
0 \,\, \ar[r] \,\, &  \,\, \operatorname{im}j_{*} \,\, \ar[r]^-{j_{*}} \,\, & \,\, \pi_{k}(R_{1},R_{2}) \,\, \ar[r]^-{{\partial}^{*}} \,\, & \,\, \operatorname{im}{\partial}^{*} \,\,  \ar[r] \,\, & \,\, 0\\
}\,, 
\label{SESPi2R1R2}
\end{equation}
by image of $\partial^{*}$ and $j_{*}$. In this case, the relative homotopy group $\pi_{k}(R_{1},R_{2})$ is called as extension of $\operatorname{im}j_{*}$ by $\operatorname{im}{\partial}^{*}$.

\subsection{LES and SES of $\pi_{n}(R_{P})$}
\label{LESAndSESofRP}
Exact sequence is a concept coming from category theory, this means there are many way to construct exact sequence of homotopy goups (of course, for others algebraic structures such as ring or domains). Here we demonstrate the exact sequence of $\pi_{n}(R_{P})$ and use this exact sequence to get Eq.~(\ref{HomotopyP}). Because $R_{P} \cong (S^2\times  U(1))/{\mathbb{Z}}_2$, the long sequence
\begin{equation}
\xymatrix@1@R=10pt@C=12pt{
... \,\, \pi_{k}(\mathbb{Z}_2) \ar[r] & \pi_{k}(S^2\times  U(1)) \ar[r] & \pi_{k}(R_{P}) \ar[r] & \,\, \pi_{k-1}(\mathbb{Z}_2) \,\,... 
}
\label{LESforRP}
\end{equation}
is exact. Thus we immediately get the SES
\begin{equation}
\xymatrix@1@R=10pt@C=13pt{
0 \,\, \ar[r] \,\, &  \,\, \mathbb{Z}^{S} \,\, \ar[r]^-{j^{*}} \,\, & \,\, \pi_{2}(R_{P}) \,\, \ar[r]^-{{m}^{*}} \,\, & \,\, 0 \,\,  \ar[r] \,\, & \,\, 0\\
}\, 
\label{SESPi2P}
\end{equation}
of $\pi_{2}(R_{P})$ and SES
\begin{equation}
\xymatrix@1@R=10pt@C=13pt{
0 \,\, \ar[r] \,\, &  \,\, \mathbb{Z}^{\Phi} \,\, \ar[r]^-{n^{*}} \,\, & \,\, \pi_{1}(R_{P}) \,\, \ar[r]^-{{f}^{*}} \,\, & \,\, \mathbb{Z}_{2} \,\,  \ar[r] \,\, & \,\, 0\\
}\, 
\label{SESPi1P}
\end{equation}
of $\pi_{1}(R_{P})$. Equation.~(\ref{SESPi2P}) simply means $\pi_{2}(R_{P}) = \mathbb{Z}^{S}$, while Eq.~(\ref{SESPi1P}) suggests $\pi_{1}(R_{P}) = \tilde{\mathbb{Z}}$ such that $\pi_{1}(R_{P})/\mathbb{Z}^{\Phi} \cong \mathbb{Z}_{2}$.

\section{Extensions with and without magnetic field}
\label{app:derivation}
\subsection{No Magnetic Field} 

In the PdB phase the explicit form of exact sequence of homomorphisms is
\begin{equation}
\xymatrix@1@R=10pt@C=12pt{
\pi_2( R_2) \ar@{-}[d] \ar[r]^{i_{*}} & \pi_2(R_1) \ar@{-}[d] \ar[r]^{j_{*}} &
\pi_2( R_1, R_2) \ar@{-}[d] \ar[r]^-{{\partial}^{k}_{*}} & \pi_1(R_2) \ar@{-}[d] \ar[r]^{m_{*}} &
\pi_1( R_1) \ar@{-}[d] \ar[r]^{n_{*}} & \pi_1(R_1,R_2) \ar@{-}[d] \ar[r]^-{{\partial}^{p}_{*}} & \pi_0( R_2) \ar@{-}[d] \ar[r]^{q_{*}} & \pi_0(R_1) \ar@{-}[d]  \\
0 \ar[r]^{i_{*}} & 0 \ar[r]^{j_{*}} & \mathbb{Z} \ar[r]^{{\partial}^{k}_{*}} &
\mathbb{Z} \ar[r]^{m_{*}} & \mathbb{Z} \times \mathbb{Z}_{2} \ar[r]^{n_{*}} & \tilde{\mathbb{Z}} \ar[r]^{{\partial}^{p}_{*}} & \mathbb{Z}_{2} \ar[r]^{q_{*}} & 0 
}
\label{sequence}
\end{equation}
where the $\partial^{k}_{*}$ and $\partial^{p}_{*}$ are boundary homomorphisms. This gives the following relative homotopy groups:
$\pi_2( R_1, R_2) \cong \mathbb{Z}$, $\pi_1( R_1, R_2) \cong  \tilde{\mathbb{Z}}$ and $\pi_0( R_1, R_2) \cong 0$ as the fibration suggests. Then following Sec.~\ref{LongAndShortES}, we get the SES of $\pi_{2}(R_{1},R_{2})$  
\begin{equation}
\xymatrix@1@R=10pt@C=13pt{
0 \,\, \ar[r] \,\, &  \,\, 0 \,\, \ar[r]^-{j^{*}} \,\, & \,\, \pi_{2}(R_{1},R_{2}) \,\, \ar[r]^-{{\partial}_{*}^{k}} \,\, & \,\, 2\mathbb{Z} \,\,  \ar[r] \,\, & \,\, 0\\
}\,, 
\end{equation}
and SES of $\pi_{2}(R_{1},R_{2})$
\begin{equation}
\xymatrix@1@R=10pt@C=13pt{
0 \,\, \ar[r] \,\, &  \,\, \mathbb{Z} \,\, \ar[r]^-{n^{*}} \,\, & \,\, \pi_{1}(R_{1},R_{2}) \,\, \ar[r]^-{{\partial}_{*}^{p}} \,\, & \,\, \mathbb{Z}_{2} \,\,  \ar[r] \,\, & \,\, 0\\
}\,. 
\end{equation}
The boundary homomorphism $\partial^{k}_{*}$ maps $S^{1} \subset S^{2}$ to $R_{2}$. Then the $\operatorname{im} \partial^{k}_{*}$ describes all classes of string defects terminated by the monopoles. We found $\operatorname{im} {\partial_{*}^{k}} = 2\mathbb{Z}\cong \mathbb{Z}$, which is the set of even numbers. This means that only the spin vortices with even winding number can form the string monopole. This situation is similar to the monopole connected with four HQVs in the A-phase, where the total winding number is 2 \cite{volovik2000}.
The topologically trivial monopole cannot connect with the string defects because of ${\ker}{\partial^{k}_{*}} \cong 0$. Actually this trivial class is identical to $\pi_{2}(R_{1})$ because ${\ker}{j_{*}} \cong \operatorname{im} {i_{*}} \cong 0$. \\[0.1cm]

The $\partial^{p}_{*}$ maps the homotopy classes of KLS string wall to homotopy classes of domain wall. From ${\ker}q_{*} \cong \operatorname{im}\partial^{p}_{*} \cong \pi_{0}(R_{2}) \cong \mathbb{Z}_{2} $, we know there are domain walls bounded by string defects. The set of half-odd integers of the group $\tilde{\mathbb{Z}}$, which come from $\operatorname{im} \partial^{p}_{*}$ describes the domain wall terminated by string defects -- 
 the KLS wall terminated by vortex with half-odd integer winding number $N=k+1/2$. The vortices, which come from ${\ker}\partial^{p}_{*}\cong\mathbb{Z}$ are vortices with integer winding number. These vortices are free. \\[0.1cm]


\subsection{In the presence of magnetic field}
\label{InThePresentOfMagneticField}

In the presence of magnetic field the corresponding exact sequence is 
\begin{equation}
\xymatrix@1@R=10pt@C=13pt{
\pi_2(S^{1} \times U(1)) \ar@{-}[d] \ar[r]^-{i_{*}} & \pi_2(R_1) \ar@{-}[d] \ar[r]^-{j_{*}} &
\pi_2( R_1, S^{1} \times U(1)) \ar@{-}[d] \ar[r]^-{{\partial}^{k}_{*}} & \pi_1(S^{1} \times U(1)) \ar@{-}[d] \ar[r]^-{m_{*}} &
\pi_1( R_1) \ar@{-}[d]\\
0 \ar[r]^{i_{*}} & 0 \ar[r]^{j_{*}} & \mathbb{Z} \ar[r]^{{\partial}^{k}_{*}} &
\mathbb{Z}\times \mathbb{Z} \ar[r]^{m_{*}} & \mathbb{Z}_{2} \times \mathbb{Z}  \\
}
\label{sequence_SkyrmionNexus}
\end{equation}
i.e. $\pi_{2}(R_{1},S^{1} \times U(1)) = 2\mathbb{Z} \cong \mathbb{Z}$ and the corresponding SES
\begin{equation}
\xymatrix@1@R=10pt@C=13pt{
0 \,\, \ar[r] \,\, &  \,\, 0 \,\, \ar[r]^-{j^{*}} \,\, & \,\, \pi_{2}(R_{1},S^{1} \times U(1)) \,\, \ar[r]^-{{\partial}_{*}^{k}} \,\, & \,\, 2\mathbb{Z} \,\,  \ar[r] \,\, & \,\, 0\\
}\,
\label{SES_SkyrmionVorticesNexus}
\end{equation}
From SES in Eq. (\ref{SES_SkyrmionVorticesNexus}) it follows that $\pi_{2}(R_{1},S^{1} \times U(1)) = 2\mathbb{Z}_{2} \cong \mathbb{Z}$. We found $\ker \partial^{k}_{*} \cong 0$ and $\operatorname{im} \partial^{k}_{*} = 2\mathbb{Z} \cong \mathbb{Z}$. That means that  only those objects, which have an even total winding number of spin rotation, are topologically protected. These objects are the $\hat{\mathbf{d}}$-vector skyrmions. Since $\pi_{2}(R_{1},S^{1} \times U(1)) \cong \pi_{2}(R_{2},R_{1})$, these $\hat{\mathbf{d}}$-skyrmions can terminate on the $\hat{\mathbf{d}}$-monopole, which in turn is the end point of spin vortices with the even number of spin rotation and is the linear analogs of the original point-like skyrmion \cite{skyrme1962,khawaja2001}. As a result one obtains the composite effect -- the nexus in Fig.~\ref{NexusFig}.  
The spin texture inside the cross-section $D_{2}$ of the skyrmion corresponds to continuous mapping to $SO_{S}(3)$, which is implemented by choosing first a direction of $\hat{\mathbf{d}}$, then making $SO_{S}(2)$ rotation of $\hat{\mathbf{e}}^{1}$ and  $\hat{\mathbf{e}}^{2}$ around this direction.  This skyrmion also represents  the spin vortex with even winding number, because of $\operatorname{im} \partial_{*}^{k} \cong 2\mathbb{Z}$ and ${\ker}\partial_{*}^{k} \cong 0$.

\section{Fibration and a theorem}
\label{FibrationAndTheorems}
Fibration is a significant concepts of Homotopy theory \cite{HatcherBook2002}. Like most of modern mathematics, fibration can be described in a quite general language with the help of category theory. The category theory concept, which be used to define fibration, is the lift property or lifting property of a given continuous mapping. When this concept be mapped into homotopy theory, it gets a more explicit name i.e., the homotopy lifting property. In order to get an intuitive understanding of this concept, let's us imagine that we have a continuous mapping $p$ between topological spaces $E$ and $B$ such that
\begin{equation}
p:E \rightarrow B,
\end{equation}    
and $B$ has a homotopy 
\begin{equation}
f_{t}:X \times [0,1] \mapsto B.
\end{equation}
Now, we introduce a new homotopy $\tilde{f}_{t}:X \times [0,1] \mapsto E$ for $E$, such that
\begin{equation}
f_{t} = p \circ \tilde{f}_{t}, \,\,\,\, f_{0} = p \circ \tilde{f}_{0}.
\end{equation}
If such $\tilde{f}$ really exists, then we say $p$ has homotopy lifting property \cite{HatcherBook2002}. Furthermore, if $p$ has homotopy lifting property respect to any topological space $X$, then $p$ is named as firbration. \\[0.1cm]

One classic example of fibration is the Hopf fibrarion between $S^{3}$ and $S^{2}$. Here the fibration $p: S^{3} \rightarrow S^{2}$ is Hopf mapping. Another example is Serre fibration between $SO(3)$ and $S^{2}$. For this example, $SO(3)$ is universal covering space of $S^{2}$. For our case, the fibration between $R_{1}$ and $R_{P}$ is quite similar with the case of Serre fibration i.e., $p:R_{1} \rightarrow R_{P}$ is covering mapping and $R_{1}$ is covering space of $R_{P}.$ As we have seen in Chapter.~\ref{Chapter3}, with the help of the third isomophsim theorem, we indeed identified the covering mapping between $R_{1}$ and $R_{P}$. \\[0.1cm]

For a given fibration, there is a theorem to describe the relations between the relative homotopy group $\pi_{n}(E,F)$ and homotopy group $\pi_{k}(B)$, where $F \subset E$ is the fiber \cite{HatcherBook2002}.
\begin{itemize}
\item \textbf{Theorem} Given a fibration $p: E \rightarrow B$, points $b \in B$ and $e \in F: p^{-1}(b)$, there is an isomorphism $p_{*} : \pi_{n}(E,F,e) \rightarrow \pi_{n}(B,b)$.
\end{itemize}


\chapter{Soblev Space, Non-Linear Optimization and Discrete Eigen-Value Problem } 

\label{AppendixB} 

In Appdeix.~\ref{AppendixB}, we discuss the mathematical background and algorithm implements of numeric minimization of functional --- the non-linear optimization. The motivation of this is using a mathematically reliable method to find out the equilibrium configurations of pseudo-random lattice of 1D nexus network. The solving domain is the unit cell of pseudo-random lattice. However, because there function $\theta$ is even respect to the $y$-direction of solving domain, the calculations are usually be practiced in one-half of unit cell. In this case, the stable configuration of $\theta$ is element of Helbert space of functions, which are defined on the solving domain \cite{CiarletBook1978}. Thanks for the Lax-Milgram lemma, there is one and only one solution for $\theta$ \cite{CiarletBook1978}. However, this is helpless to solve our problem, because it says nothing about how and where to find out this solution. \\[0.1cm] 

In this case, we introduce the new concept about function space i.e., the Sobolev space, which is subset of $C^{m}(\Omega)$ within domain $\Omega$. The significant property of Sobolev space is it supports the weak derivatives of its elements. This means all elements of Sobolev space satisfy the Green integral formula --- the high dimensional analog of integral by parts \cite{CiarletBook1978}. Based on the Sobolev space, the Ritz strategy and Galerkin strategy are developed, such that the former is suitable for functional minimization, while the latter is suitable for eigen-equation problem \cite{CiarletBook1978}. We start our discussion from Ritz strategy, then introduce the finite element partition of elements of Sobolev space. Finally we discuss how the Garlkin strategy be conducted for NMR eigen-equations. The main algorithm will be shown by pseudo-code and the Matlab script. The C++ code for calculation of sparse matrices of the square of derivative terms will be shown in last section .
\begin{figure*}
\centerline{\includegraphics[width=0.87\linewidth]{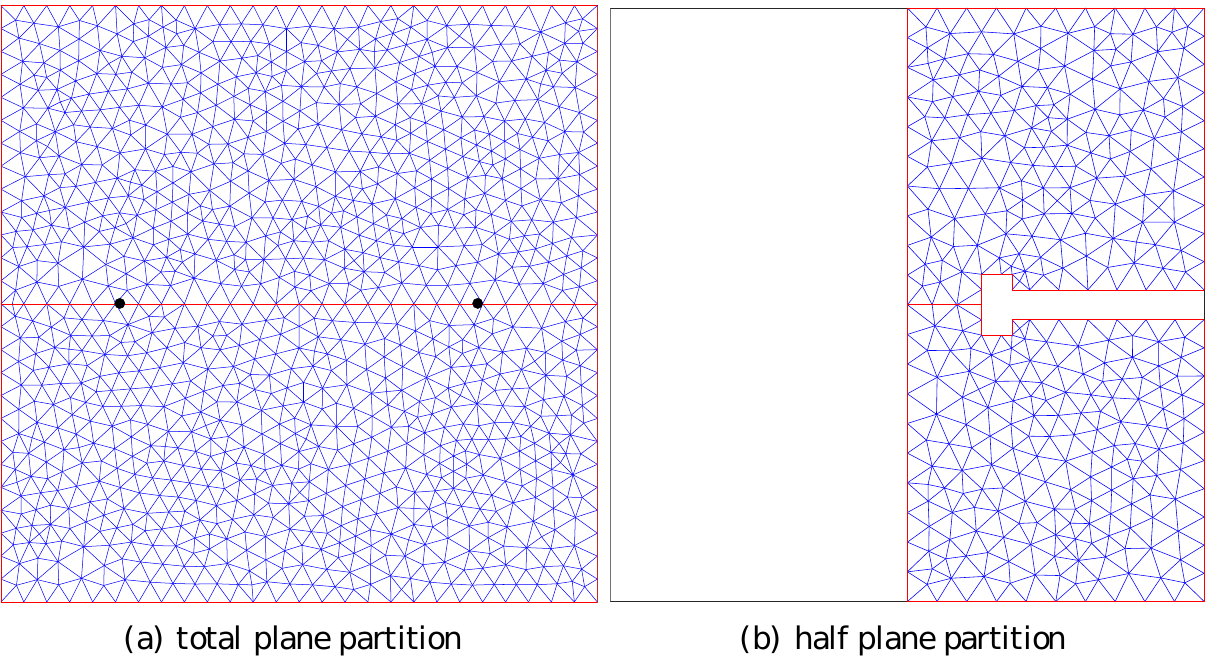}}
\caption{Triangular elements partitions in unit cell of pseudo-random lattice. (a) The partition may be done for the whole unit cell, while (b) it also can be done for one-half of the unit cell. In most case, I used the partition (b) because it provide more higher points density than the manner (a).
\label{Partation}   
} 
\end{figure*}

\section{Ritz strategy and finite elements partition}
\label{RitzStrategyAndFiniteElementsPartition}
The Hilbert space on the solving domain is an infinite dimensional space, then it is impossible to deal with it by computer. The basic idea of Ritz strategy is choosing a subset $S_{N}$ of Hilbert space to substitute the Hilbert space. $S_{N}$ is the $N$-elements Sobolev space, then the $\theta$ function is expanded as 
\begin{equation}
\theta = {\sum_{i}^{N}} c_{i} \phi_{i}(\mathbf{x}),
\label{ThetaExpansion}
\end{equation}
where $\phi_{i}(\mathbf{x}) \in S_{N}$, $c_{i} \in \mathbb{C}$ and  $\mathbf{x}$ is space coordinates. Pluging Eq.~(\ref{ThetaExpansion}) into the dimensionless  London limit free energy Eq.~(\ref{FreeEnergyThetaDimensionless}), we get the finite-dimensional approximation 
\begin{align}
\tilde{F}(c_{i})_{London} 
 = & {\frac{1}{\xi_{D}}}  \sum_{ij}^{N} \{ \int\nolimits_{\Sigma} c_{i}c_{j} [\frac{1}{2} (\gamma_{1}+2\gamma_{2}) \partial_{x}\phi_{i} \partial_{x}\phi_{j} + \frac{1}{2}\gamma_{1}\partial_{y}\phi_{i} \partial_{y}\phi_{j} \} d\Sigma  \notag \\
 & + \frac{1}{\xi_{D}^{2}} \int\nolimits_{\Sigma} (-\frac{1}{2}{\gamma_{4}}cos2{\sum_{i}^{N} c_{i} \phi_{i}}-\gamma_{3}sin{\sum_{i}^{N} c_{i} \phi_{i}})] d\Sigma
\label{RitzFreeEnergyThetaDimensionless}
\end{align}
of the continuous London limit free energy. Eq.~(\ref{RitzFreeEnergyThetaDimensionless}) actually transfers the continuous solution $\theta$ over $\Sigma$ to a discrete set of coefficients $c_{1}$, $c_{2}$, $c_{3}$ ...$c_{i}$... $c_{N}$. We can get $\theta$ from Eq.~(\ref{ThetaExpansion}) as long as we get $c_{i}$. Moreover, there is additional benefit of Eq.~(\ref{RitzFreeEnergyThetaDimensionless}), that is Eq.~(\ref{RitzFreeEnergyThetaDimensionless}) converts the free energy functional to a high dimensional function, which can be numerically minimized by non-linear optimization. In order to get $c_{i}$ by using computer, we still need a properly chosen set of $\phi_{i} \in S_{N}$. And we hope $\phi_{i}$ could be as simple as possible. \\[0.1cm]

This special set of $\phi_{i}$ is set of interpolation functions on triangular elements of finite elements (FE) partition \cite{CiarletBook1978}. In this manner, the solving domain is partitioned into small triangles as shown in Fig.~\ref{Partation}. Then Eq.~(\ref{RitzFreeEnergyThetaDimensionless}) can be written as
\begin{align}
\tilde{F}(c_{i})_{London} 
 = & {\frac{1}{\xi_{D}}}  \sum_{ij}^{N} \sum_{c} \{ \int\nolimits_{c} c_{i}c_{j} [\frac{1}{2} (\gamma_{1}+2\gamma_{2}) \partial_{x}\phi_{i} \partial_{x}\phi_{j} + \frac{1}{2}\gamma_{1}\partial_{y}\phi_{i} \partial_{y}\phi_{j} \} d\Sigma  \notag \\
 & + \sum_{c} \frac{1}{\xi_{D}^{2}} \int\nolimits_{c} (-\frac{1}{2}{\gamma_{4}}cos2{\sum_{i}^{N} c_{i} \phi_{i}}-\gamma_{3}sin{\sum_{i}^{N} c_{i} \phi_{i}})] d\Sigma,
\label{RitzFreeEnergyThetaDimensionlessI}
\end{align}
where $c$ represents the triangular element (cell) of the partition. The benefit of converting free energy functional into the form og Eq.~(\ref{RitzFreeEnergyThetaDimensionlessI}) is we only need to deal the explicit form of $\phi_{i}$ on a small cell, other than on the whole solving domain. In this situation, we can expect any element (e.g., $\theta$ for our problem) of $S_{N}$ is a linear function on every $c$ as long as they connect continuously with each other and $c$ keeps be small. In most case, this element linear function is given as \cite{CiarletBook1978}
\begin{equation}
\theta^{c} = \sum_{\alpha}^{3} c_{\alpha} \phi_{\alpha}^{c},
\label{ElementIntepolationFunction}
\end{equation} 
where 
\begin{equation}
\phi_{\alpha}^{c} = \frac{1}{D} (\eta_{\alpha} x -\xi_{\alpha} y + \omega_{\alpha}), \,\, D = \sum_{\alpha}^{3} \omega_{\alpha} 
\end{equation}   
with
\begin{align}
\xi_{1} = x_{2} -x_{3}, & \, \xi_{2} = x_{3} -x_{1}, \, \xi_{3} = x_{1} -x_{2}, \notag \\\
\eta_{1} = y_{2} -y_{3}, & \, \eta_{2} = y_{3} - y_{1}, \, \eta_{3} = y_{1} -y_{2}, \\
\omega_{1} = x_{2}y_{3} -x_{3}y_{2}, & \, \omega_{2} = x_{3}y_{1} -x_{1}y_{3}, \, \xi_{3} = x_{1}y_{2} -x_{2}y_{1}. \notag
\label{Coefficients}
\end{align}
Plugging Eq.~(\ref{ElementIntepolationFunction}) into Eq.~(\ref{RitzFreeEnergyThetaDimensionlessI}), and keep in mind that the coefficients $c_{\alpha}$ of cell should be reorganized as $c_{i}$ of whole domain, we get 
\begin{equation}
\tilde{F}(c_{i})_{London} 
 = {\frac{1}{\xi_{D}}}  \sum_{ij}^{N} a_{ij} c_{i}c_{j} \, + \, \sum_{c} \frac{1}{\xi_{D}^{2}} \int\nolimits_{c} (-\frac{1}{2}{\gamma_{4}}cos2{\sum_{\alpha}^{3} c_{\alpha} \phi_{\alpha}^{c}}-\gamma_{3}sin{\sum_{\alpha}^{3} c_{\alpha} \phi_{\alpha}^{c}})] d\Sigma,
\label{RitzFreeEnergyThetaDimensionlessII}
\end{equation}
where $a_{ij}$ is a sparse matrix and contains different summations of the quadratic integration in Eq.~(\ref{RitzFreeEnergyThetaDimensionlessI}).

\section{BFGS algorithm and implements}
\label{BFGSAlgorithmAndImplements}
After the conversion of the continuous free energy functional to high-dimensional function in Eq.~(\ref{RitzFreeEnergyThetaDimensionlessII}), we must consider how to minimize this function, which usually has $4000$ to $6000$ $c_{i}$. What we chosen is the popular BFGS algorithm \cite{JorgeBook2006}. As a quasi-Newtonian method, BFGS algorithm merely uses first order derivative and fast enough. The algorithm is given as
\begin{algorithm}
\caption{BFGS iteration}
\begin{algorithmic} 
\REQUIRE Given vector $c_{i}^{0}$, $epsilon > 0$ and unit matrix $H^{0} = I$, and calculate the zeroth gradient $\partial_{i} \tilde{F}(c_{i})$ 
\WHILE{$\text{norm}(\partial_{i} \tilde{F}(c_{i}) > \epsilon$} 
\STATE $d_{k} \leftarrow -H_{k} \partial_{i} \tilde{F}(c_{i})$
\STATE Take single variable minimization for $\tilde{F}(c_{i}^{k}+\lambda_{k} d_{k})$ to search $\lambda_{k}$
\STATE $c_{i}^{k+1} \leftarrow c_{i}^{k}+\lambda_{k} d_{k}$
\STATE calculate the $k+1$th gradient $\partial_{i} \tilde{F}(c_{i}^{k+1})$
\STATE make the BFGS modification for $H^{k+1}$
\ENDWHILE
\end{algorithmic}
\label{BFGSAlgorithm}
\end{algorithm}
The $a_{ij}$ matrix is used for calculating the gradient $\partial_{i} \tilde{F}(c_{i})$ and $\lambda_{k}$-search, then $a_{ij}$ actually has been gotten before the iteration starting. The iteration.~\ref{BFGSAlgorithm} is conducted on Matlab and the $a_{ij}$ is calculated by self-developed C++ library. The C++ code must be complied to a sharing library in Matlab environment with supported g++ compiler, this is because Matlab has its own API for C++ program. In the last part of Sec.~\ref{BFGSAlgorithmAndImplements}. we show he matlab script and C++ code. \\[0.1cm]

Here is the Matlab script of iteration.~\ref{BFGSAlgorithm}:
\begin{lstlisting}[language=Matlab]
gkn1=gradient2DdSolitonWallQN(Cv,CbIv,A,...
     Thetakn1,elements,nodes,Np,gamma1,gamma);                             
while norm(gkn1)>tol
      dkn1=-Hkn1*gkn1;
      lambdakn1=fminsearch(@(lambda) PiThetaLambda(lambda,...
                           A,Thetakn1,dkn1, Cv,CbIv,...
                           elements,Nelem,nodes,Np,gamma1,...
                           gamma),0);    
                           % search \lambda_{k}  
      CV=[Cv CbIv]; thetakn1t=zeros(length(CV),1); 
      for k=1:length(CV) 
          thetakn1t(k,1)=Thetakn1(CV(k));
      end
      % take out all variables of $c_{i}$
           
      thetakt=thetakn1t+lambdakn1.*dkn1; % iteration
      Thetak=TransferVtoTheta2DSolitonWallQN(thetabw,...
             thetabu,thetabd,thetabiu,...
             thetabid,thetakt,CbW,Cbu,Cbd,..
             Cbiu,Cbid,CbIv,Cv,Thetakn1,Np,2);
             % refresh the solution
             
      %%%%%%%% BFGS modification %%%%%%%                                    
      skn1=thetakt-thetakn1t;
      gk=gradient2DdSolitonWallQN(Cv,CbIv,A,Thetak,...
          elements,nodes,Np,gamma1,gamma);
      ykn1=gk-gkn1; rhokn1=1./((ykn1')*skn1);
      Hk=InHassian2DSolitonQN(Is,rhokn1,skn1,ykn1,Hkn1);
      %%%%%%%%%%%%%%%%%%%%%%%%%%%%%%%%%%
      Hkn1=Hk; Thetakn1=Thetak; gkn1=gk;     
      % refresh H^{k} and gradient for next iteration           
end    
\end{lstlisting}
In this script, the functions \textit{gradient2DdSolitonWallQN}, \textit{PiThetaLambda} and \textit{TransferVtoTheta2DSolitonWallQN} are all self-developed scripts, they can be found in my Github repository \cite{MyGithub}. The implement of calculation for $a_{ij}$ is a C++ class, which was named as "CalculateAMatrix". The head file is shown as
\begin{lstlisting}[language=c++]
/*
 * this code is the header file of 
 * c++ class AMtrixCalculator, which 
 * gets elemts and nodes array to
 * calculate the "A matrix"
 * the inputs are the pointer arraies 
 * of element array (2D) and node array (2D)
 */
 
#ifndef calculateAMatrix_hpp
#define calculateAMatrix_hpp
#include <cstddef>
#include <iostream>

class CalculateAMatrix {
    int ** elementArrayPtr;
    double ** nodeArrayPtr;
    double ** AMatrixArrayPtr;     
public:
  size_t lengthRowElement;
  size_t lengthColElement;
  size_t lengthRowNode;
  size_t lengthColNode;

  CalculateAMatrix(): lengthRowElement(0), lengthColElement(0), 
                      elementArrayPtr(NULL), lengthRowNode(0), 
                      lengthColNode(0), nodeArrayPtr(NULL) { }; 
  // default constructor to initilaze data member

  // pass in the elements array and node array
  void getAllArray(int ** array1ptr, double ** array2ptr, 
                  const size_t rowDim1, const size_t colDim1, 
                  const size_t rowDim2, const size_t colDim2);
  
  // Show what array have been passed in this "A-Matrix" calculator
  void showOriginalArr(int ** array1ptr, double ** array2ptr, 
                      const size_t rowDim1, const size_t colDim1,
                      const size_t rowDim2, const size_t colDim2);
   
  void showObjectLocalArray();

  void AMatrixCalculatorRun();

  void showObjectLocalAMatrix();

  double ** const returnAMatrixPtr() const;
  
  // implements of member functions 
  void CalculateAMatrix::getAllArray(int ** array1ptr, 
       double ** array2ptr, const size_t rowDim1,
       const size_t colDim1, const size_t rowDim2, 
       const size_t colDim2)
{
  // make a object copy of Element array
  if (this->lengthRowElement == 0 
     && this->lengthColElement == 0 
      && this->elementArrayPtr == NULL){ 
      // allocate the row pointers array, 
      // and allocate column array for every
         this->elementArrayPtr = new int * [rowDim1];
         for(size_t idx_row = 0; idx_row < rowDim1; ++idx_row)
	     {
	         elementArrayPtr[idx_row] = new int[colDim1];
	     }
	  }
  // copy array element into the Object-local Element array 
  for (size_t idx_row = 0; 
       idx_row < rowDim1; ++idx_row)
	  {
	    for (size_t idx_col = 0; 
	         idx_col < colDim1; ++idx_col)
                this->elementArrayPtr[idx_row][idx_col] 
                   = array1ptr[idx_row][idx_col];
	  }  
  this->lengthRowElement = rowDim1;
  this->lengthColElement = colDim1;
  // make a object copy of Node array
  if (this->lengthRowNode == 0 
      && this->lengthColNode == 0 
      && this->nodeArrayPtr == NULL)
        {
	 
      // allocate the row pointers array, 
      // and allocate column array for every
	  // pointer elments of row pointer array.
             this->nodeArrayPtr = new double * [rowDim2];
             for(size_t idx_row = 0; idx_row < rowDim2; ++idx_row)
	     {
	         nodeArrayPtr[idx_row] = new double[colDim2];
	     }
        }

  // copy array element into the Object-local Node array 
  for (size_t idx_row = 0; idx_row < rowDim2; ++idx_row)
	  {
	    for (size_t idx_col = 0; idx_col < colDim2; ++idx_col)
            this->nodeArrayPtr[idx_row][idx_col]
             = array2ptr[idx_row][idx_col];
	  }  
  this->lengthRowNode = rowDim2;
  this->lengthColNode = colDim2;
}

void CalculateAMatrix::showOriginalArr(int ** array1ptr, 
     double ** array2ptr,const size_t rowDim1,
     const size_t colDim1, const size_t rowDim2, 
     const size_t colDim2)
{
  std::cout << " first pointer to pointer 
       of orignal Element array: " << array1ptr << std::endl;
  std::cout << " second pointer to pointer
       of original Node array: " << array2ptr << std::endl;

  std::cout << " Show the first array passed in " << std::endl;
  for(size_t rowidx = 0; rowidx < rowDim1; ++rowidx)
    {
      for(size_t colidx = 0; colidx < colDim1; ++colidx)
	std::cout << array1ptr[rowidx][colidx] << "  ";
      std::cout << std::endl;
    }

  std::cout << " Show the second array passed in " << std::endl;
  for(size_t rowidx = 0; rowidx < rowDim2; ++rowidx)
    {
      for(size_t colidx = 0; colidx < colDim2; ++colidx)
	  std::cout << array2ptr[rowidx][colidx] << "  ";
      std::cout << std::endl;
    }
}

void CalculateAMatrix::showObjectLocalArray()   
{
  std::cout << " pointer to pointer of Object-local Element array: " 
       << elementArrayPtr << std::endl; 
  for (size_t idx_row = 0; idx_row < lengthRowElement; ++idx_row)
	  {
	    for (size_t idx_col = 0; 
	    idx_col < lengthColElement; ++idx_col)
	    std::cout << 
	     this->elementArrayPtr[idx_row][idx_col] << " ";
	    std::cout << std::endl;
	  }  

  std::cout << " pointer to pointer of Object-local Node array: "
            << nodeArrayPtr << std::endl; 
  for (size_t idx_row = 0; idx_row < lengthRowNode; ++idx_row)
	  {
	    for (size_t idx_col = 0; idx_col < lengthColNode; ++idx_col)
	    std::cout <<  this->nodeArrayPtr[idx_row][idx_col] << " ";
	    std::cout << std::endl;
	  }  
  
}

// memeber function AMatrixCalculatorRun() 
// is  the main functionality of this class
// i.e., calculate the A Matrix with object-local arraies
void CalculateAMatrix::AMatrixCalculatorRun()
{
  std::cout << " calculator starts " << std::endl;
  // dynamically allocate the array of A-Matrix, 
  // using pointer array for store the Row
  AMatrixArrayPtr = new double * [lengthRowNode];
  std::cout << " ptr array allocate successfully " << std::endl;
 
  for(size_t idx_row = 0; idx_row < lengthRowNode; ++idx_row)
    {
      AMatrixArrayPtr[idx_row] = new double[lengthRowNode];
    }
  std::cout << " A-Mtrix columns 
      allocate successfully " << std::endl;
  
  // put 0.0 into the allocated A-Matrix, innitilization
  for(size_t idx_row = 0; idx_row < lengthRowNode; ++idx_row)
    {
      std::cout << " assign loop 
      1-level, idx_row is  " 
      << idx_row << std::endl;
      for(size_t idx_col = 0; 
      idx_col < lengthRowNode; ++idx_col)
	{
      std::cout << " assign loop 2-level,
       idx_col is  " << idx_col << std::endl;
	  AMatrixArrayPtr[idx_row][idx_col] = 0.0;
	}  
    }

  std::cout << " dynamical allocated successfull " << std::endl;

  // calculate the Matrix-elements and save 
  // the results into the initilized A-Matrix array
  size_t idx_element = 0;
  while( idx_element < lengthRowElement )
    {
     std::cout << " loop start: 
       idx_element is " << idx_element << std::endl;
      
     // data prerparations  
     int nodeIndex0Elem = elementArrayPtr[idx_element][0];
     int nodeIndex1Elem = elementArrayPtr[idx_element][1];
     int nodeIndex2Elem = elementArrayPtr[idx_element][2];
     int nodeIndex[] = {nodeIndex0Elem,
              nodeIndex1Elem, nodeIndex2Elem};

     std::cout << " node indcies 
          with idx_element " << idx_element << std::endl;
     std::cout << " nodeIndex in element " << idx_element
      << " are " << nodeIndex[0] << " " << nodeIndex[1] 
      << " " << nodeIndex[2] << std::endl;

     double x0 = nodeArrayPtr[nodeIndex0Elem][0];
     double y0 = nodeArrayPtr[nodeIndex0Elem][1];
     
     double x1 = nodeArrayPtr[nodeIndex1Elem][0];
     double y1 = nodeArrayPtr[nodeIndex1Elem][1];

     double x2 = nodeArrayPtr[nodeIndex2Elem][0];
     double y2 = nodeArrayPtr[nodeIndex2Elem][1];

     // xi, eta, omega
     double xi[] = {x1-x2, x2-x0, x0-x1};
     double eta[] = {y1-y2, y2-y0, y0-y1};
     double omega[] = {x1*y2-x2*y1, x2*y0-x0*y2, x0*y1-x1*y0};
     
     double Di = 0; // initilization Di
       for(size_t i = 0; i < 3; ++i) { Di += omega[i]; }
     double Si = (std::abs (Di))/2;
     
     // Wall effect
     // double gx = gamma1 + 2*gamma2;
     // double gy = gamma1;
        double gx = 0.1;
	double gy = 0.05;

	/*
      all data have been prepared, next is build 
      the element matrix ax2ij, ay2ij, a2ij
     */

     //	define a 3 by 3 array for ax2ij
	double ax2ij[3][3] = {
	  {0.0, 0.0, 0.0},
	  {0.0, 0.0, 0.0},
	  {0.0, 0.0, 0.0}
	};

     //	define a 3 by 3 array for ay2ij
	double ay2ij[3][3] = {
	  {0.0, 0.0, 0.0},
	  {0.0, 0.0, 0.0},
	  {0.0, 0.0, 0.0}
	};
	
     //	define a 3 by 3 array for a2ij
	double a2ij[3][3] = {
	  {0.0, 0.0, 0.0},
	  {0.0, 0.0, 0.0},
	  {0.0, 0.0, 0.0}
	};

     // calculate ax2ij and ay2ij
	for(size_t o = 0; o < 3; ++o)
	  {
	    ax2ij[o][0] = (1.0/(2.0*(Di*Di)))*eta[o]*eta[0]*Si*gx;
	    ax2ij[o][1] = (1.0/(2.0*(Di*Di)))*eta[o]*eta[1]*Si*gx;
	    ax2ij[o][2] = (1.0/(2.0*(Di*Di)))*eta[o]*eta[2]*Si*gx;

	    ay2ij[o][0] = (1.0/(2.0*(Di*Di)))*xi[o]*xi[0]*Si*gy;
	    ay2ij[o][1] = (1.0/(2.0*(Di*Di)))*xi[o]*xi[1]*Si*gy;
	    ay2ij[o][2] = (1.0/(2.0*(Di*Di)))*xi[o]*xi[2]*Si*gy;
	  }
     //	a2ij = ax2ij + ay2ij
	for(size_t i = 0; i < 3; ++i)
	  {
	    for(size_t j = 0; j < 3; ++j)
	      {
		a2ij[i][j] = ax2ij[i][j] + ay2ij[i][j];
	      }
         
	  }

	std::cout << " a2ij matrix now is " << std::endl;
        for(size_t i = 0; i < 3; ++i)
	  {
	    for(size_t j = 0; j < 3; ++j)
	      std::cout << a2ij[i][j] << " ";
	    std::cout << std::endl;
	  }

    // traspose the a2ij
	double a2ji[3][3] = {
	  {0.0, 0.0, 0.0},
	  {0.0, 0.0, 0.0},
	  {0.0, 0.0, 0.0}
	};

       for(size_t i = 0; i < 3; ++i)
	 {
	   for(size_t j = 0; j < 3; ++j)
	     {
	       a2ji[j][i] =a2ij[i][j]; // matrix transpose 
	     }
	 }

       std::cout << " a2ji matrix now is " << std::endl;
        for(size_t i = 0; i < 3; ++i)
	  {
	    for(size_t j = 0; j < 3; ++j)
	      std::cout << a2ji[i][j] << " ";
	    std::cout << std::endl;
	  }

    // symmetrize the element array
	double A2ij[3][3] = {
	  {0.0, 0.0, 0.0},
	  {0.0, 0.0, 0.0},
	  {0.0, 0.0, 0.0}
	};

        for(size_t i = 0; i < 3; ++i)
	  {
	    for(size_t j = 0; j < 3; ++j)
	      {
	        A2ij[i][j] = (1.0/2.0)*(a2ij[i][j] + a2ji[i][j]);
		std::cout << A2ij[i][j] << " and " 
		   << (1/2)*(a2ij[i][j]+a2ji[i][j]) << std::endl;
	      } 
	  }
	  {
	    for(size_t j = 0; j < 3; ++j)
	      std::cout << A2ij[i][j] << " ";
	    std::cout << std::endl;
	  }
        	
    // add the element reslults onto the A-Matrix array
	for(size_t row = 0; row < 3; ++row)
	  {
       std::cout << " row is " << row << std::endl;
	    
	   AMatrixArrayPtr[nodeIndex[row]][nodeIndex[0]] 
	      = A2ij[row][0] 
	        + AMatrixArrayPtr[nodeIndex[row]][nodeIndex[0]];
	   AMatrixArrayPtr[nodeIndex[row]][nodeIndex[1]] 
	      = A2ij[row][1] 
	        + AMatrixArrayPtr[nodeIndex[row]][nodeIndex[1]];
	   AMatrixArrayPtr[nodeIndex[row]][nodeIndex[2]] 
	      = A2ij[row][2]
	       + AMatrixArrayPtr[nodeIndex[row]][nodeIndex[2]];
	    
	  }
      ++idx_element;
    }
}

void  CalculateAMatrix::showObjectLocalAMatrix()
{
  std::cout << " pointer to pointer of Object-local 
    A-Matrix array: " << AMatrixArrayPtr << std::endl; 
  for (size_t idx_row = 0; idx_row < lengthRowNode; ++idx_row)
	  {
	    for (size_t idx_col = 0; 
	        idx_col < lengthRowNode; ++idx_col)
	      std::cout << 
	       this->AMatrixArrayPtr[idx_row][idx_col] << " ";
	    std::cout << std::endl;
	  }  
}
// Here, you can define the return 
// function as "const double **", but
// you can difine it as double ** const,
// i.e., you can return const pointer,
// but you can not require pointer to point const object
// double ** const Mat::returnPointerOfMatrix() const
// {
//  return this->arr;
// }

double ** const CalculateAMatrix::returnAMatrixPtr() const
{
  return this->AMatrixArrayPtr;
}

};
#endif
\end{lstlisting}

\section{Garlkin strategy with finite element partition}
\label{GarlkinStretargy}
In order to solve the eign-equation Eq.~(\ref{NMREigenEquationDimensonless}) by suing the expanded coefficients of $\theta$ in Sobolev space. We must expand $\delta{S_{+}}$ in Sobolev space as well. Similar with what we have talked in Sec.~\ref{RitzStrategyAndFiniteElementsPartition}, this operation coverts the partial differential Equation into a discrete form. Moreover, this discrete form of Eq.~(\ref{NMREigenEquationDimensonless}) is a high-dimensional linear equation systems, which looks like a matrix representation of quantum observable. The lowest eigen-value and eigen-mode of this linear system are what we want. This process is called Garlkin strategy \cite{CiarletBook1978}. With the FE partition in Sec.~\ref{RitzStrategyAndFiniteElementsPartition}, Eq.~(\ref{NMREigenEquationDimensonless}) can be rewritten as 
\begin{align}
\lambda \sum_{c}\sum_{\alpha} & b_{\alpha \beta}^{c} \xi_{\alpha} = \notag \\
& \sum_{c}\sum_{\alpha} \{ [6{\gamma}_{2}^{2}+(\gamma_{1}^{2}+1)] a_{\alpha \beta}^{c,y}  \\
& + [3\gamma_{1}^{2} + (2\gamma_{2}^{2} + 1)] a_{\alpha \beta}^{c,x} 
+ 2i a_{\alpha \beta}^{c,3} + 2i(1 + \gamma_{1}^{2}) a_{\alpha \beta}^{c,4} 
+ a_{\alpha \beta}^{c,5}\} \xi_{\alpha}, \notag
\end{align}
where cell matrices $b_{\alpha \beta}^{c}$, $a_{\alpha \beta}^{c,y}$, $a_{\alpha \beta}^{c,x}$, $a_{\alpha \beta}^{c,3(4,5)}$ correspond to the cell integrals of terms on the right hand side of Eq.~(\ref{NMREigenEquationDimensonless}). After reorganizing all cell matrices and cell expansion coefficients $\xi_{\alpha}$, we get the linear system
\begin{equation}
\lambda B_{ij} \xi_{j} = A_{ij} \xi_{j},
\end{equation}
where $A_{ij}$ and $B_{ij}$ are high dimensional spares matrices, which can be calculated by the same C++ library in Sec.~\ref{BFGSAlgorithmAndImplements} and they form general eigen-value system. This kinds linear system can be directly solved by Matlab build-in function.

%
%
%
%
%
%
%
%
%


\chapter{Simplifications of Eign-Equations} 

\label{AppendixC} 


\section{The derivation of the first order dynamic equation}
\label{For1stOrderEqations}
Using Eq.~(\ref{GradientEnergy}), Eq.~(\ref{PdBOrderParameter}), Eq.~(\ref{MagneticEnergy}) and Eq.~(\ref{EnergyDensityOfSOC}), All terms of energy densities in hydrodynamic free energy $F_{hydrodynamics}$ are given as
\begin{align}
f_{grad} = & \frac{1}{2}\{K_{1}{\Delta_{P}^{2}}{\partial_{i}{\hat{d}}_{\alpha}}{\partial_{i}{\hat{d}}_{\alpha}} + K_{1}{\Delta_{\bot1}^{2}}{\partial_{i}{\hat{e}}^{1}_{\alpha}}{\partial_{i}{\hat{e}}^{1}_{\alpha}} + K_{1}{\Delta_{\bot2}^{2}}{\partial_{i}{\hat{e}}^{2}_{\alpha}}{\partial_{i}{\hat{e}}^{2}_{\alpha}}   \\ 
& + K_{2}{\Delta_{P}^{2}}{\partial_{j}{\hat{d}}_{\alpha}}{\hat{z}_{j}}{\partial_{i}{\hat{d}}_{\alpha}}{\hat{z}_{i}} + K_{2}{\Delta_{\bot1}^{2}}{\partial_{j}{\hat{e}^{1}}_{\alpha}}{\hat{x}_{j}}{\partial_{i}{\hat{e}^{1}}_{\alpha}}{\hat{x}_{i}} +  K_{2}{\Delta_{\bot2}^{2}}{\partial_{j}{\hat{e}^{2}}_{\alpha}}{\hat{y}_{j}}{\partial_{i}{\hat{e}^{2}}_{\alpha}}{\hat{y}_{i}} \notag \\
& + K_{3}{\Delta_{P}^{2}}{\partial_{i}{\hat{d}}_{\alpha}}{\hat{z}_{i}}{\partial_{j}{\hat{d}}_{\alpha}}{\hat{z}_{j}} + K_{3}{\Delta_{\bot1}^{2}}{\partial_{i}{\hat{e}^{1}}_{\alpha}}{\hat{x}_{i}}{\partial_{j}{\hat{e}^{1}}_{\alpha}}{\hat{y}_{j}} +  K_{3}{\Delta_{\bot2}^{2}}{\partial_{i}{\hat{e}^{2}}_{\alpha}}{\hat{y}_{i}}{\partial_{j}{\hat{e}^{2}}_{\alpha}}{\hat{y}_{j}} \notag \\
& + K_{2}[{\partial_{j}}{\hat{d}_{\alpha}}{\hat{z}_{i}}{\partial_{j}}{\hat{e}^{1}_{\alpha}}{\hat{x}_{j} + {\partial_{j}}{\hat{e}^{1}_{\alpha}}{\hat{x}_{i}}{\partial_{i}}{\hat{d}_{\alpha}}{\hat{z}_{j}}}]{\Delta_{P}}{\Delta_{\bot1}} + K_{2}[{\partial_{j}}{\hat{d}_{\alpha}}{\hat{z}_{i}}{\partial_{i}}{\hat{e}^{2}_{\alpha}}{\hat{y}_{j} + {\partial_{j}}{\hat{e}^{2}_{\alpha}}{\hat{y}_{i}}{\partial_{i}}{\hat{d}_{\alpha}}{\hat{z}_{j}}}]{\Delta_{P}}{\Delta_{\bot2}}  \notag \\
& + K_{2}[{\partial_{j}}{\hat{e}^{1}_{\alpha}}{\hat{x}_{i}}{\partial_{i}}{\hat{e}^{2}_{\alpha}}{\hat{y}_{j} + {\partial_{j}}{\hat{e}^{2}_{\alpha}}{\hat{y}_{i}}{\partial_{i}}{\hat{e}^{1}_{\alpha}}{\hat{x}_{j}}}]{\Delta_{\bot1}}{\Delta_{\bot2}} + K_{3}[{\partial_{i}}{\hat{d}_{\alpha}}{\hat{z}_{i}}{\partial_{j}}{\hat{e}^{1}_{\alpha}}{\hat{x}_{j} + {\partial_{i}}{\hat{e}^{2}_{\alpha}}{\hat{x}_{i}}{\partial_{j}}{\hat{d}_{\alpha}}{\hat{z}_{j}}}]{\Delta_{P}}{\Delta_{\bot1}} \notag \\
& + K_{3}[{\partial_{i}}{\hat{d}_{\alpha}}{\hat{z}_{i}}{\partial_{j}}{\hat{e}^{2}_{\alpha}}{\hat{y}_{j} + {\partial_{i}}{\hat{e}^{2}_{\alpha}}{\hat{y}_{i}}{\partial_{j}}{\hat{d}_{\alpha}}{\hat{z}_{j}}}]{\Delta_{P}}{\Delta_{\bot2}} \notag + K_{3}[{\partial_{i}}{\hat{e}^{1}_{\alpha}}{\hat{x}_{i}}{\partial_{j}}{\hat{e}^{2}_{\alpha}}{\hat{y}_{j} + {\partial_{i}}{\hat{e}^{2}_{\alpha}}{\hat{y}_{i}}{\partial_{j}}{\hat{e}^{1}_{\alpha}}{\hat{x}_{j}}}]{\Delta_{\bot1}}{\Delta_{\bot2}}\}, \notag
\label{DensityOfGrad}
\end{align}
\begin{align}
f_{soc} = & \frac{3g_{D}}{5} \{{\Delta_{P}^{2}}({\hat{d}_{i}{\hat{z}_{i}}})^{2} + {\Delta_{P}}{\Delta_{\bot1}}({\hat{d}_{i}{\hat{z}_{i}}})({\hat{e}^{1}_{j}{\hat{x}_{j}}}) + {\Delta_{P}}{\Delta_{\bot2}}({\hat{d}_{i}{\hat{z}_{i}}})({\hat{e}^{2}_{j}{\hat{y}_{j}}}) + {\Delta_{P}}{\Delta_{\bot1}}({\hat{e}^{1}_{i}{\hat{x}_{i}}})({\hat{d}_{j}{\hat{z}_{j}}}) \notag \\
& + {\Delta_{\bot1}^{2}}({\hat{e}^{1}_{i}{\hat{x}_{i}}})^{2} +  {\Delta_{\bot1}}{\Delta_{\bot2}}({\hat{e}^{1}_{i}{\hat{x}_{i}}})({\hat{e}_{j}{\hat{y}_{j}}}) +  {\Delta_{\bot2}}{\Delta_{\bot1}}({\hat{e}^{2}_{i}{\hat{y}_{i}}})({\hat{d}_{j}{\hat{z}_{j}}}) + {\Delta_{\bot2}}{\Delta_{\bot1}}({\hat{e}^{2}_{i}{\hat{y}_{i}}})({\hat{e}^{1}_{j}{\hat{x}_{j}}}) \notag \\ 
& + {\Delta_{\bot2}^{2}}({\hat{e}^{2}_{i}{\hat{y}_{i}}})^{2} + {\Delta_{P}^{2}}({\hat{d}_{i}{\hat{z}_{i}}})^{2} + {\Delta_{P}}{\Delta_{\bot1}}({\hat{d}_{i}{\hat{x}_{i}}})({\hat{e}^{1}_{j}{\hat{z}_{j}}}) \notag \\
& + {\Delta_{P}}{\Delta_{\bot2}}({\hat{d}_{i}{\hat{z}_{j}}})({\hat{e}^{2}_{j}{\hat{y}_{i}}}) + {\Delta_{P}}{\Delta_{\bot1}}({\hat{e}^{1}_{i}{\hat{x}_{j}}})({\hat{d}_{j}{\hat{z}_{i}}}) + {\Delta_{\bot1}^{2}}({\hat{e}^{1}_{i}{\hat{x}_{i}}})^{2}  \\ 
& +  {\Delta_{\bot1}}{\Delta_{\bot2}}({\hat{e}^{1}_{i}{\hat{x}_{j}}})({\hat{e}^{2}_{j}{\hat{y}_{j}}}) +  {\Delta_{\bot2}}{\Delta_{P}}({\hat{e}^{2}_{i}{\hat{y}_{j}}})({\hat{d}_{j}{\hat{z}_{i}}}) + {\Delta_{\bot1}}{\Delta_{\bot2}}({\hat{e}^{2}_{i}{\hat{y}_{j}}})({\hat{e}^{1}_{j}{\hat{x}_{i}}}) + {\Delta_{\bot2}^{2}}({\hat{e}^{2}_{j}{\hat{y}_{j}}})^{2} \notag \\
& - \frac{2}{3}(\Delta^{2}_{P} + \Delta^{2}_{\bot1} + \Delta^{2}_{\bot2})\}, \notag
\label{DensityOfSOC}
\end{align}
\begin{equation}
f_{H} = -{\gamma}H_{\beta}S_{\beta} + \frac{\gamma^{2}}{2}{\chi_{\bot}^{-1}}[({\hat{d}}_{\beta}S_{\beta})^{2}\delta + S_{\beta}S_{\beta}].
\label{DensityOfMagnatic}
\end{equation}
Then we have all of functional derivatives
\begin{equation}
\frac{\delta{F_{hydrodynamics}}}{\delta{S_{\beta}}}(\mathbf{r}') = \frac{\partial{f_{H}}}{\partial{S_{\beta}}}(\mathbf{r}') = -{\gamma}H_{\beta} + \gamma^{2}{\chi^{-1}_{\bot}}[(S_{\gamma}\hat{d}_{\gamma}){\hat{d}_{\beta}}{\delta} + S_{\beta}],
\label{FunctionalDerivaticveOfS}
\end{equation}
\begin{align}
\frac{\delta{F_{hydrodynamics}}}{\delta{{\hat{d}}_{\beta}}}(\mathbf{r}') = & \frac{\partial{f_{H}}}{\partial{{\hat{d}}_{\beta}}}(\mathbf{r}') + \frac{\partial{f_{soc}}}{\partial{{\hat{d}}_{\beta}}}(\mathbf{r}') - {\partial_{i}}\frac{f_{grad}}{{\partial}{\partial_{i}}{\hat{d}_{\beta}}}(\mathbf{r}') \notag \\
= & {\gamma^{2}}{\chi^{-1}_{\bot}}{\delta}S_{\gamma}{\hat{d}_{\gamma}}S_{\beta} + \frac{3g_{D}}{5} ( 2{\Delta_{P}^{2}}{\hat{d}_{\gamma}}{\hat{z}_{\gamma}}{\hat{z}_{\beta}} + {\Delta_{P}}{\Delta_{\bot1}}{\hat{z}_{\beta}}{\hat{e}^{1}_{i}}{\hat{x}_{i}} + {\Delta_{P}}{\Delta_{\bot2}}{\hat{z}_{\beta}}{\hat{e}^{2}_{i}}{\hat{y}_{i}} \notag \\
& + {\Delta_{P}}{\Delta_{\bot1}}{\hat{z}_{\beta}}{\hat{e}^{1}_{i}}{\hat{x}_{i}} + {\Delta_{P}}{\Delta_{\bot2}}{\hat{z}_{\beta}}{\hat{e}^{2}_{i}}{\hat{y}_{i}} + 2{\Delta_{P}^{2}}{\hat{d}_{\gamma}}{\hat{z}_{\gamma}}{\hat{z}_{\beta}} + {\Delta_{P}}{\Delta_{\bot1}}{\hat{z}_{i}}{\hat{e}^{1}_{i}}{\hat{x}_{\beta}} \notag \\
& + {\Delta_{P}}{\Delta_{\bot2}}{\hat{z}_{i}}{\hat{e}^{2}_{i}}{\hat{y}_{\beta}} + {\Delta_{P}}{\Delta_{\bot1}}{\hat{x}_{\beta}}{\hat{e}^{1}_{i}}{\hat{z}_{i}} + {\Delta_{P}}{\Delta_{\bot2}}{\hat{z}_{i}}{\hat{e}^{2}_{i}}{\hat{y}_{\beta}})  \notag \\
& - \frac{1}{2} ( K_{1}{\Delta^{2}_{P}}2{\partial_{i}}{\partial_{i}}{\hat{d}_{\beta}} + K_{2}{\Delta^{2}_{P}}2{\partial_{i}}{\partial_{j}}{\hat{d}_{\beta}}{\hat{z}_{i}}{\hat{z}_{j}} \label{FunctionalDerivaticveOfd} \\
& + 2K_{2}{\Delta_{P}}{\Delta_{\bot1}}{\partial_{i}}{\partial_{j}}{\hat{e}^{1}_{\beta}}{\hat{z}_{j}}{\hat{x}_{i}} + 2K_{2}{\Delta_{P}}{\Delta_{\bot2}}{\partial_{i}}{\partial_{j}}{\hat{e}^{2}_{\beta}}{\hat{z}_{j}}{\hat{y}_{i}} \notag \\
& + 2K_{3}{\Delta^{2}_{P}}{\partial_{i}}{\partial_{j}}{\hat{d}_{\beta}}{\hat{z}_{j}}{\hat{z}_{i}} + 2K_{3}{\Delta_{P}}{\Delta_{\bot1}}{\hat{z}_{i}}{\partial_{i}}{\partial_{j}}{\hat{e}_{\beta}}{\hat{x}_{j}}  + 2K_{3}{\Delta_{P}}{\Delta_{\bot1}}{\partial_{i}}{\partial_{j}}{\hat{e}^{2}_{\beta}}{\hat{y}_{j}}{\hat{z}_{i}}), \notag
\end{align}
\begin{align}
\frac{\delta{F_{hydrodynamics}}}{\delta{{\hat{e}^{1}}_{\beta}}}(\mathbf{r}') = & \frac{\partial{f_{soc}}}{\partial{{\hat{e}^{1}}_{\beta}}}(\mathbf{r}') - {\partial_{i}}\frac{f_{grad}}{{\partial}{\partial_{i}}{\hat{e}^{1}_{\beta}}}(\mathbf{r}') \notag \\
= & \frac{3g_{D}}{5}({\Delta_{P}}{\Delta_{\bot1}}{\hat{d}_{\gamma}}{\hat{z}_{\gamma}}{\hat{x}_{\beta}} + {\Delta_{P}}{\Delta_{\bot1}}{\hat{d}_{\gamma}}{\hat{z}_{\gamma}}{\hat{x}_{\beta}} + 2{\Delta^{2}_{\bot1}}{\hat{e}^{1}_{\gamma}}{\hat{x}_{\gamma}}{\hat{x}_{\beta}} \notag \\
& + {\Delta_{\bot1}}{\Delta_{\bot2}}{\hat{e}^{2}_{\gamma}}{\hat{y}_{\gamma}}{\hat{x}_{\beta}} + {\Delta_{P}}{\Delta_{\bot1}}{\hat{d}_{\gamma}}{\hat{x}_{\gamma}}{\hat{z}_{\beta}} + {\Delta_{P}}{\Delta_{\bot1}}{\hat{d}_{\gamma}}{\hat{x}_{\gamma}}{\hat{z}_{\beta}} \label{FunctionalDerivaticveOfe1} \\
& + {\Delta_{\bot1}^{2}}2{\hat{e}^{1}_{\gamma}}{\hat{x}_{\gamma}}{\hat{x}_{\beta}} + {\Delta_{\bot1}}{\Delta_{\bot2}}{\hat{e}^{2}_{\gamma}}{\hat{x}_{\gamma}}{\hat{y}_{\beta}} \notag + {\Delta_{\bot2}}{\Delta_{\bot1}}{\hat{e}^{2}_{\gamma}}{\hat{y}_{\gamma}}{\hat{x}_{\beta}} + {\Delta_{\bot2}}{\Delta_{\bot1}}{\hat{e}^{2}_{\gamma}}{\hat{x}_{\gamma}}{\hat{y}_{\beta}}) \notag  \\
& - \frac{1}{2}(2K_{1}{\Delta^{2}_{\bot1}}{\partial_{i}}{\partial_{i}}{\hat{e}^{1}_{\beta}}{\hat{x}_{j}}{\hat{x}_{j}} + 2K_{2}{\Delta^{2}_{\bot1}}{\partial_{i}}{\partial_{j}}{\hat{e}^{1}_{\beta}}{\hat{x}_{j}}{\hat{x}_{i}} + 2K_{2}{\Delta_{P}}{\Delta_{\bot1}}{\partial_{i}}{\partial_{j}}{\hat{d}_{\beta}}{\hat{x}_{j}}{\hat{z}_{i}} \notag  \\
& + 2K_{2}{\Delta_{\bot1}}{\Delta_{\bot2}}{\partial_{i}}{\partial_{j}}{\hat{e}^{2}_{\beta}}{\hat{x}_{j}}{\hat{y}_{i}} \notag  \\ 
& + 2K_{3}{\Delta_{\bot1}^{2}}{\partial_{i}}{\partial_{j}}{\hat{e}^{1}_{\beta}}{\hat{x}_{j}}{\hat{x}_{i}} + 2K_{3}{\Delta_{P}}{\Delta_{\bot1}}{\partial_{i}}{\partial_{j}}{\hat{d}_{\beta}}{\hat{z}_{j}}{\hat{x}_{i}} + 2K_{3}{\Delta_{\bot1}}{\Delta_{\bot2}}{\partial_{i}}{\partial_{j}}{\hat{e}^{2}_{\beta}}{\hat{y}_{j}}{\hat{x}_{i}}), \notag
\end{align}
\begin{align}
\frac{\delta{F_{hydrodynamics}}}{\delta{{\hat{e}^{2}}_{\beta}}}(\mathbf{r}') = & \frac{\partial{f_{soc}}}{\partial{{\hat{e}^{2}}_{\beta}}}(\mathbf{r}') - {\partial_{i}}\frac{f_{grad}}{{\partial}{\partial_{i}}{\hat{e}^{2}_{\beta}}}(\mathbf{r}') \notag \\
= & \frac{3g_{D}}{5} ({\Delta_{P}}{\Delta_{\bot2}}{\hat{d}_{\gamma}}{\hat{z}_{\gamma}}{\hat{y}_{\beta}} + {\Delta_{P}}{\Delta_{\bot2}}{\hat{e}^{1}_{\gamma}}{\hat{x}_{\gamma}}{\hat{y}_{\gamma}} + {\Delta_{\bot2}}{\Delta_{P}}{\hat{y}_{\beta}}{\hat{d}_{\gamma}}{\hat{z}_{\gamma}} \notag \\
& + {\Delta_{\bot2}}{\Delta_{\bot1}}{\hat{e}^{1}_{\gamma}}{\hat{x}_{\gamma}}{\hat{y}_{\beta}} + 2{\Delta_{\bot2}^{2}}{\hat{e}^{2}_{\gamma}}{\hat{y}_{\gamma}}{\hat{y}_{\beta}} + {\Delta_{P}}{\Delta_{\bot2}}{\hat{d}_{\gamma}}{\hat{y}_{\gamma}}{\hat{z}_{\beta}} \notag \\
& + {\Delta_{\bot1}}{\Delta_{\bot2}}{\hat{e}^{1}_{\gamma}}{\hat{y}_{\gamma}}{\hat{x}_{\beta}} + {\Delta_{\bot2}}{\Delta_{P}}{\hat{d}_{\gamma}}{\hat{y}_{\gamma}}{\hat{z}_{\beta}} + {\Delta_{\bot1}}{\Delta_{\bot2}}{\hat{e}^{1}_{\gamma}}{\hat{y}_{\gamma}}{\hat{x}_{\beta}} + {\Delta^{2}_{\bot2}}{2\hat{e}^{2}_{\gamma}}{\hat{y}_{\gamma}}{\hat{y}_{\beta}})  \notag \\
& - \frac{1}{2}(2K_{1}{\Delta^{2}_{\bot2}}{\partial_{i}}{\partial_{i}}{\hat{e}^{2}_{\beta}}{\hat{y}_{j}}{\hat{y}_{j}} + 2K_{2}{\Delta^{2}_{\bot2}}{\partial_{i}}{\partial_{j}}{\hat{e}^{2}_{\beta}}{\hat{y}_{j}}{\hat{y}_{i}} + 2K_{2}{\Delta_{P}}{\Delta_{\bot2}}{\partial_{i}}{\partial_{j}}{\hat{d}_{\beta}}{\hat{y}_{j}}{\hat{z}_{i}} \notag \\
& + 2K_{2}{\Delta_{\bot1}}{\Delta_{\bot2}}{\partial_{i}}{\partial_{j}}{\hat{e}^{1}_{\beta}}{\hat{y}_{j}}{\hat{x}_{i}}  \notag \\
& + 2K_{3}{\Delta_{\bot2}^{2}}{\partial_{i}}{\partial_{j}}{\hat{e}^{2}_{\beta}}{\hat{y}_{j}}{\hat{y}_{i}} + 2K_{3}{\Delta_{P}}{\Delta_{\bot1}}{\partial_{i}}{\partial_{j}}{\hat{d}_{\beta}}{\hat{z}_{j}}{\hat{y}_{i}} \label{FunctionalDerivaticveOfe2} \\
& + 2K_{3}{\Delta_{\bot1}}{\Delta_{\bot2}}{\partial_{i}}{\partial_{j}}{\hat{e}^{1}_{\beta}}{\hat{x}_{j}}{\hat{y}_{i}}). \notag
\end{align}
Plugging Eq.~(\ref{FunctionalDerivaticveOfS}), Eq.~(\ref{FunctionalDerivaticveOfd}), Eq.~(\ref{FunctionalDerivaticveOfe1}) and Eq.~(\ref{FunctionalDerivaticveOfe2}) into Eq.~(\ref{LiouvilleEquations1}) and Eq.~(\ref{LiouvilleEquations2}), we get Eq.~ (\ref{1stOderEquationsA}) and Eq.~(\ref{1stOderEquationsB}).

\section{The derivation of the second order dynamic response equation}
\label{For2stOrderEqations}
Firstly we take time-derivative to Eq.~(\ref{1stOderEquationsA}) and get
\begin{align}
{\gamma}{\epsilon_{{\alpha}{\beta}{\gamma}}} & ({H^{(0)}_{\beta}}\frac{\partial}{{\partial}t}{\delta{S_{\gamma}}} + \frac{\partial}{\partial{t}}{{\delta}H_{\beta}}{{S^{(0)}_{\gamma}}}) = \notag \\ 
& {\epsilon_{{\alpha}{\beta}{\gamma}}}[-\frac{6g_{D}}{5}{Q^{bd}_{{\beta}j}} ({V^{d(0)}_{j}}\frac{\partial}{\partial{t}}{\delta{V^{b}_{\gamma}}} + \frac{\partial}{\partial{t}}{\delta}{V^{d}_{j}}{V^{b(0)}_{r}}) \notag \\ 
& + {K^{ba}_{ij}} ({\partial_{i}}{\partial_{j}}{V^{b(0)}_{\beta}}{\frac{\partial}{\partial{t}}}{\delta}{V^{a}_{\gamma}} + {\partial_{i}}{\partial_{j}}{\frac{\partial}{\partial{t}}}{\delta{V_{\beta}^{b}}}{V^{a(0)}_{\gamma}})],  \label{TimeDerivativeOf1stOrderEqOfS}
\end{align} 
where
\begin{align}
\frac{\partial}{\partial{t}}{\delta}{V_{\alpha}^{a}} = & \{{H_{\beta}^{(0)}}{V_{\gamma}^{a(0)}}{\gamma} + {\gamma}{H_{\beta}^{(0)}}{\delta}{V_{\gamma}^{a}} + {\gamma}{\delta}{H_{\beta}}{V_{\gamma}^{a(0)}} \notag \\
& - {\gamma^{2}}{\chi^{-1}_{\bot}}{\delta}({S_{\eta}^{(0)}}{V_{\eta}^{3(0)}}) ({V_{\beta}^{3(0)}}{\delta}{V_{\gamma}^{a}} + {V_{\gamma}^{3(0)}}{\delta}{V_{\beta}^{3}}) \notag \\ 
& + {\gamma^{2}}{\chi_{\bot}^{-1}}(S_{\beta}^{(0)}{V_{\gamma}^{a(0)}} + S_{\beta}^{(0)}{\delta}{V_{\gamma}^{a}} + {\delta}{S_{\beta}}{V_{\gamma}^{a(0)}})\}{\epsilon_{{\alpha}{\beta}{\gamma}}}.
\label{DerivativeOfV}
\end{align}
Taking into account the relations:
\begin{equation}
{\gamma}{S_{\beta}^{(0)}}={H_{\alpha}^{(0)}}{\chi_{{\alpha}{\beta}}},\,\,{V^{3(0)}_{\eta}}{S^{(0)}_{\eta}} = {\hat{\mathbf{d}}^{(0)}}{\cdot}{\mathbf{S}^{(0)}} = 0,
\end{equation}
where magnetic susceptibility $\chi_{{\alpha}{\beta}} = {\chi_{\parallel}}{\delta_{\alpha\beta}} - (\chi_{\bot}-\chi_{\bot}){\hat{d}_{\alpha}^{(0)}}{\hat{d}_{\beta}^{(0)}}$,
Eq.~(\ref{DerivativeOfV}) is simplified to
\begin{equation}
\frac{\partial}{\partial{t}}{\delta}{V_{\alpha}^{a}} = {\epsilon_{{\alpha}{\beta}{\gamma}}}{V_{\gamma}^{a(0)}}({\gamma}{\delta}{H_{\beta}^{a(0)}} - \gamma^{2}{\chi^{-1}_{\bot}}{\delta}{S_{\beta}})
\label{DerivativeOfVSimplifed}
\end{equation}
Taking Eq.~(\ref{DerivativeOfVSimplifed}) back into Eq.~(\ref{TimeDerivativeOf1stOrderEqOfS}), we get
\begin{align}
 \frac{\partial^{2}} {\partial{t^{2}}} {\delta}{S_{\alpha}} 
= {\gamma}{\epsilon_{\alpha\beta\gamma}}(H^{(0)}_{\beta} \frac{\partial}{\partial{t}}{\delta}{S_{\gamma}} + \frac{\partial}{\partial{t}}{\delta}{H_{\beta}}S_{\gamma}^{(0)}) + {\Xi_{{\alpha}{\lambda}}}{\delta}{S_{\lambda}} + {C_{{\alpha}{\eta}}}{\delta}{H_{{\eta}}}. \label{BeforeFourier}
\end{align}
The last step is taking time Fourier transformation for dynamic variables $\delta{S_{\alpha}}$ and $\delta{H_{\beta}}$ as well their derivatives
\begin{align}
{\delta}H_{\beta}(\omega) = & \frac{1}{\sqrt{2\pi}}\int dt {\delta}H_{\beta} e^{-i{\omega}t}, \,\,
{\delta}S_{\alpha}(\omega) = \frac{1}{\sqrt{2\pi}}\int dt {\delta}S_{\alpha} e^{-i{\omega}t}, \,\, \notag \\
\mathfrak{F}(\frac{\partial}{\partial{t}} & {\delta{H_{\beta}}}) = i\omega {\delta}H_{\beta}(\omega), \,\, 
\mathfrak{F}(\frac{\partial}{\partial{t}} {\delta{S_{\alpha}}}) = i\omega {\delta}S_{\alpha}(\omega), \,\,
\mathfrak{F}(\frac{\partial^{2}}{\partial{t^{2}}} {\delta{S_{\alpha}}}) = -\omega^{2} {\delta}S_{\alpha}(\omega)
\end{align}
in Eq.~(\ref{BeforeFourier}), this gives out Eq.~(\ref{2ndOderEquations}).

\section{The derivation of transverse NMR response equation}
\label{TransverseNMRResponseEquation}
In the limit of $|\omega-\omega_{L}| \ll \omega_{L}$ and under parametrization Eq.~(\ref{PARA1}), Eq.~(\ref{2ndOderEquations}) within components form are
\begin{align}
i\omega{\delta}{S_{2}}(\omega) & = - \gamma{S_{3}^{(0)}}{\delta}{H_{1}}(\omega), \notag \\
i\omega{\delta}{S_{1}}(\omega) & = \gamma{H_{2}^{(0)}}{\delta}{S_{3}}(\omega) + \frac{\Xi_{11}}{i\omega}{\delta}{S_{1}}(\omega) 
 + \frac{\Xi_{13}}{i\omega}{\delta}{S_{3}}(\omega) + \frac{C_{31}}{i\omega}{\delta}{H_{1}}(\omega),  \label{ResponseEq1} \\ 
i\omega{\delta}{S_{3}}(\omega) & =  \gamma[S_{2}^{(0)}{\delta}{H_{1}}(\omega)- {H_{2}^{(0)}}{\delta}{S_{1}}(\omega)]
 + \frac{\Xi_{31}}{i\omega}{\delta}{S_{1}}(\omega) + \frac{\Xi_{33}}{i\omega}{\delta}{S_{3}}(\omega) + \frac{C_{31}}{i\omega}{\delta}{H_{1}}(\omega). \notag
\end{align} 
We expand $\omega$ around $\omega_{L}$ as $\omega = \omega_{L} + \epsilon + O^{2}(\epsilon)$, then we get
\begin{equation}
{\delta}S_{1}(\omega) = \frac{{\delta}S_{3}(\omega)}{i(1+\epsilon)}, \,\, {\delta}S_{3}(\omega) = \frac{1}{i(1+\epsilon)}(\frac{\chi_{\bot}}{\gamma}{\delta}{H_{1}}(\omega) - {\delta}S_{1}(\omega)),
\label{RelationBetweenS1S3}
\end{equation}
where $\epsilon = (\omega - \omega_{L})$. By multiplying $i\omega$ and utilizing Eq.~(\ref{RelationBetweenS1S3}), Eq.~ (\ref{ResponseEq1}) can be reorganized as 
\begin{align}
(\omega^{2} - \omega^{2}_{L})(\delta{S_{1}}(\omega) + i \delta{S_{3}}(\omega)) = & i[- {\Xi_{31}}(\frac{{\delta}{S_{1}}}{1+2\epsilon} + i \frac{{\delta}{S_{3}}}{1+2\epsilon}) + \Xi_{13}(\frac{\delta{S_{1}}}{1+2\epsilon} + i{\delta}S_{3})] \notag \\
& + \frac{\Xi_{11}}{1+\epsilon}(\delta{S_{1}} + i {\delta}S_{3}) + \frac{\Xi_{33}}{1+\epsilon}(\delta{S_{1}} + i {\delta}S_{3}) \\
& - (C_{11} +i C_{31})\delta{H_{1}} - \frac{\chi_{\bot}}{\gamma}(\frac{\Xi_{33}}{1+\epsilon} + i\frac{\Xi_{13}}{1+2\epsilon} -i\frac{\Xi_{31}}{1+2\epsilon}) \delta{H_{1}}. \notag
\end{align}
In the case of $\epsilon \rightarrow 0$, this gives Eq.~(\ref{TransverseResponseEquation}).

\section{All $\Xi_{{\alpha}{\lambda}}$ terms in Eq.~(\ref{NMREigenEquation})}
\label{SimplifyTheNMREigenEquation}
By utilizing Eq.~(\ref{XiAndC}) and paramentrization Eq.~(\ref{PARA1}), we have
\begin{align}
\Xi_{11} + \Xi_{33} = & [2c_{1}(K_{1}+K_{2}+K_{3})\Delta^{2}_{\bot2} + c_{1}K_{1}(\Delta^{2}_{\bot1} + \Delta^{2}_{P})]{\partial_{y}}{\partial_{y}}{\delta}S_{+} \notag \\
& + [c_{1}(K_{1}+K_{2}+K_{3})\Delta^{2}_{\bot1} + c_{1}K_{1}(2\Delta^{2}_{\bot2} + \Delta^{2}_{P}] {\partial_{x}}{\partial_{x}}{\delta}S_{+}  \\
& + c_{2}(\Delta_{P} + \Delta_{\bot1})[-(\Delta_{\bot1} + \Delta_{P})cos2{\theta} - 5{\Delta_{\bot2}}sin{\theta}) + c_{2}(\Delta^{2}_{P} + \Delta^{2}_{\bot1} + 4\Delta^{2}_{\bot2}), \notag 
\label{Xi11Xi33}
\end{align}
\begin{align}
\Xi_{13} + \Xi_{31} = & -2c_{1}K_{1}{\Delta^{2}_{P}}{\partial_{i}}{\delta}{S_{+}}{\partial_{i}}{\theta} - 2c_{1}{\Delta^{2}_{\bot}}[K_{1}{\partial_{y}}{\delta}{S_{+}}{\partial_{y}}{\theta} + (K_{1}+K_{2}+K_{3}){\partial_{x}}{\delta}S_{+}{\partial_{x}}{\theta}cos2{\theta}] \notag \\
& - c_{1}\{[K_{1}{\Delta^{2}_{P} + (K_{1} + K_{2} + K_{3}){\Delta^{2}_{\bot1}}}]{\partial_{x}}{\partial_{x}}\theta + K_{1}(\Delta^{2}_{P} + \Delta^{2}_{\bot1}){\partial_{y}}{\partial_{y}}\theta\}, 
\end{align}
where 
\begin{equation}
c_{1} =\frac{\gamma^{2}}{\chi_{\bot}},\,\,\,c_{2} = \frac{6g_{D}{\gamma^{2}}}{5\chi_{\bot}}.
\label{c1c2}
\end{equation}
plugging Eq.~(\ref{Xi11Xi33}), Eq.~(\ref{c1c2}) into Eq.~(\ref{NMREigenEquation}) and multiplying $\tilde{\Omega}^{-2}$ on both sides, we get
\begin{align}
\frac{\omega^{2}-\omega^{2}_{L}}{\tilde{\Omega}^{2}}{\delta}S_{+} = & \{ \frac{5}{6g_{D}} [6K_{1}\rho^{2}_{2} + K_{1}(\rho_{1}^{2} +1)] {\partial_{y}}{\partial_{y}} + \frac{5}{6g_{D}} [3K_{1}\rho^{2}_{1} + K_{1}(2\rho_{2}^{2} +1)] {\partial_{x}}{\partial_{x}} \notag \\
& -i\frac{10}{6g_{D}}[(K_{1} + 3{\rho_{1}^{2}}K_{1}cos2{\theta}){\partial_{x}}\theta{\partial_{x}} -K_{1}(1 + \rho_{1}^{2}) {\partial_{y}}\theta{\partial_{y}}]\} {\delta}S_{+} \label{OriginalResponseEq} \\
& -i\frac{5}{6g_{D}}[K_{1}(1+3{\rho_{1}^{2}}){\partial_{x}}{\partial_{x}}\theta + K_{1}(1 + \rho_{1}^{2}){\partial_{y}}{\partial_{y}}\theta]\delta{S_{+}} \notag \\
& + \{(1 + \rho_{1})[-(1 + \rho_{1})cos2\theta - 5 \rho_{2}sin\theta] + (1 + \rho^{2}_{1} + 4\rho^{2}_{2})\}{\delta}{S_{+}}. \notag
\end{align}
To simplify Eq.~(\ref{OriginalResponseEq}), we need the Lagrangian equation of $\theta$
\begin{align}
\frac{{\delta}{F_{London}}(\theta)}{\delta{\theta}} = & c_{1}\{[K_{1}{\Delta^{2}_{P}} + (K_{1}+K_{2}+K_{3}){\Delta^{2}_{\bot1}}]{\partial_{x}}{\partial_{x}}{\theta} + K_{1}({\Delta^{2}_{P}} + {\Delta^{2}_{\bot1}}){\partial_{y}}{\partial_{y}}{\theta}\} \notag \\ & - c_{2}\{{\Delta_{\bot2}}(\Delta_{P} + \Delta_{\bot1})cos{\theta} + ({\Delta_{P} + {\Delta_{\bot1}}})^{2}sin2{\theta}\} = 0. 
\label{LagrangianEqOftheta}
\end{align}
This equation can be simplified to 
\begin{equation}
(1+3\rho_{1}^{2}){\partial_{x}}{\partial_{x}}\theta + (1+\rho_{1}^{2}){\partial_{y}}{\partial_{y}}\theta = \frac{1}{\xi_{D}^{2}}[(1 + \rho_{1})^{2}sin2{\theta} - (1 + \rho_{1})\rho_{2}cos{\theta}].
\end{equation}
Then Eq.~(\ref{OriginalResponseEq}) can be written as 
\begin{align}
\frac{\omega^{2}-\omega^{2}_{L}}{\tilde{\Omega}^{2}}{\delta}S_{+} = & \{ \frac{5}{6g_{D}} [6K_{1}\rho^{2}_{2} + K_{1}(\rho_{1}^{2} +1)] {\partial_{y}}{\partial_{y}} + \frac{5}{6g_{D}} [3K_{1}\rho^{2}_{1} + K_{1}(2\rho_{2}^{2} +1)] {\partial_{x}}{\partial_{x}} \notag \\
& -i\frac{10}{6g_{D}}[(K_{1} + 3{\rho_{1}^{2}}K_{1}cos2{\theta}){\partial_{x}}\theta{\partial_{x}} -K_{1}(1 + \rho_{1}^{2}) {\partial_{y}}\theta{\partial_{y}}]\} {\delta}S_{+} \label{ReducedlResponseEqAppendices} \\
& -i\frac{5K_{1}}{6g_{D}}{\xi^{-2}_{D}}[(1 + \rho_{1})^{2}sin2{\theta} - (1 + \rho_{1})\rho_{2}cos{\theta}]\delta{S_{+}} \notag \\
& + \{(1 + \rho_{1})[-(1 + \rho_{1})cos2\theta - 5 \rho_{2}sin\theta] + (1 + \rho^{2}_{1} + 4\rho^{2}_{2})\}{\delta}{S_{+}}. \notag
\end{align}
This is Eq.~(\ref{NMREigenEquationDimensonless}).

%
%
%
%
%
%
%
%
%


\chapter{Separable spin soliton with $\theta_{KLS} = \pi$ and its NMR } 
\label{AppendixD} 


\begin{figure*}
\centerline{\includegraphics[width=1.0\linewidth]{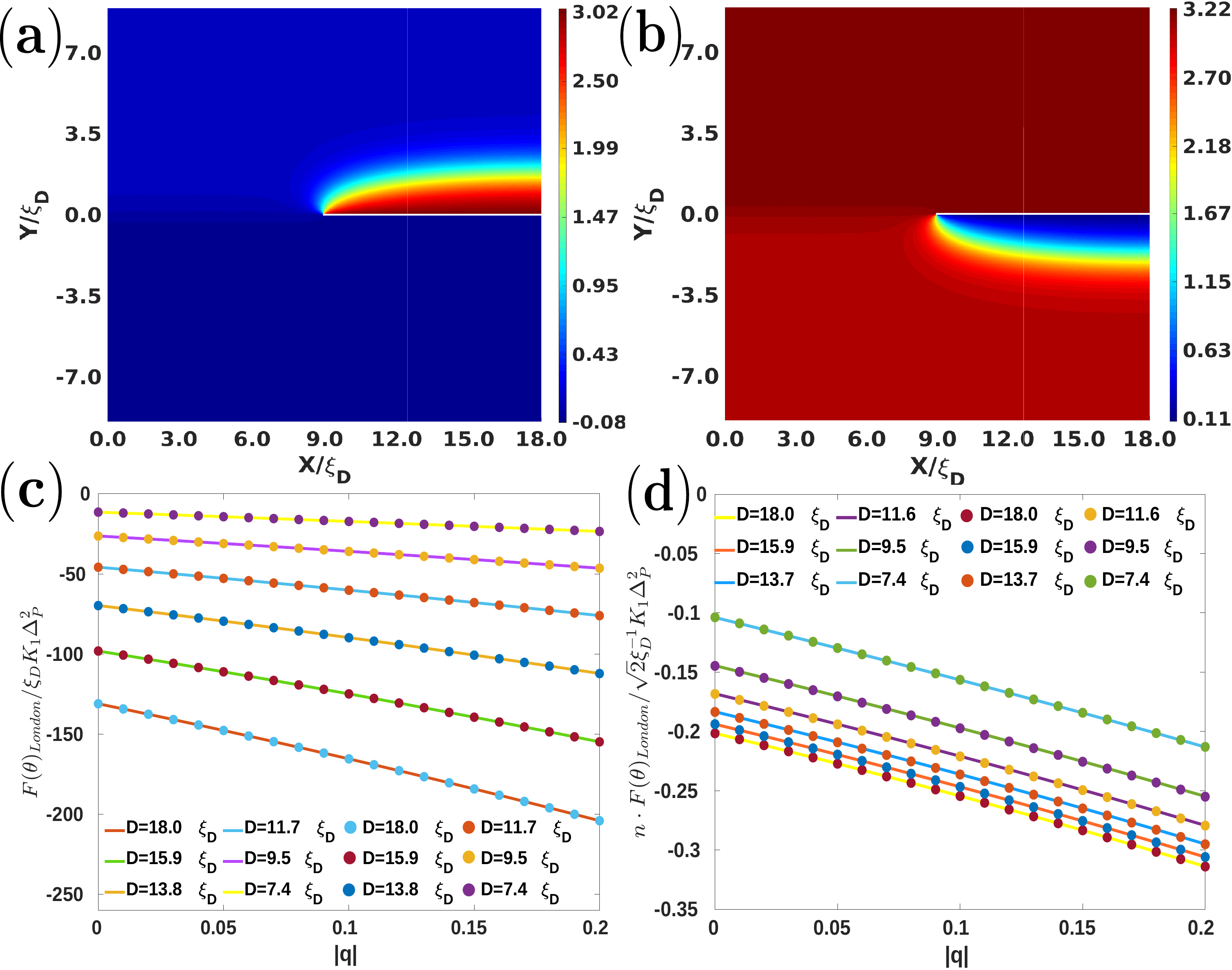}}
\caption{Equilibrium spin textures and equilibrium London limit free energies of one-half unit cell consisting of $1/4+1/4$ separable spin solitons with $\theta_{KLS}=0$ and $\theta_{KLS}=\pi$. The dots represent data of equilibrium configurations with $\theta_{KLS}=0$, while the solid lines represent data of equilibrium configurations with $\theta_{KLS}=\pi$. (a) is the equilibrium spin textures of one-half unit cell with $\theta_{KLS}=0$, $|q|=0.2$ and $D=18\xi_{D}$. (b) is the equilibrium spin textures of one-half unit cell with $\theta_{KLS}=\pi$, $|q|=0.2$ and $D=18\xi_{D}$. (a) and (b) have same $|\Delta{\theta}|=\pi$ and are related by a $\pi$-rotation around $\hat{x}$ axis. (c) depicts the equilibrium London limit free energies  of spin textures with $\theta_{KLS}=0$ and $\theta_{KLS}=\pi$ respectively. (d) depicts the surface densities  of equilibrium London limit free energies of spin textures with $\theta_{KLS}=0$ and $\theta_{KLS}=\pi$ respectively. (c) and (d) demonstrate the pseudo-random lattices consisting of separable spin solitons have same equilibrium London limit free energies for boundary conditions $\theta_{KLS}=0$ and $\theta_{KLS}=\pi$. \label{FiguresOfSeparableSolitonTwoBoundaryConditions}}
\end{figure*}

\section{Pseudo-random lattices with two different domain wall boundary conditions -- $\theta_{KLS}=0$ and $\theta_{KLS} =\pi$}
\label{SeparableSolitionTwoBoundryConditions}
Here, we demonstrate the numeric results the spin textures of one-half unit cell of the lattices consisting of separable spin solitons with topological invariant $1/4+1/4$ under these two boundary conditions. The London limit free energies of one-half unit cell and the surface densities of London limit free energies of the pseudo-random lattices were calculated. In Fig.~\ref{FiguresOfSeparableSolitonTwoBoundaryConditions} (a) and (b), we show the equilibrium spin textures in one-half unit cell. We can see these two textures are related by $\pi$-rotation about $x$-axis. They have same London limit free energies as well as same surface densities of free energy as shown in Fig.~\ref{FiguresOfSeparableSolitonTwoBoundaryConditions} (c) and (d).

\begin{figure*}
\centerline{\includegraphics[width=1.0\linewidth]{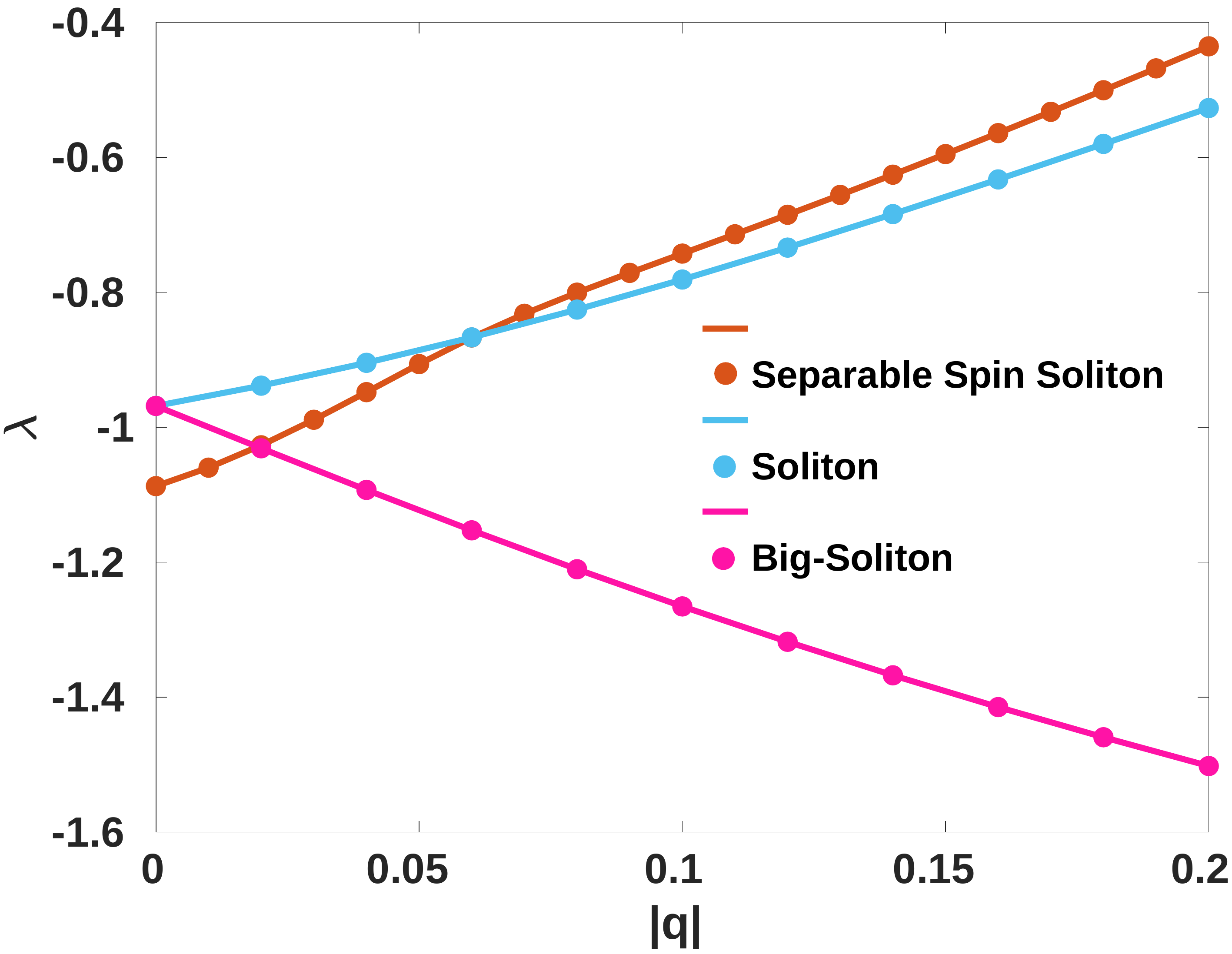}}
\caption{NMR frequency shifts of unit cell of pseudo-random lattices consisting of separable spin soliton, NMR frequency shifts of soliton ($|\Delta{\theta}|=\pi-2\theta_{0}$) and big-soliton ($|\Delta{\theta}|=\pi+2\theta_{0}$). All colored dots are original numeric data, while colored lines are the linear interpolations of numeric data. $D=14.1\xi_{D}$. We found the NMR frequency shifts $\lambda$ of soliton ($|\Delta{\theta}|=\pi-2\theta_{0}$) monotonically  increase when $|q|$ increases. The typical values of $\lambda$ of soliton (blue line) are larger than $-0.7$ when $|q|\geq 0.14$. In contrast, the NMR frequency shifts of big-soliton (pink line) monotonically decrease when $|q|$ increases and the typical values of $\lambda$ in this case are smaller than $-1.35$ when $|q|\geq0.14$. On the other side, the frequency shifts induced by pseudo-random lattices of $2/4$ separable spin solitons (brown lines) monotonically increase as $|q|$ increase. The typical values of $\lambda$ are larger than $-0.65$ when $|q|>0.14$. We find this range of $\lambda$ are very close to those induced by soliton ($|\Delta{\theta}|=\pi-2\theta_{0}$). This is because only the soliton ($|\Delta{\theta}|=\pi-2\theta_{0}$) of $2/4$ separable spin soliton responds to the continuous wave transverse magnetic drive. Moreover, we can see that the frequency shifts of pseudo-random lattices consisting of $2/4$ separable spin solitons is smaller than $-1$ when $|q|=0$. This deviates from the experimental observations of the frequency shifts of spin soliton in polar phase with $|q|=0$ \cite{Autti2016}.  \label{FrequencyShiftSolitonBigSolitonSeparableSoliton}}
\end{figure*}
\section{NMR frequency shifts Of soliton ($|\Delta{\theta}|=\pi-2\theta_{0}$) and big-soliton ($|\Delta{\theta}|=\pi+2\theta_{0}$)}
\label{SpinDynamicsSolionAndBigSoliton}
Here we discuss the transverse NMR frequency shifts of soliton ($|\Delta{\theta}|=\pi-2\theta_{0}$) and big-soliton ($|\Delta{\theta}|=\pi+2\theta_{0}$) in the absence of KLS string wall. The frequency shifts $\lambda$ are numeric results of Eq.~(\ref{NMREigenEquationDimensonless}) with equilibrium spin textures of soliton ($|\Delta{\theta}|=\pi-2\theta_{0}$) and big-soliton ($|\Delta{\theta}|=\pi+2\theta_{0}$) which we got in Sec.~\ref{AbsenceOfHQVs}. 
We depict the results with $|q|$ from $0.0$ to $0.2$ in Fig.~\ref{FrequencyShiftSolitonBigSolitonSeparableSoliton}. It shows the transverse NMR frequency shift of soliton ($|\Delta{\theta}|=\pi-2\theta_{0}$) increases when $|q|$ increases,  while the transverse NMR frequency shift of big-soliton ($|\Delta{\theta}|=\pi+2\theta_{0}$) decreases when $|q|$ increases. The typical values of $\lambda$ of soliton and big-soliton are $\lambda \geq -0.7$ and $\lambda \leq -1.3$ respectively when $|q| \geq 0.14$. \\[0.1cm]

Because the unit cell of pseudo-random lattices of separable spin solitons with topological invariant $1/4+1/4$ contains KLS-soliton and soliton, the transverse NMR frequency shift of unit cell is determined by the equilibrium spin texture of soliton. As are result, $\lambda$ of pseudo-random lattices consisting of separable spin soliton with topological invariant $1/4+1/4$ is very close to those induced by soliton ($|\Delta{\theta}|=\pi-2\theta_{0}$).


\printbibliography[heading=bibintoc]



\end{document}